\documentclass[fleqn,usenatbib]{mnras}

\usepackage[T1]{fontenc}

\DeclareRobustCommand{\VAN}[3]{#2}
\let\VANthebibliography\thebibliography
\def\thebibliography{\DeclareRobustCommand{\VAN}[3]{##3}\VANthebibliography}

\usepackage{graphicx}	
\usepackage{amsmath}	
\usepackage{amssymb}	

\title[The origin of ETNOs within MOND]{On the origin of extreme trans-Neptunian objects within Modified Newtonian Dynamics}

\author[Cezary Migaszewski]{
Cezary Migaszewski$^{1}$\thanks{E-mail: cezary.migaszewski@gmail.com}
\\
$^{1}$Faculty of Physics, Astronomy and Informatics, Nicolaus Copernicus University, Grudzi\k{a}dzka 5, Toru\'n, Poland\\
}

\date{Accepted 2023 July 20. Received 2023 July 06; in original form 2023 March 23}

\pubyear{2022}

\usepackage{newtxtext,newtxmath}

\def\idm#1{{\mbox{\scriptsize #1}}}
\def\vec#1{{\pmb #1}}
\def\corr#1{#1}
\newcommand{\au}{\mbox{au}}

\newcommand{\mSun}{\mbox{m}_{\odot}}

\newcommand{\mE}{\mbox{m}_{\oplus}}

\newcommand{\MG}{M_{\idm{G}}}
\newcommand{\apert}{\vec{a}_{\idm{pert}}}
\newcommand{\atidal}{\vec{a}_{\idm{tidal}}}
\newcommand{\rpert}{\hat{\vec{r}}_{\idm{pert}}}
\newcommand{\Atidal}{A_{\idm{tidal}}}
\newcommand{\mpert}{m_{\idm{pert}}}
\newcommand{\distpert}{\Delta_{\idm{pert}}}

\newcommand{\psj}{The Planetary Science Journal}

\begin{document}
\label{firstpage}
\pagerange{\pageref{firstpage}--\pageref{lastpage}}
\maketitle

\begin{abstract}
In this work, we investigate the dynamical origin of extreme trans-Neptunian objects (ETNOs) under the action of the External Field Effect (EFE), which is a consequence of Modified Newtonian Dynamics (MOND) applied to gravity around the Sun embedded in the gravitational field of the Galaxy. We perform N-body integrations of known ETNOs treated as massless particles and perturbed by four giant planets and EFE. Backward integrations show that these objects originated in the giant planet region, from where they were scattered and then evolved to their current orbits. A striking example of such evolution is Sedna, which may have been temporarily in a horseshoe orbit with Jupiter and Saturn only $30$~Myr ago. Another interesting example is the newly discovered retrograde ETNOs, whose dynamical connection with prograde ETNOs and Centaurs is shown. The EFE is considered as an alternative to Planet Nine in explaining the anomalous distribution of ETNO orbits, namely the orbital plane clustering and apsidal confinement. We also analyse the effect of MOND on the obliquity of the solar spin with respect to the invariant plane of the solar system.
\end{abstract}




\section{Introduction}

The existence of an unobserved distant planet in the solar system, referred to as Planet Nine, has been proposed in response to the unexplained structure of the trans-Neptunian solar system. The anomalous features are the perihelion detachment of Sedna \citep{Brown2004} and other objects with perihelion distances $q$ beyond the gravitational influence of Neptune, highly inclined or even retrograde (the inclination $i > 90~$deg) orbits of Centaurs \citep{Batygin2016b}, the apsidal confinement as well as the clustering of orbital planes of ETNOs with $a \gtrsim 150\,\au$ and $q > 30\,\au$ \citep{Trujillo2014} or, based on better statistics, for $a \gtrsim 250\,\au$ and $q \gtrsim 42\,\au$ \citep{Batygin2016a,Brown2021}. Following the discovery of new objects, the apsidal confinement proved to be bimodal \citep{Sheppard2016}, with the two modes differing by $180~$degrees. For a thorough review of the Planet Nine hypothesis, see \citep{Batygin2019}.

These features, however, are not yet fully confirmed. While \cite{Brown2017} and \cite{Brown2019} argue that the probability that the clustering is due to random chance combined with observational bias is very low, \cite{Bernardinelli2020} and \cite{Napier2021} point out that there is no evidence for the clustering of the observed orbits and \cite{Clement2020} conclude that further new discoveries of ETNOs are needed to confirm the necessity of an additional planet. We do not attempt to settle the dispute in this article, and we treat the orbital plane clustering and the apsidal confinement as presumed anomalous features of the ETNO orbits. However, even if they are not real, two facts remain problematic for solar system formation scenarios, namely the existence of ETNOs with large $q$ and the high inclinations of both the Centaurs and the recently discovered retrograde ETNOs with $q > 50\,\au$ and $i > 160~$deg.

However, an additional planet on the periphery of the solar system is not the only possible explanation for the anomalous features of trans-Neptunian solar system. Modified Newtonian Dynamics \citep{Milgrom1983} was originally proposed as an alternative to dark matter to explain the flat rotation curves of galaxies. However, it turned out that MOND is able to explain or even predict other properties of galaxies. The proposed gravity model explains the Tully-Fisher law \citep{Milgrom1983b,Tully1977} with a power-law coefficient consistent with observations \citep{Milgrom1983b,TorresFlores2011}. It was also predicted that there would be a significant discrepancy between the dynamical mass and the luminous mass for galaxies with low surface brightness \citep{Milgrom1983b,deBlok1997}. It was also shown that the discrepancy between the centripetal acceleration and the gravitational acceleration caused by the luminous matter measured for many galaxies has regularities as expected from MOND \citep{Milgrom1983b,Lelli2017}.

The original formulation of MOND \citep{Milgrom1983} was further developed into a potential theory \citep{Bekenstein1984,Milgrom2010} that allows dynamical studies of disc galaxies. It was shown that MOND can solve the problem of the stability of disc galaxies \citep{Milgrom1989,Brada1999}. For a long time, MOND had a problem with the complete description of the motions of galaxies within galaxy clusters \citep{Sanders1999}, which brought it some criticism. However, recent studies \citep{LopezCorredoira2022} have shown that the discrepancy between luminous mass and virial mass in galaxy clusters can be explained with MOND if appropriate assumptions about hydrostatic equilibrium are used. For a thorough review of dark matter and MOND, see \citep{Sanders2010,Famaey2012}.

However, the effects of MOND are not limited to galactic and extra-galactic scales, they are important close to the Sun as well. The External Field Effect \citep{Milgrom1983} appears when a massive body, e.g., Sun, is embedded in an external gravitational field, e.g., the field of the Galaxy. It was shown in \citep{Milgrom2009} that the effect can play a role in the Solar system even in the giant planets region. Even more significant is the effect in the trans-Neptunian region. \cite{Pauco2016} considered formation of ETNOs as well as highly inclined Centaurs within EFE and found that such objects can be naturally formed. They also demonstrated that both the groups belong to the same trans-Neptunian objects, TNO\footnote{We use the term TNOs for objects with $a > 30\,\au$, with no criterion for $q$.}, population, but in different stages of their evolution.

The other two anomalous features of the ETNOs proved more difficult to explain in the context of EFE. \cite{Pauco2017} showed that the reconstruction of the apsidal confinement as well as the clustering of the orbital planes would require an improbable value of the EFE quadrupole strength parameter, $Q$, with respect to the Cassini data \citep{Blanchet2011,Hees2014,Hees2016}.

In this paper we reconsider the possible explanation of the ETNO characteristics within EFE. Our approach differs in some respects from that presented in \citep{Pauco2017}. We compute the EFE perturbing acceleration by directly solving the modified Poisson equation \citep{Milgrom2010} instead of using an analytical model as in \citep{Pauco2017}. We include all giant planets in the $N$-body model of motion, not just Neptune as in \citep{Pauco2017}. We have mainly focused on known objects rather than synthetic populations.

The reason for the latter is that a comparison between the observed and the synthetic distribution of the orbital elements is problematic. The first difficulty is that the final synthetic distribution depends on the initial distribution, which is not known. The second difficulty is the epoch of comparison. Since the EFE-induced acceleration depends on the position of the Galaxy centre in the ecliptic reference frame, one should compare the synthetic distribution with the observations only in the vicinity of the current phase of the solar motion in the Galaxy. For example, it is possible that the orbits are not clustered around a fixed plane, but around the plane whose orientation changes as the Sun orbits around the centre of the Galaxy.

Through forward and backward integrations of known Centaurs and ETNOs, we have shown that the transition between the two groups of objects occurs on typical timescales of tens to hundreds of Myrs, but can even be as short as a few Myrs for the widest orbits. Their evolution is chaotic due to close encounters with all giant planets. The simulations show that most of these objects in the past had semi-major axes $a$ below $\sim 30\,\au$ as well as moderate eccentricities $e$ and inclinations $i$. This suggests that the ETNOs and Centaurs originate from the unstable region of the solar system, with their initial orbits lying between the orbits of the giant planets. The objects may also return to this region from time to time.

In addition, we investigate the possible origin of the three newly discovered retrograde ETNOs (2022~FN12, 2022~FM12, 2019~EE6) whose $a \sim 150\,\au$, $q > 50\,\au$ and $i > 160\,$deg within EFE, showing that they had lower inclinations as well as $q < 30\,\au$ in the past. This shows that they belong to the same population as Centaurs and prograde ETNOs.

The paper is structured as follows. In Section~\ref{sec:efe} we compute the EFE-induced acceleration for the Sun embedded in the gravitational field of the Galaxy. In Section~\ref{sec:simple_model} we analyse the evolution of the test particles in a simple model without the planetary perturbations and with a fixed direction towards the centre of the Galaxy. In Section~\ref{sec:kuiper} we discuss the distribution of orbital elements of TNOs in the context of the Planet Nine hypothesis. In Section~\ref{sec:known_objects} we introduce the $N$-body model of motion and examine the evolution of the known TNOs. In Section~\ref{sec:clustering} we analyse the clustering of orbits and apsidal confinement of ETNOs within EFE. Section~\ref{sec:discussion} is devoted to the discussion of the possible existence of Planet Nine within MOND, the influence of EFE on the obliquity of the solar spin with respect to the invariant plane of the solar system, and possible implications of the results presented in this paper on the dark matter problem. The conclusions can be found in Section~\ref{sec:conclusions}.

\section {External Field Effect for the Sun--Galaxy system}
\label{sec:efe}

We use a quasi-linear formulation of MOND \citep{Milgrom2010} in which the modified gravitational potential $\phi$ is given by the solution of the equation:
\begin{equation}
\nabla^2 \phi = \nabla \cdot \big[ \nu\left( |\nabla \phi_N|/a_0 \right) \nabla \phi_N \big],
\label{eq:Poisson}
\end{equation}
where $\phi_N$ is a Newtonian potential given by the standard Poisson equation $\nabla^2 \phi_N = 4 \pi G \rho$, where $\rho$ denotes the baryonic mass density. The above equation is linear in $\phi$ and can be solved in the same way as the standard equation.

The transition between the regime of high acceleration (Newtonian) and the regime of low acceleration (Milgromian) is described by a function $\nu$ whose particular form is not given by the theory. There are several different classes of functions \citep[see e.g.][]{Famaey2012}. In our analysis, the following function is used
\begin{equation}
\nu(\tilde{y}) = \left( \frac{1 + \sqrt{1 + 4 \tilde{y}^{-\alpha}}}{2} \right)^{1/\alpha},
\label{eq:nu_function}
\end{equation}
where $\tilde{y} \equiv |\nabla \phi_N|/a_0$ and $a_0$ is a fundamental acceleration of the order of $\sim 10^{-10}\,$m/s$^2$. Both the value of $a_0$ and the coefficient $\alpha$ must be determined from observations. We use $a_0 = 1.2 \times 10^{-10}\,$m/s$^2$ \citep{McGaugh2016} as well as different values of $\alpha = 1.5, 2$ or $2.5$. The coefficient $\alpha$ determines how smooth the transition between the Newtonian and Milgromian regimes is. For smaller $\alpha$, the transition is smoother.

\corr{\cite{Blanchet2011} constrained the MOND transition function by analysing the solar system data and for the class of functions defined in Eq.~\ref{eq:nu_function} they found that $\alpha \geq 8$ in order to fit the precession rates of the planets. Similarly, \cite{Hees2016} constrained the parameter to be $>6$. For the "standard" value of $\alpha = 2$, the expected EFE-induced precession rates make only $0.1-5\,$ppm (it is larger for the outer planets) of the precession rates caused by the Newtonian planet-planet perturbations. The parameters of the solar system are, however, known with significantly worse precision. The relative standard deviation of the planets' masses $\sim 46\,$ppm (slightly more for Mercury, $\sim 64\,$ppm)\footnote{The mass uncertainties were taken from the JPL database, accessed 6 July 2023}. The EFE-induced precession can be then balanced by changing the planets' masses within the uncertainties. Due to the difficulties in determining the $\alpha$ parameter, we use the "standard" value of $2$ and two other close values, $1.5$ and $2.5$.}

Although MOND was formulated to model the rotation curves of galaxies, its consequences extend to smaller scales. When a massive body is embedded in an external gravitational field, e.g. the Sun in the Galaxy's field, the so-called External Field Effect appears \citep[EFE,][]{Milgrom2009}. It stems from the fact that between the Sun and the Galaxy centre there is a point where the resulting Newtonian gravitational acceleration disappears and in a certain area around it the Newtonian acceleration $g_N \equiv |\nabla \phi_N| \lesssim a_0$. Solving Eq.~\ref{eq:Poisson} for such a system, it turns out that the gravitational field is modified not only in the region where $g_N \lesssim a_0$, but also near the Sun where $g_N \gg a_0$. The perturbing acceleration has the form of a quadrupole field in the vicinity of the Sun \citep{Milgrom2009}.

In this work we solve Eq.~\ref{eq:Poisson} without the assumption of a constant external field and beyond the limit $g_N \gg a_0$, which would allow us to follow the evolution of trans-Neptunian objects thousands of astronomical units away from the Sun. We treat the field of the Galaxy as a point-source field. There are numerous estimates for the distance and rotational velocity of the Sun with respect to the centre of the Galaxy. We use one of the most recent estimates from \citep{Hunt2016}, in which the velocity $V_{\odot} = 239 \pm 9\,$km/s and the distance $R_{\odot} = 7.9 \pm 0.3\,$kpc.

The rotational velocity and distance can be translated into the mass of the Galaxy inside $R_{\odot}$, denoted $\MG$. Since the centripetal acceleration of the Sun $a_{\idm{cp}} \equiv V_{\odot}^2/R_{\odot} \approx 2.3 \times 10^{-10}\,$m/s$^2$, which is only $\approx 2 a_0$, we need MOND prescription to find $\MG$, thus
\[
\frac{G \MG}{R_{\odot}^2} = g_N = \mu(\tilde{x}) g = \mu(\tilde{x}) a_{\idm{c}},
\]
where $\mu(\tilde{x})$ is a transition function of MOND\footnote{There are two equivalent formulations of MOND, $\mu(\tilde{x}) g = g_N$ and $g = \nu(\tilde{y}) g_N$, therefore there are two equivalent transition functions $\mu(g/a_0)$ and $\nu(g_N/a_0)$.} where $\tilde{x} \equiv g/a_0$ \citep{Milgrom1983} of the form
\[
\mu(\tilde{x}) = \frac{\tilde{x}}{\left( 1 + \tilde{x}^{\alpha} \right)^{1/\alpha}}.
\]
The value of $\mu$ depends on $\alpha$ and for $\tilde{x} \approx 2$ and $\alpha = 2$ this gives $\mu \approx 0.89$, resulting in $\MG \approx 9.34 \times 10^{10}\,\mSun$, while for other $\alpha$ the mass may be slightly larger or smaller. Naturally, different $R_{\odot}$ and $V_{\odot}$ would give different $\MG$. We have tested \corr{various} values of these quantities and found that they are not crucial for the \corr{overall} dynamics of trans-Neptunian objects, \corr{although, the evolution of a particular object would change in detail for different $R_{\odot}$ and $V_{\odot}$.}

Another aspect of the Galaxy's gravitational field must be considered. The Sun does not move in an exactly circular and planar orbit. Its velocity has non-zero components both in the radial direction and in the direction perpendicular to the Galaxy disc. An analysis of the periodicity of the comet flux of the Oort cloud \citep{Matese1995} shows that the period of the radial motion $T_R \sim 170\,$Myr and an amplitude of the radial oscillations $\sim 0.5\,$kpc. The variation of $g_N$ with an amplitude of $\sim 14\,$per~cent is obtained. The vertical oscillation occurs with a period of $70\,$Myr. Since the current perpendicular velocity $V_z \sim 7\,$km/s \citep{Karim2017}, the maximum deviation of the Sun above or below the midplane of the Galaxy disc, $z_{\idm{max}} \sim 100\,$pc. Assuming harmonic oscillations, the maximum acceleration perpendicular to the disc is $g_{z,\idm{max}} \sim 2.5 \times 10^{-11}\,$m/s$^2$, which is an order of magnitude weaker than the centripetal acceleration. The maximum angle by which the total acceleration deviates from the radial direction is $\sim 7\,$degrees, while in the current epoch it is $\sim 1\,$degree.

A more realistic model of the Galaxy should also take into account the dependence of $\MG$ on $R_{\odot}$, which changes with time. Because of the uncertainties in determining $R_{\odot}$ and $V_{\odot}$, we omit all these corrections from our analysis and treat the Galaxy's potential as a point-mass potential and the Sun as moving in a circular orbit around the centre of the Galaxy.

If the masses $\mSun$ and $\MG$ are known, the total potential at the position $\vec{r}$ is given by
\begin{equation}
\phi_N(\vec{r}) = -\frac{G \mSun}{|\vec{r} - \vec{r}_{\odot}|} - \frac{G \MG}{|\vec{r} - \vec{r}_{\idm{G}}|},
\label{eq:potential}
\end{equation}
where $\vec{r}_{\odot}$ and $\vec{r}_{\idm{G}}$ are the position vectors of the Sun and the Galactic centre. It can be used to calculate the right-hand side of Eq.~\ref{eq:Poisson}. Since $\phi_N$ has axial symmetry, the potential $\phi$ we are looking for has the same symmetry. We can therefore use a cylindrical coordinate system whose origin is in the Sun. One of the axes, $z$, points towards the centre of the Galaxy, the second, $x$, is perpendicular to $z$. We solve Eq.~\ref{eq:Poisson} numerically using the standard five-point method \citep[e.g.][]{Hoffman2001}. The grid is non-uniform, the size of the grid cells ranging from $\sim 2.6\,\au$ near the Sun and the critical point (defined by $g_N = 0$) to $\sim 670\,\au$ in the outer parts of the domain. The size of the area is $x \in [0, 90]\,$kau and $z \in [-90, +90]\,$kau. The boundary condition at $x = 0$ results from the cylindrical symmetry, i.e.,
\[
\frac{\partial \phi}{\partial x} \bigg|_{x = 0} = 0,
\]
while at the other three limits the potential corresponds to the Milgromian potential of the point-mass Galaxy alone.

\begin{figure*}
\centerline{
\vbox{
\hbox{
\includegraphics[width=0.49\textwidth]{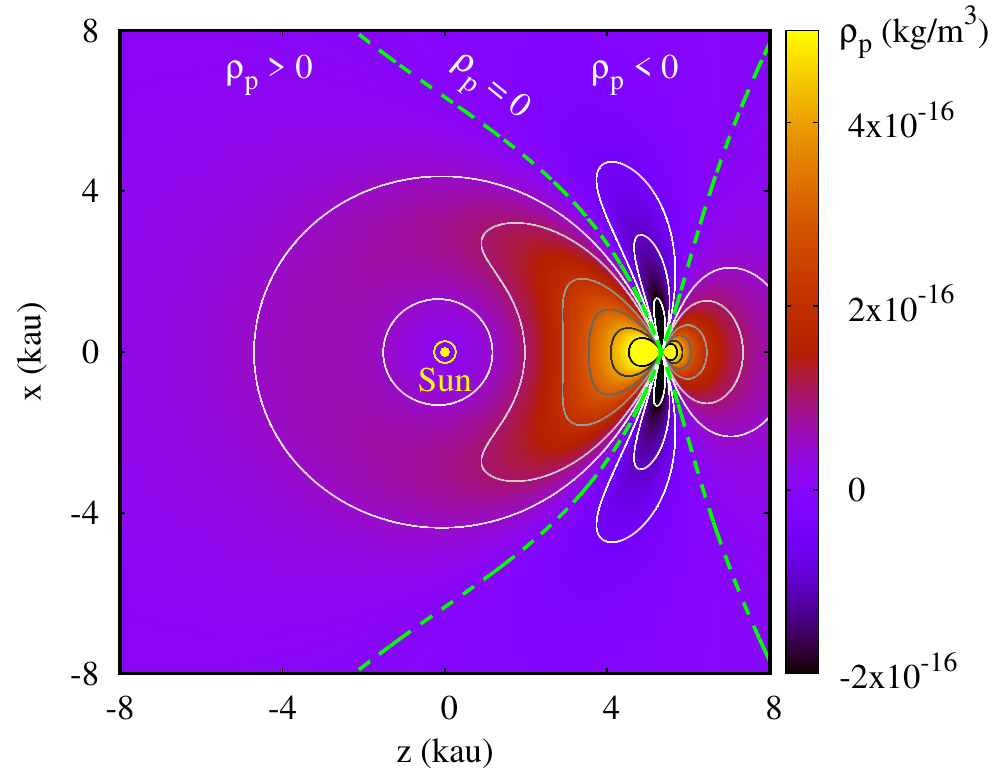}
\includegraphics[width=0.49\textwidth]{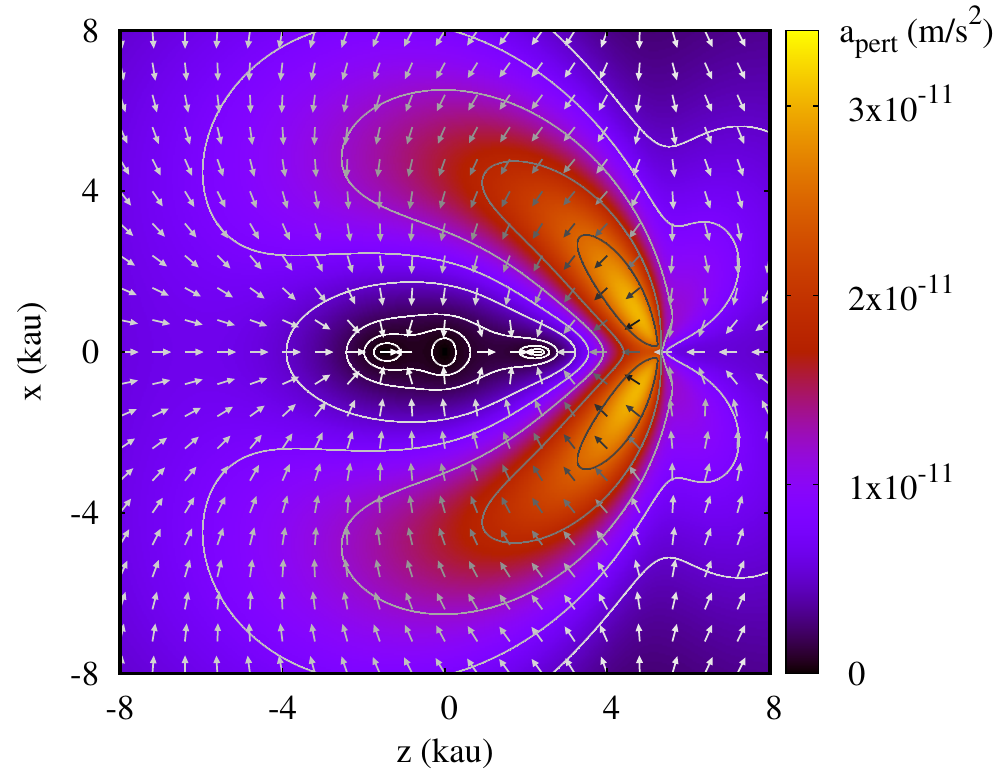}
}
\hbox{
\includegraphics[width=0.49\textwidth]{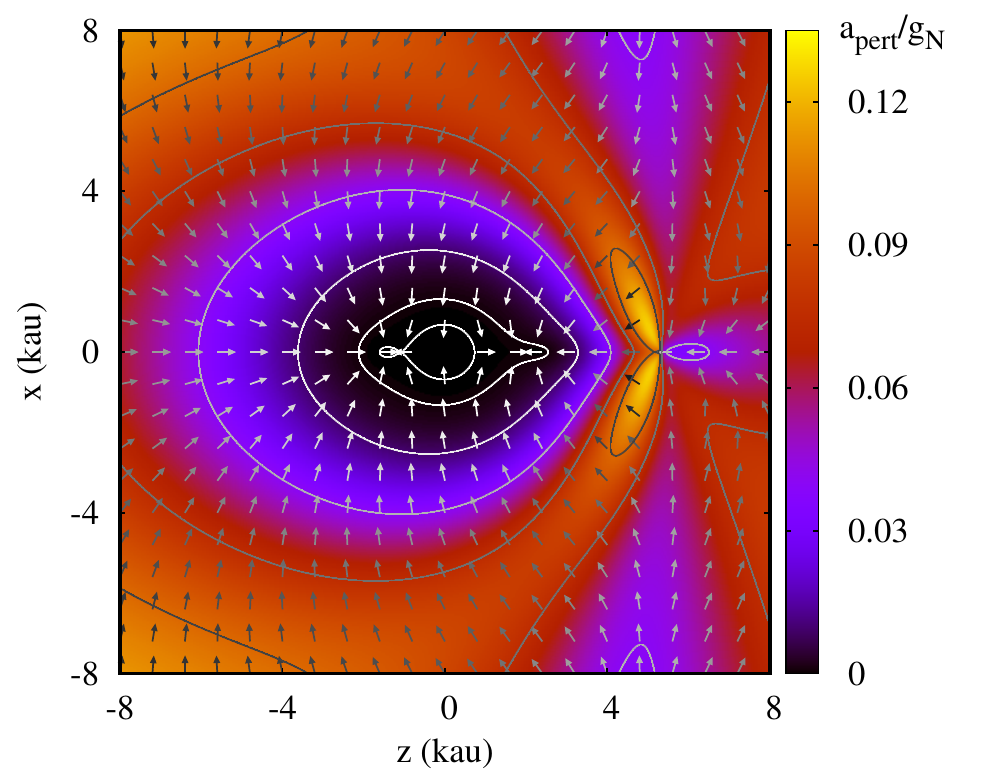}
\includegraphics[width=0.49\textwidth]{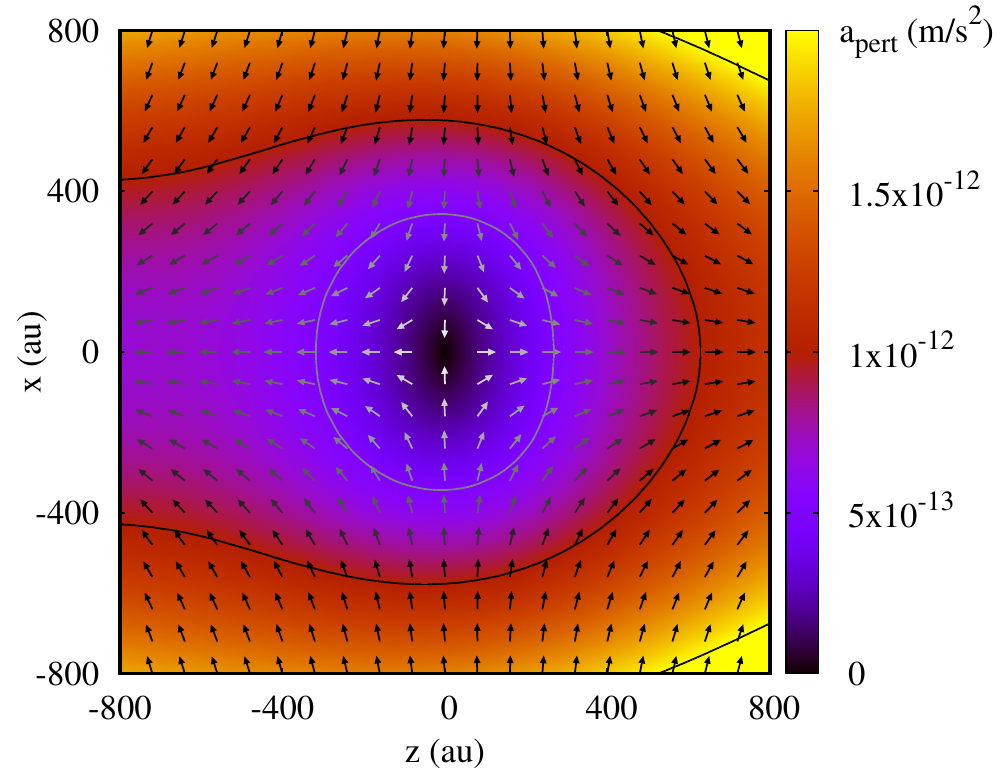}
}
}
}
\caption{Top-left: Phantom mass density as a function of position around the Sun (colour-coded). The solid curves denote the contours of constant densities. The dashed green curves denote zero density. Top-right: a vector field of perturbative acceleration (arrows). Its magnitude is colour-coded (with contours for constant values). Bottom-left: the vector field of perturbative acceleration (arrows) and the magnitude of the perturbative acceleration relative to Newtonian acceleration from the Sun (colour-coded). Bottom-right: the same as top-right, but for a smaller area around the Sun. The transition function coefficient $\alpha = 2$.}
\label{fig:Fig1}
\end{figure*}

Figure~\ref{fig:Fig1} shows the solution of Eq.~\ref{eq:Poisson} for the transition function coefficient $\alpha=2$. The so-called phantom mass density is shown in the upper left panel. With respect to the dark matter hypothesis, the modification of the standard Poisson equation, Eq~\ref{eq:Poisson}, can be interpreted as the addition of dark matter to the right-hand side of the equation, i.e.,
\begin{equation}
\nabla^2 \phi = 4 \pi G \left( \rho + \rho_p \right) = \nabla \cdot \big[ \nu\left( |\nabla \phi_N|/a_0 \right) \nabla \phi_N \big],
\label{eq:phantom}
\end{equation}
where $\rho_p$ would be the density of dark matter. In terms of MOND, the additional mass is called phantom dark matter, whose distribution is determined by the baryonic mass distribution according to Eq.~\ref{eq:phantom} \citep{Milgrom2010}.

The phantom mass density can be positive or negative, while it disappears at the position of the Sun \citep[which agrees with the analytical results in][]{Milgrom2009}, as well as at the axially symmetric surface marked with green dashed curves. The remaining three panels of Fig.~\ref{fig:Fig1} show the perturbative acceleration $\apert$ given by the partial derivatives of $\phi$ (computed numerically), completed with the centrifugal acceleration. The latter results from the fact that the reference frame is not inertial as it relates to the Sun orbiting the centre of the Galaxy. The addition of the centrifugal acceleration is equivalent to the substruction of the Milgromian gravitational acceleration of the Sun by the Galaxy. The total acceleration of a test particle with respect to the Sun is then given by
\begin{equation}
\ddot{\vec{r}} = -\frac{G \mSun}{r^3} \vec{r} + \apert
\label{eq:acc}
\end{equation}
where the components of $\apert = (a_x, a_y, a_z)$ read\footnote{Note that $(x,y,z)$ are the cylindrical coordinates usually denoted $(\rho, \phi, z)$ and should not be confused with the Cartesian coordinates. The use of such a non-standard notation is due to the fact that in this work $\rho$ and $\phi$ are used for density and potential, respectively.}
\[
a_x = -\frac{\partial \phi}{\partial x}, \quad
a_y = 0, \quad
a_z = -\frac{\partial \phi}{\partial z} - \frac{V_{\odot}^2}{R_{\odot}}.
\]

The acceleration $\apert$ disappears at the Sun and at two other points on the $z$-axis. It is strongest in the lobes extending around the critical point, where its magnitude reaches $3 \times 10^{-11}\,$m/s$^2$. The lower left panel shows the magnitude of the acceleration relative to the Newtonian acceleration. In the region of $\sim 5000\,\au$ the perturbation can be up to $10~$percent of $g_N$, which means that the elongated orbits of the semi-major axes $\sim 2500\,\au$ are strongly perturbed. \corr{For even wider orbits, $a \gtrsim 10\,$kau, the relative perturbation can be as high as $20~$percent (this region is not shown). The dynamics of the Oort cloud is thus supposed to be significantly affected by EFE \citep{Iorio2010,Pauco2017b}.}

The bottom-right panel shows the perturbative acceleration in the smaller region around the Sun. The vector field shows that the perturbing force is acting along the $x$ axis towards the Sun and along the $z$ axis away from it. It has the form of a quadrupole field, as it should be for EFE near the Sun \citep{Milgrom2009}. However, in this plot both EFE and the enhanced gravity effect (EGE; the effect of MOND for an isolated Sun) are shown together. The latter effect is radially symmetric and acts in the direction of the Sun.

\begin{figure*}
\centerline{
\vbox{
\hbox{
\includegraphics[width=0.49\textwidth]{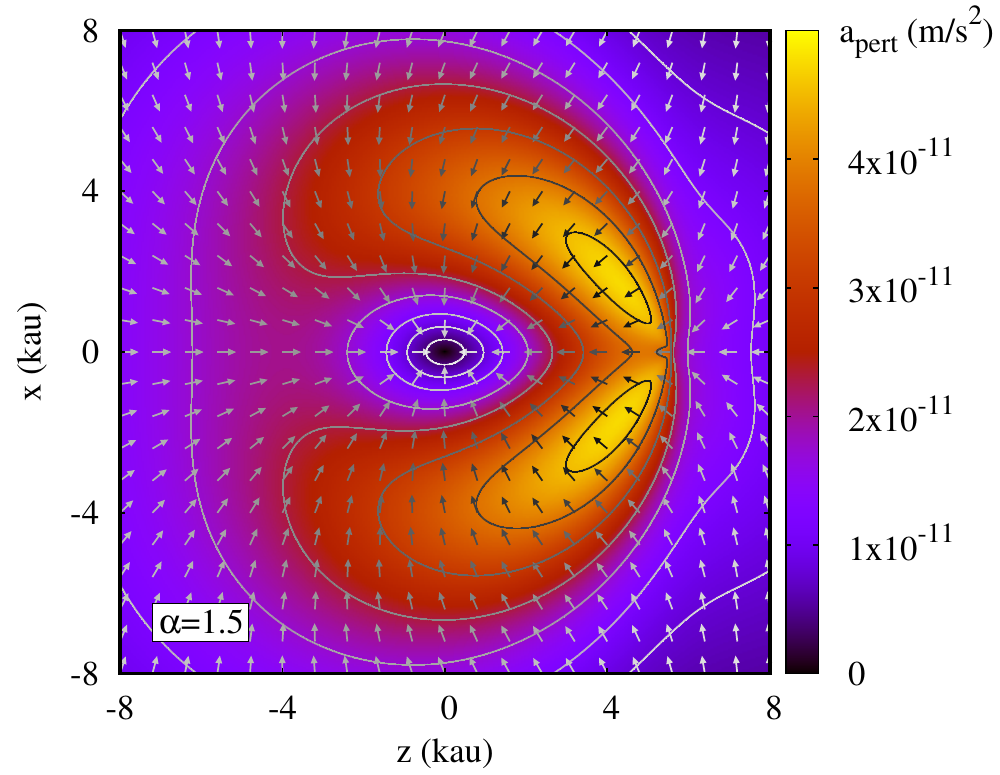}
\includegraphics[width=0.49\textwidth]{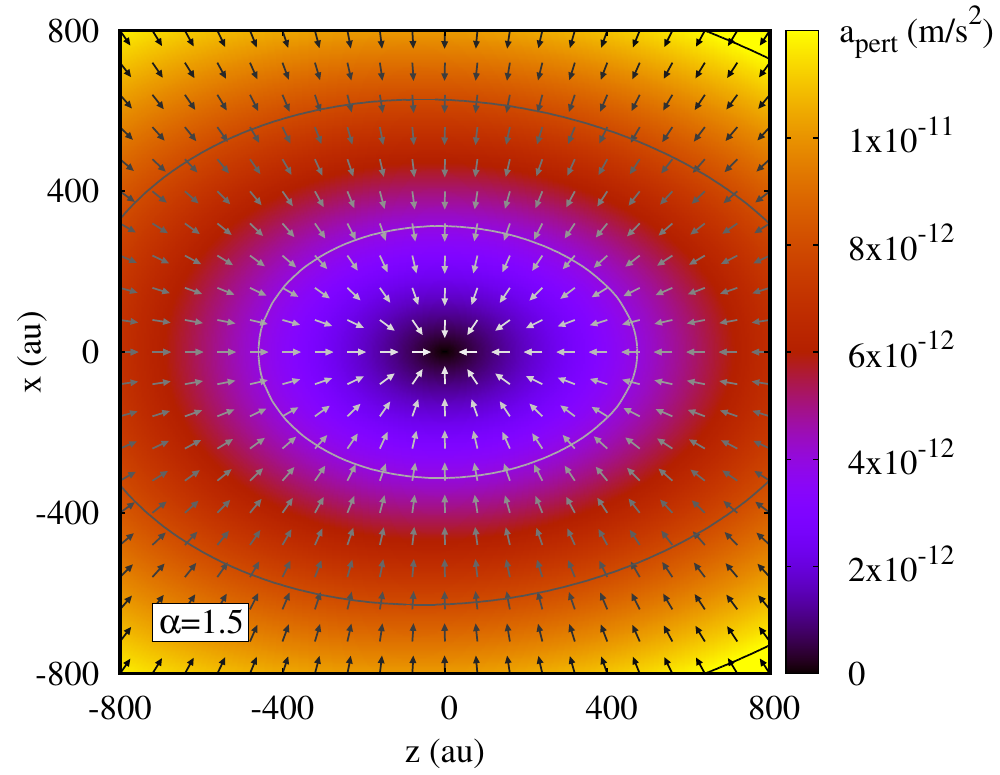}
}
\hbox{
\includegraphics[width=0.49\textwidth]{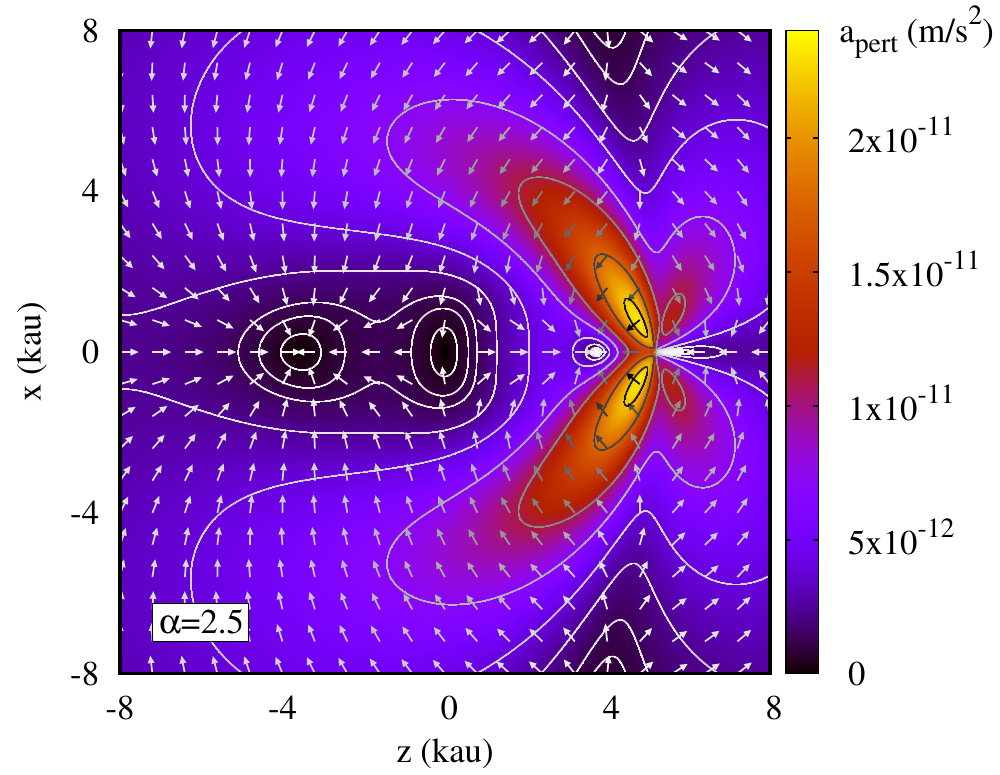}
\includegraphics[width=0.49\textwidth]{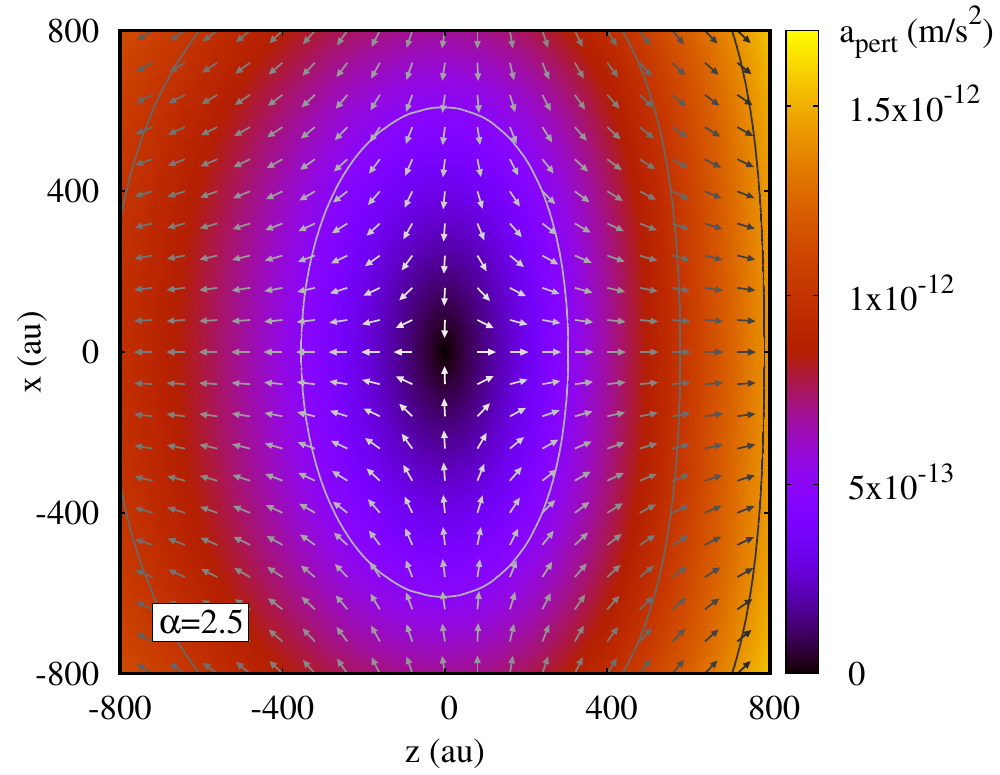}
}
}
}
\caption{Vector fields of perturbative acceleration, shown in the same way as in the right-hand panels of Fig.~\ref{fig:Fig1}. Each row shows the results obtained for different $\alpha = 1.5$ (top) and $2.5$ (bottom). The right column shows a close-up of the region around the Sun.}
\label{fig:Fig2}
\end{figure*}

For different values of $\alpha$, the perturbative acceleration differs both near the Sun and farther from it. Figure~\ref{fig:Fig2} shows the results for $\alpha = 1.5$ (top row) and $\alpha = 2.5$ (bottom row) at two different scales. For $\alpha = 1.5$ the perturbation is generally stronger, but near the Sun it acts as an additional gravitational pull regardless of direction. This means that the EGE dominates the EFE. For $\alpha = 2.5$ the perturbation is weaker and the picture is more complex. Similar to the case of $\alpha = 2$, EFE is more important than EGE.

Near the Sun, EFE has the same functional form as tidal acceleration at position $\vec{r}$ by a distant perturber, i.e.,
\[
\atidal = -G \Atidal \left[ \vec{r} - 3\left( \vec{r} \cdot \rpert\right) \rpert \right],
\]
where $\Atidal$ is the tidal parameter that can be expressed with the mass $\mpert$ and the distance $\distpert$ of the perturber as $\Atidal = \mpert/\distpert^3$. The unit vector pointing from the Sun to the perturber is denoted by $\rpert$. Within the EFE, $\rpert$ would point towards the centre or anticentre of the Galaxy. Using the numerically solved modified Poisson equation, for $\alpha = 2$ we can find the tidal parameter $\Atidal = 1.77 \times 10^{-13}\,\mSun/\au^3$, while for $\alpha = 2.5$ $\Atidal = 1.30 \times 10^{-13}\,\mSun/\au^3$. As mentioned earlier, for $\alpha = 1.5$ the EGE dominates the EFE, hence the total perturbation deviates from the tidal form.

\cite{Holman2016} analysed the astrometry of Pluto and a few TNOs to constrain the tidal parameter due to the hypothetical Planet Nine and found that the upper limit of $\Atidal$ is between $10^{-12}$ and $10^{-10}\,\mSun/\au^3$, depending on the position of the perturber in the sky. Below this limit, the perturbation cannot be detected based on the currently available observations. Both values of the EFE--induced $\Atidal$ are well below the upper limit.

The values can be expressed in terms of $\mpert$ and $\distpert$. Taking a representative mass of Planet Nine, $\mpert = 10\,\mE$ \citep{Batygin2016a}, we find that the values of $\Atidal = 1.77 \times 10^{-13}\,\mSun/\au^3$ and $\Atidal = 1.30 \times 10^{-13}\,\mSun/\au^3$ can be obtained for $\distpert = 554\,\au$ and $\distpert = 614\,\au$, respectively. For smaller $\mpert = 6.2\,\mE$ \citep{Brown2019} the distances are $472\,\au$ and $524\,\au$ respectively. All distances are consistent with the current constraints for Planet Nine's orbit. According to \citep{Brown2019}, the most likely position of P9 in the sky is near the anticentre of the Galaxy and the most likely distance is $\sim 400-500\,\au$.

While the long-term dynamics of TNOs within the two hypotheses, i.e. Planet Nine and MOND, differ from each other, when modelling an additional unexplained acceleration in the outer parts of the solar system, Milgrom's gravity corrections cannot be distinguished from the acceleration due to Planet Nine. \cite{Fienga2016} analysed the {\em Cassini} radio ranging data to constrain the true anomaly of Planet Nine, assuming the orbital elements according to \citep{Batygin2016a}. They found that for the true anomaly $\nu \in [108, 129]\,$deg the {\em Cassini} residuals are reduced. This range of $\nu$ corresponds to the direction being shifted by $50-60\,$deg with respect to the direction towards the centre of the Galaxy. The latter lies within the uncertainty zone for which the perturbative acceleration is below the detection limit. \corr{\cite{Iorio2017} analysed the perturbation of the Saturn's orbit inferred from the {\em Cassini} data and found that the true anomaly of Planet Nine could be constrained to $\nu \in [130, 240]\,$deg, i.e. close to the aphelion, which itself lies close to the anticentre of the Galaxy.} \cite{Holman2016b} has relaxed constraints on the mass and orbit of Planet Nine in their analysis of the {\em Cassini} data. It turns out that the direction towards the centre of the Galaxy is on the boundary between decreasing and increasing residuals. Therefore, the MOND perturbation of gravity cannot be ruled out by the {\em Cassini} data.

\section{A secular model of isolated and fixed EFE}
\label{sec:simple_model}

Once the modified Poisson equation is solved for given parameters of the Galaxy, the perturbative acceleration for any $(z,x)$ point can be determined by interpolation between the grid nodes. Before turning to the $N$-body integrations of an asteroid under the combined effect of Newtonian acceleration due to the Sun, the Galaxy and the giant planets, as well as the MOND perturbation, in this section we examine the secular dynamics of an asteroid without the gravitational attraction of the planets and assuming that the position of the Galaxy centre, which generally varies in time due to the motion of the Sun around the Galaxy centre, is fixed in the ecliptic reference frame.

\begin{figure*}
\centerline{
\hbox{
\includegraphics[width=0.32\textwidth]{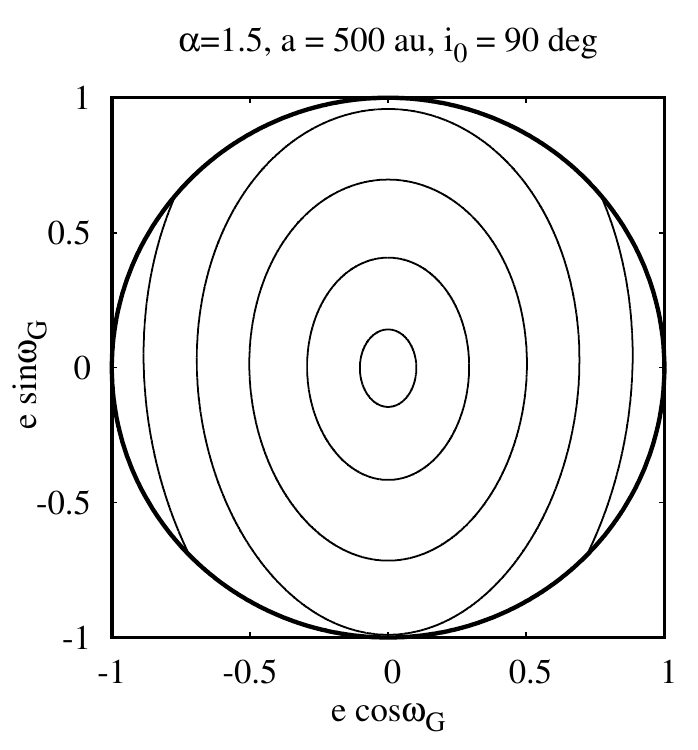}
\includegraphics[width=0.32\textwidth]{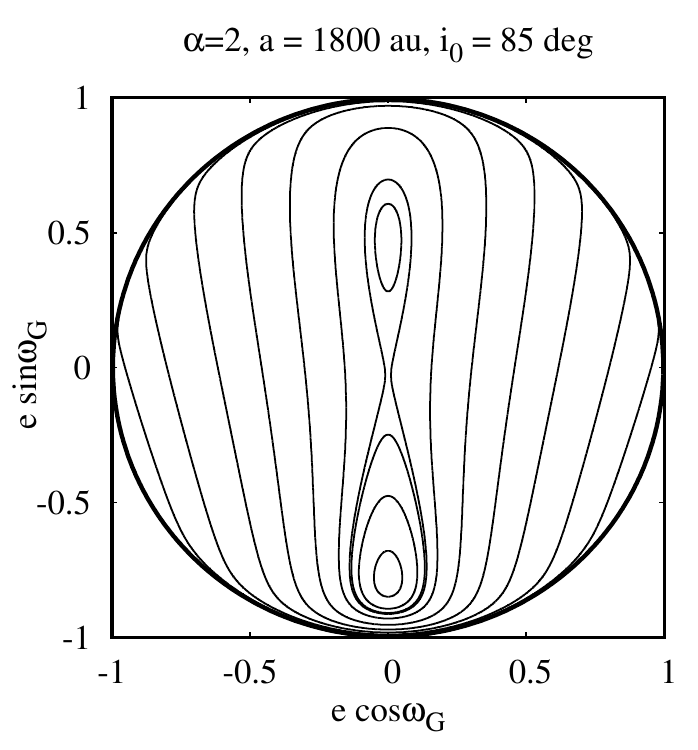}
\includegraphics[width=0.32\textwidth]{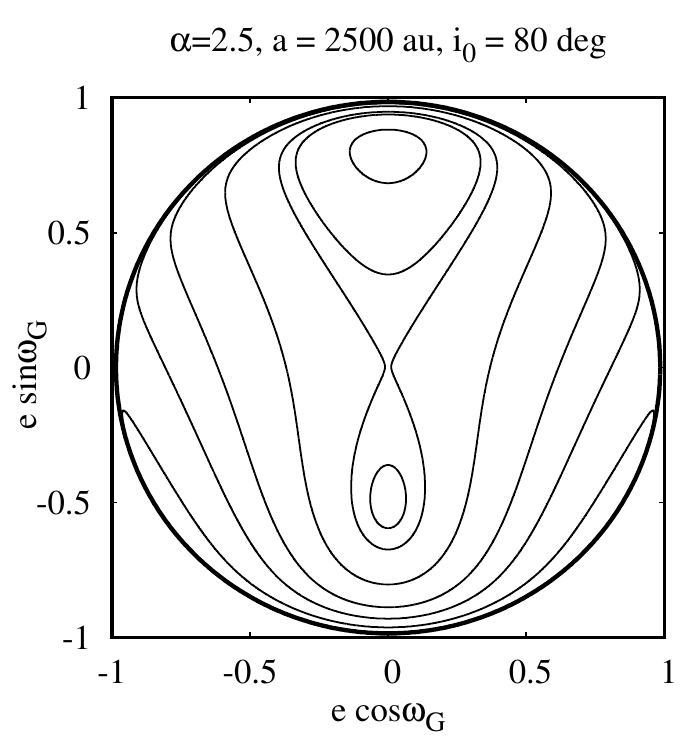}
}
}
\caption{Contours of the constant secular Hamiltonian shown in the $(e \cos\omega_G, e \sin\omega_G)$--diagram.}
\label{fig:Fig3}
\end{figure*}

The Hamiltonian of a test particle under the MOND effect of the Sun and the Galaxy is as follows
\begin{equation}
\mathcal{H} = \mathcal{H}_1 + \mathcal{H}_2,
\label{eq:ham}
\end{equation}
where $\mathcal{H}_1$ is the Keplerian part, while $\mathcal{H}_2$ is the perturbative Hamiltonian. The perturbative Hamiltonian is equal to
\begin{equation}
\mathcal{H}_2 = \phi + \phi_{\idm{cf}} + \mbox{const},
\label{eq:Hpert}
\end{equation}
where $\phi$ is the solution of Eq.~\ref{eq:Poisson} and $\phi_{\idm{cf}}$ is the centrifugal potential, which is as follows
\[
\phi_{\idm{cf}} = \frac{V_{\odot}^2}{R_{\odot}} z.
\]
The constant value is added to the Hamiltonian in Eq.~\ref{eq:Hpert} so that $\mathcal{H}_2 = 0$ at the Sun.

The canonical Delaunay angle-action variables are defined as \citep[e.g.][]{Brouwer1961}
\begin{eqnarray}
\varphi_1 = \mathcal{M}, && I_1 = \sqrt{G \mSun a},\nonumber\\
\varphi_2 = \omega, && I_2 = I_1 \sqrt{1 - e^2},\nonumber\\
\varphi_3 = \Omega, && I_3 = I_2 \cos i,\nonumber
\end{eqnarray}
where $\mathcal{M}, \omega, \Omega$ are the mean anomaly, the argument of the perihelion and the longitude of the ascending node, respectively, while $a, e, i$ denote the semi-major axis, the eccentricity and the inclination of the orbit. The Keplerian Hamiltonian depends only on $I_1$, i.e.,
\[
\mathcal{H}_1 = -\frac{\left( G \mSun \right)^2}{2 I_1^2}.
\]
The perturbing Hamiltonian generally depends on all variables. However, because of the axial symmetry of the perturbation, it is useful to choose the $z$-axis of the reference frame to coincide with the axis of symmetry (also denoted $z$). With such a choice, $\mathcal{H}_2$ does not depend on $\varphi_3$, so $I_3$, the projection of angular momentum onto the $z$-axis, is the integral of motion. The Hamiltonian therefore has two degrees of freedom. To further simplify the Hamiltonian, the method of averaging \citep[e.g.][]{Arnold2006} is used. The perturbation $\mathcal{H}_2$ is much smaller than the Keplerian part, so we can average the Hamiltonian over the mean anomaly, which is a fast variable
\[
\langle \mathcal{H} \rangle = \frac{1}{2\pi} \int_0^{2\pi} \mathcal{H} \mathrm{d}\mathcal{M}.
\]
Averaging over the mean anomaly can be replaced by averaging over the eccentric anomaly $E$, using the Kepler equation. We obtain
\begin{equation}
\langle \mathcal{H} \rangle = \frac{1}{2\pi} \int_0^{2\pi} \mathcal{H} \left(1 - e \cos E \right) \mathrm{d}E.
\label{eq:meanH}
\end{equation}

After the averaging, $\langle \mathcal{H} \rangle$ does not depend on $\varphi_1$, thus $I_1$ is an integral of motion. The secular Hamiltonian of a test particle thus has one degree of freedom and is integrable. The averaging is done numerically using the $100$-th order Gauss-Legendre quadrature. For a given position of a test particle in its Keplerian orbit, the value of $\mathcal{H}_2$ ($\mathcal{H}_1$ does not need to be averaged as it is constant after averaging) is determined using bicubic interpolation on the grid nodes.

Once the integrals $I_1$ and $I_3$ are set, the secular Hamiltonian depends on $(\varphi_2, I_2)$, which translates into the dependence on the argument of pericentre and eccentricity. Since the reference frame whose $z$-axis points towards the centre of the Galaxy (the G-frame from now on\footnote{The G-frame should not be confused with the galactic reference frame used in astrometry, whose $x$-axis points towards the centre of the Galaxy.}) differs from the ecliptic reference frame normally used to describe the orbits of planets and asteroids in the solar system, the argument of pericentre and the inclination are referred to as $\omega_G$ and $i_G$ respectively when expressed in the reference frame related to the Galaxy.

Since the secular Hamiltonian has one degree of freedom, it is sufficient to construct a phase diagram using contours with constant values of $\langle \mathcal{H} \rangle$. Figure~\ref{fig:Fig3} shows such contours for different $I_1$ and $I_3$, which can be translated into $a$ and $i_0$, the latter being the inclination of a circular orbit, which gives $I_3 = I_1 \cos i_0$. Each panel was obtained for different $\alpha$. The plots show representative phase diagrams. The dynamics of the system from the left panel is simple. There are oscillations of $e$ accompanied by rotations of $\omega_G$. There is a stable equilibrium near the origin of the diagram. In the middle panel the dynamics is more complex as there are two stable equilibria and one unstable equilibrium. The diagram is asymmetric, i.e. the equilibrium for $\omega_G = \pi/2$ occurs for lower $e$ than that for $\omega_G = -\pi/2$. In the right-hand panel, the situation is reversed. Due to the axial symmetry of the problem, the equilibria only exist for $\omega_G = \pm \pi/2$.

\begin{figure*}
\centerline{
\includegraphics[width=0.99\textwidth]{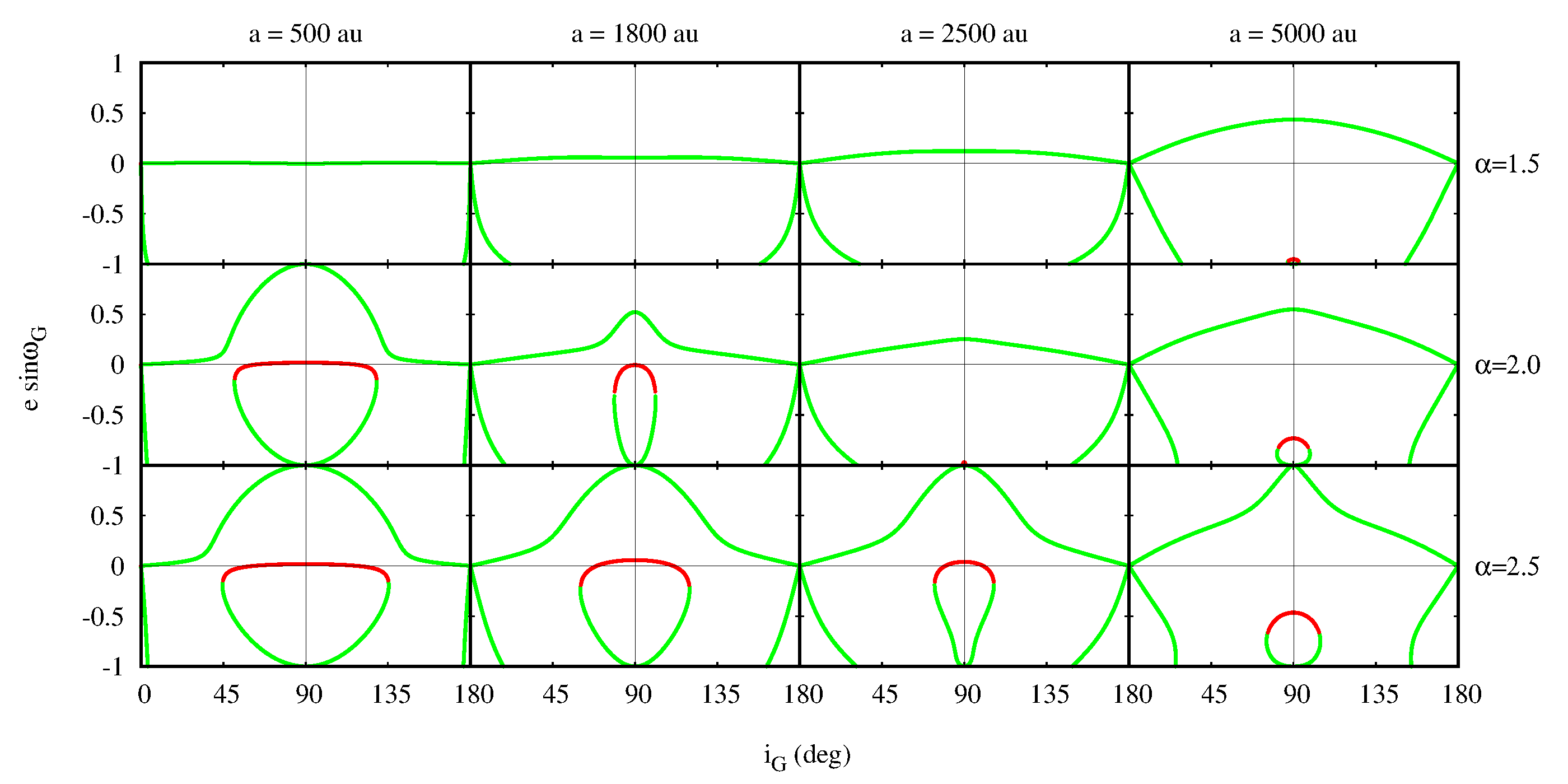}
}
\caption{Secular equilibria: stable (green) and unstable (red). Each column corresponds to different $a = 500, 1800, 2500, 5000\,\au$ (from left to right), while each row represents different $\alpha = 1.5, 2, 2.5$ (top and bottom respectively).}
\label{fig:Fig4}
\end{figure*}

The equilibria are examined more systematically in Fig.~\ref{fig:Fig4}. Each panel corresponds to a particular pair $(a, \alpha)$ and shows families of equilibria parameterised by $i_0 \in [0, \pi]$. Since $\omega_G = \pm \pi/2$, the ordinate axis is $e \sin\omega_G$ and a positive value means that $\omega_G = \pi/2$, while a negative value means that $\omega_G = -\pi/2$.

For $\alpha = 1.5$ there are only stable equilibria (apart from a tiny part of the rightmost panel, i.e., $a = 5000\,\au$ and $e \sin\omega_G \approx -1, i_G \approx \pi/2$), the one for $e \sim 0$ (increasing for larger $a$) in the whole range of $i_G$\footnote{Note: While $i_0$ is used to parameterise the equilibrium families, the abscissa axis is $i_G$.} and for $i_G \sim 0$ or $\pi$ (or $\sin i_G \sim 0$) in the whole range of $e$. The first family is characterised by $\omega_G = \pi/2$, while the other families have $\omega_G = -\pi/2$. The families with $\sin i_G \sim 0$ are not seen in the diagrams in Fig.~\ref{fig:Fig3}.

For $\alpha = 2$ and $\alpha = 2.5$ there is a family of unstable equilibria. They exist for sufficiently high inclinations, e.g. for $a = 500\,\au$ the unstable equilibria appear for $i_G \gtrsim 40-50~$degrees depending on $\alpha$. They appear when the stable equilibrium $e \sim 0$ bifurcates. As $i_0$ (and $i_G$) increases, the stable equilibria move towards higher eccentricities, reaching $1$ for $i_G = \pi/2$. For larger $a$, the families of equilibria change. The unstable equilibria exist for a narrower range of $i_G$ or may even disappear.

Since the centre of the Galaxy is near the ecliptic, the orbits with low inclination $i$ (in the ecliptic reference frame) have a high inclination $i_G$ in the reference frame defined by the centre of the Galaxy. Furthermore, the orbits whose perihelions point in the direction of the Galaxy centre or anticentre have $\omega_G \sim \pi/2$ and $-\pi/2$ respectively.

\begin{figure*}
\centerline{
\vbox{
\hbox{
\includegraphics[width=0.32\textwidth]{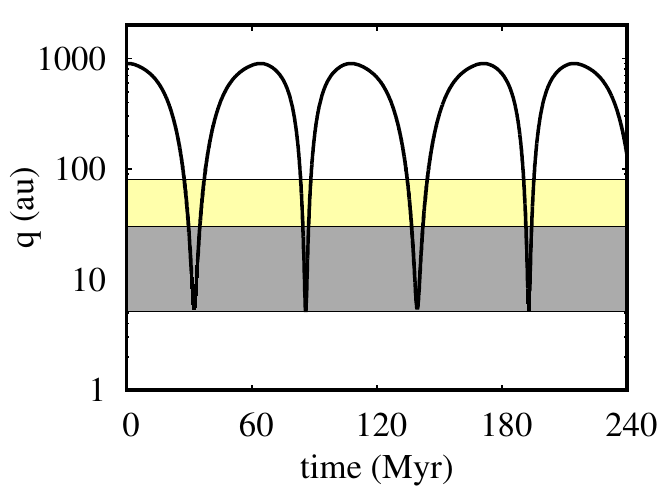}
\includegraphics[width=0.32\textwidth]{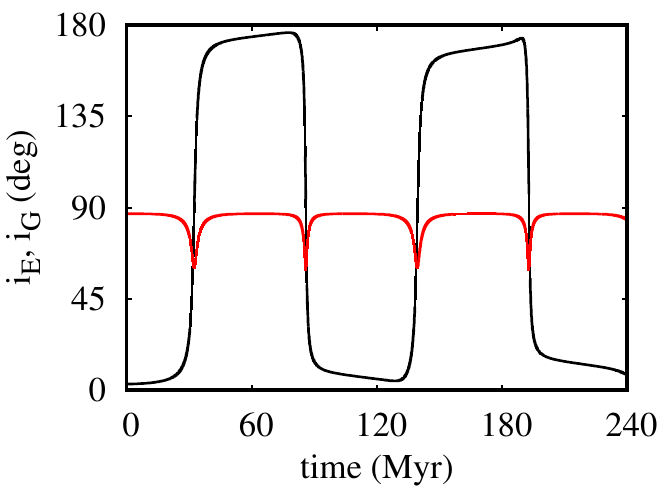}
\includegraphics[width=0.32\textwidth]{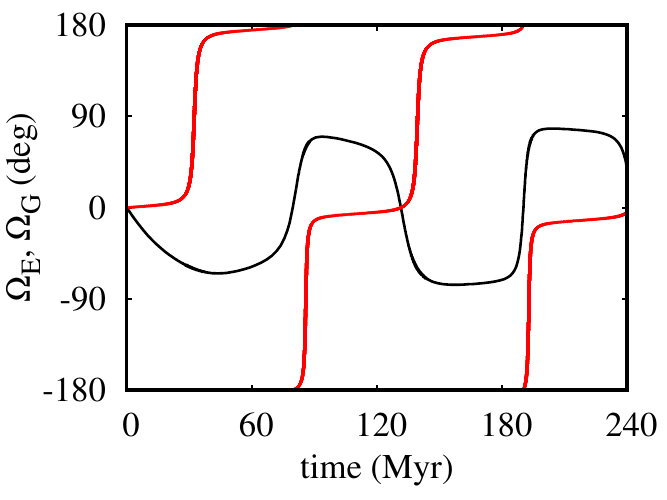}
}
\hbox{
\includegraphics[width=0.32\textwidth]{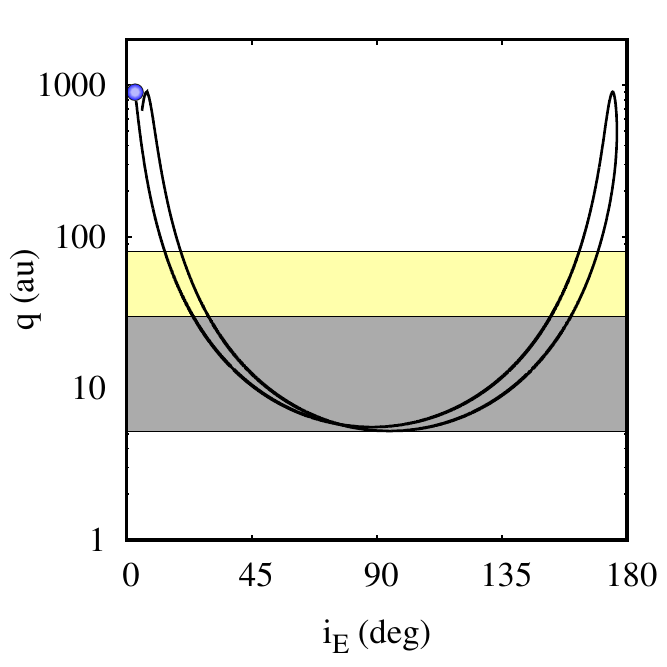}
\includegraphics[width=0.32\textwidth]{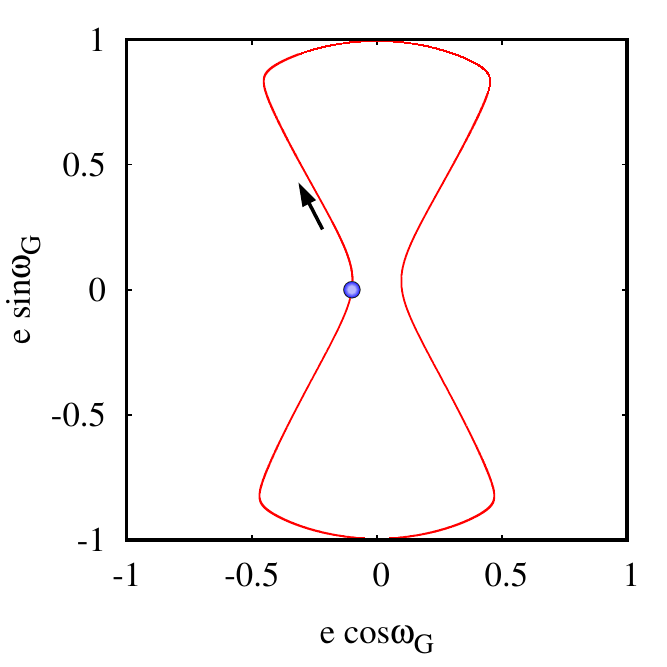}
\includegraphics[width=0.32\textwidth]{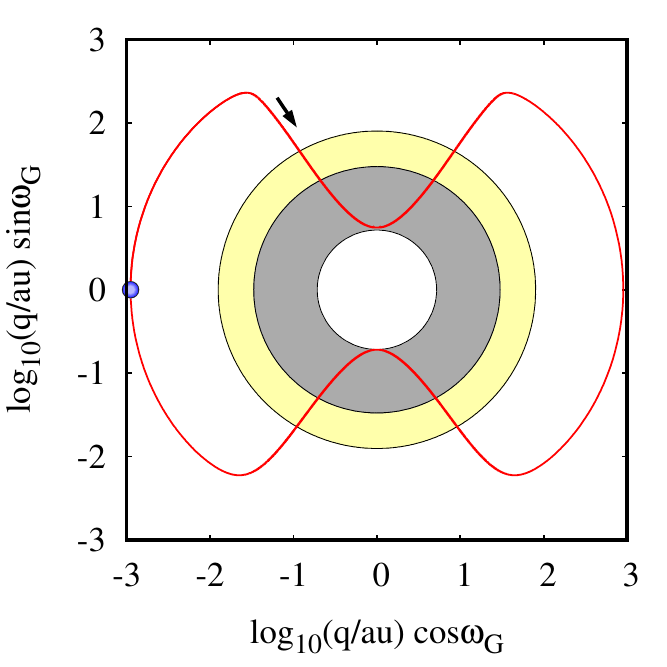}
}
}
}
\caption{An example evolution of a system with initial parameters: $a=1000\,\au, e=0.1$, $i = 3\,$deg, $i_G = 87\,$deg, $\Omega = \Omega_G = 0$, $\omega = 0$, $\omega_G = 180\,$deg. The MOND parameter $\alpha = 2$. The Euler angles are given either in the G frame (red curves) or in the E frame (black curves). The initial state of the system is marked with a blue symbol. The grey and yellow shaded areas mark the limits of $q$, i.e. $q \in [5.2, 30]$ and $q \in [30, 80]\,\au$ respectively}
\label{fig:Fig5}
\end{figure*}

The presence of an unstable equilibrium for $e \sim 0$ leads to high-amplitude oscillations of $e$ between $\sim 0$ and $\sim 1$, which means that an asteroid with a large $a$ can have its perihelion during evolution both close to the Sun and very far from it.

In order to follow the evolution of the mean system, one can use one of two approaches. In the first case, the secular Hamiltonian, Eq.~\ref{eq:meanH}, is used to construct the canonical equations of motion. In this case, one needs to calculate the partial derivatives of the Hamiltonian numerically. In the second approach, the Gauss's planetary equations \citep[e.g.][]{Murray1999} are averaged over the Keplerian motion. We have chosen the second method, although both are good.

An example evolution is shown in Fig.~\ref{fig:Fig5}. The perihelion distance (the upper left panel) varies between $\sim 1000\,\au$ and $\lesssim 5\,\au$. The asteroid remains detectable over a short part of the periodic oscillations of $q$ ($q \lesssim 80\,\au$ for the current statistics of known TNOs), in particular the asteroid can be classified as a Centaur (the grey area denotes $q \in [5.2, 30]\,\au$) or as an ETNO (the yellow area denotes $q \in [30, 80]\,\au$).

The upper middle panel of Fig.~\ref{fig:Fig5} shows the evolution of the inclination in two different reference frames, i.e. the red curve corresponds to the G-frame, while the black curve corresponds to the frame in which the coordinates of the Galaxy centre are $(0, -1, 0)$. The latter are close to the current position of the Galaxy centre in the ecliptic frame and can be referred to as the E-frame for short. The inclination in the G frame, $i_G$, varies between $\sim 90$ and $\sim 60~$degrees, while the inclination in the E frame, $i_E$, varies throughout the whole range. The variation is faster when $q$ is small. The orbit is alternately prograde and retrograde. The rapid reorientation of the orbital plane results from the precession of the orbit in the G-frame (the red curve in the upper right panel). For a given torque acting on the orbit, the precession is faster with a lower angular momentum of the orbit, i.e. higher $e$/lower $q$ when $a$ is constant. In the E-frame, the longitude of the ascending node oscillates.

The bottom left panel of Fig.~\ref{fig:Fig5} shows the evolution in the $(i_E, q)$ diagram. The initial position is marked with the blue symbol. The system evolves along a U-shaped trajectory, visiting the regions with $i_E$ either $\sim 0$ or $\sim 180~$degrees for moderate and large $q$, and with $i_E$ between these values for small $q$. The evolution in this diagram is not periodic since there are two fundamental frequencies of motion. One is related to the variation of $\omega$, the second to $\Omega$. While $i_G$ and $q$ are related by the conservation of projection of angular momentum on the $z$ axis, this relationship does not exist in the E-frame. As a result, the phase trajectory is not closed and with a longer integration (more cycles of the $q$ oscillations), the trajectory would fill the diagram more evenly and the U-shaped structure might be less clear.

The two remaining panels of Fig.~\ref{fig:Fig5} show the evolution of the argument of pericentre $\omega_G$ and $e$ (bottom-middle) and $q$ (bottom-right). The direction of the motion is marked with arrows. The phase trajectory is closed since $\Omega_G$ is a cyclic variable in the G-frame. The eccentricity varies between low and high values along an $8$-shaped trajectory. The variation of the perihelion distance is shown in the logarithmic scale. As was mentioned earlier, the maximum $e$/minimum $q$ is reached for $\omega_G = \pm \pi/2$. The example system has a large initial $q$, but we could also start the simulation with a small $q$. If the initial parameters are $q \sim 5\,\au$ (which would correspond to an asteroid scattered by Jupiter) and $\omega_G \sim 90~$degrees, the value of $\omega_G$ reached for $q \in [30, 80]\,\au$ (the upper limit corresponds to the current detection limit) would be close to $\sim 45~$degrees. Other values of $\sim -45, -135, 135~$degrees are reached over the whole cycle. This would mean that the apsidal line could be shifted by about $45~$degrees with respect to the Galaxy centre-anticentre line.

\begin{figure}
\centerline{
\vbox{
\hbox{\includegraphics[width=0.3\textwidth]{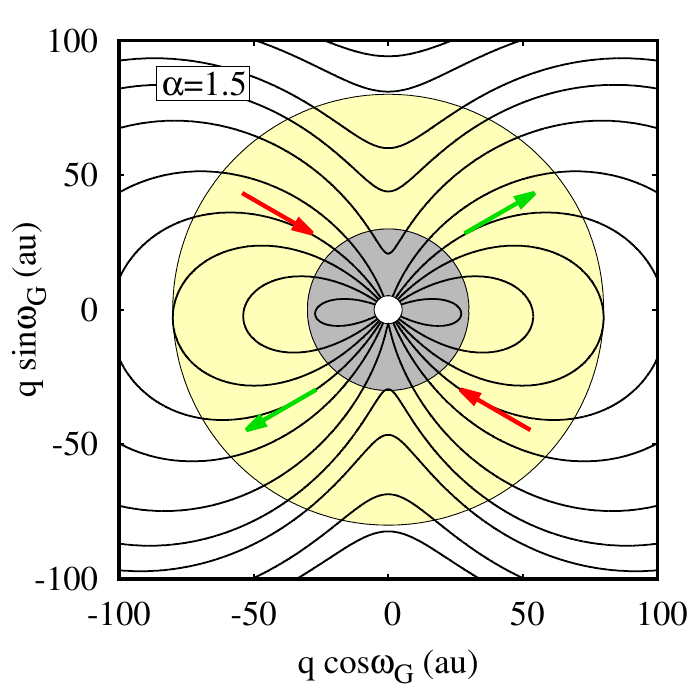}}
\hbox{\includegraphics[width=0.3\textwidth]{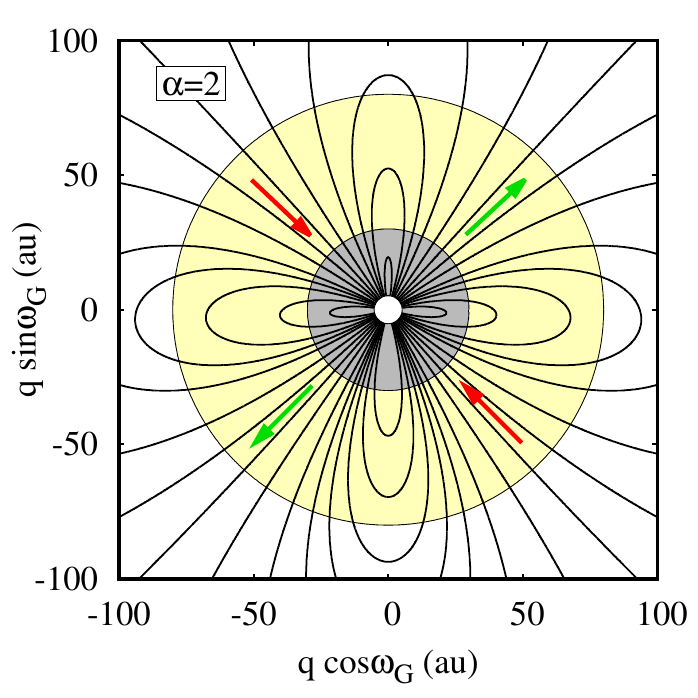}}
}
}
\caption{Contours of the constant secular Hamiltonian shown in the $(q \cos\omega_G, q \sin\omega_G)$--diagram for $a = 500\,\au$, $i_0 = 90\,$deg and $\alpha = 1.5$ (top) and $\alpha = 2$ (bottom). The grey colour denotes the area between the orbits of Jupiter and Neptune, while the yellow colour denotes $q \in (30, 80)\,\au$.}
\label{fig:Fig6}
\end{figure}

The presence of the unstable equilibrium at low eccentricity leads to a large-amplitude variation of $q$. For $\alpha = 1.5$ there is no such equilibrium and the variation occurs with smaller amplitude. However, it is still possible to drift perihelia of scattered asteroids outside the orbit of Neptune. Figure~\ref{fig:Fig6} shows the energy levels for $\alpha = 1.5$ and $\alpha = 2$, which is for a moderate semi-major axis $a = 500\,\au$ and $i_0 = 90\,$deg (implying that $I_3 = 0)$. For both values of $\alpha$, the perihelia, which are initially close to Jupiter's orbit ($q \sim 5.2\,\au$), are driven to greater distances. Therefore, it can be difficult to establish a specific value of $\alpha$ based on the observed properties of ETNOs. Most of the simulations presented further in this paper were performed for $\alpha = 2$.

\begin{figure}
\centerline{
\hbox{\includegraphics[width=0.4\textwidth]{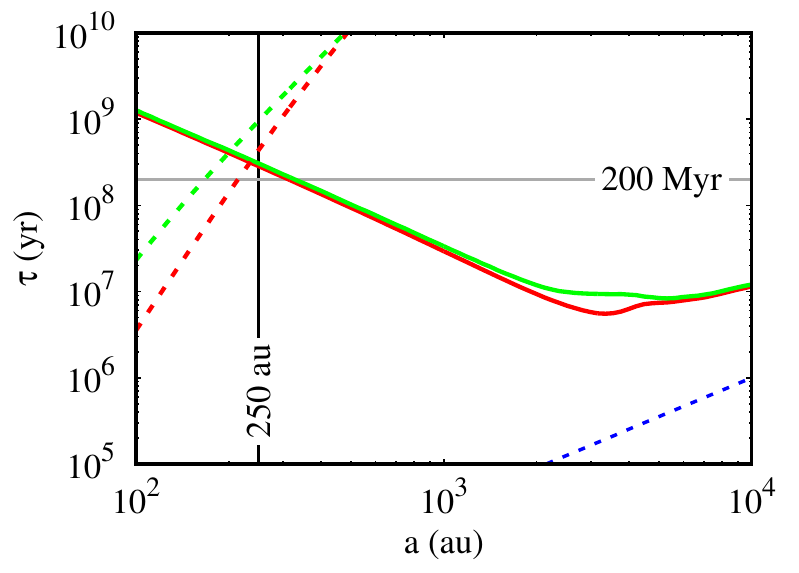}}
}
\caption{Timescales of secular evolution of $\omega$ (red) and $\Omega$ (green) resulting from MOND (solid curves) and planetary perturbation (dashed curves) as a function of $a$. The blue dashed curve indicates the Keplerian period. The MOND parameter $\alpha=2$. The initial orbital parameters $e = 0.8$, $i_G = 70\,$deg, $\omega_G = \Omega_G = 0$.}
\label{fig:Fig7}
\end{figure}

In a more realistic model, the EFE-induced cycle competes with planetary perturbations. While the former speeds up with larger orbits, the latter depends inversely on $a$. Figure~\ref{fig:Fig7} illustrates these dependencies. The timescales for $\omega$ and $\Omega$ are shown as functions of $a$ with the red and green curves, respectively. The solid curves denote the EFE, while the dashed curves correspond to the planetary perturbation. The timescales depend not only on $a$, but also on $e$ and $i$, as well as $\omega_G$ for the EFE perturbation. Therefore, for illustration, we have chosen $e = 0.8$ and $i_G = 70~$degrees (which corresponds to $i_E = 20~$degrees). These numbers correspond to typical values for known ETNOs. The initial values for $\omega_G$ and $\Omega_G$ are set to $0$.

The orbits of the planetary perturbers are assumed to be circular and coplanar with each other. The perturbations are expanded to third order in a Taylor series with respect to a small quantity $a/a_P$, where $a_P$ is the size of planets' orbits, and then averaged over the Keplerian motions. As is known, planetary perturbations lead to the precession of an asteroid orbit with respect to the axis perpendicular to the orbital plane of the planets, as well as to the rotation of the apsidal line.

It is clear that for $a \sim 250\,\au$ the timescales for the two effects are approximately equal. For smaller $a$ the planetary perturbations dominate, while for larger $a$ the MOND effect is more significant. For the semi-major axis beyond a few hundred astronomical units, the MOND-induced evolution dominates over the planetary perturbations by several orders of magnitude. Moreover, for $a \gtrsim 350\,\au$, the Milgromian evolution occurs over a period shorter than the Sun's orbital period around the centre of the Galaxy, which is $\sim 200\,$Myr according to \citep{Hunt2016}. For $a \gtrsim 3000\,\au$, the EFE-induced cycle is only an order of magnitude slower than Keplerian motion, so the secular model is not sufficient.

In this paper we extend the model in two ways. We abandon averaging because we want to model possible close approaches between the planets and an asteroid. Thus, the interaction between planets and asteroids is modelled within the restricted $N$-body model. We also take into account the motion of the Sun in the Galaxy (a circular orbit is assumed), which causes the axis of symmetry of EFE to change in time. Before that, in the next section, we describe observational evidence for the hypothesis of an unseen planet in the periphery of the solar system.

\section{Extreme trans-Neptunian objects and the Planet Nine hypothesis}
\label{sec:kuiper}

The hypothesis of Planet Nine is justified by the anomalous structure of trans-Neptunian solar system. The first indication of the planet's existence was the detachment of the perihelia of Sedna and several other ETNOs from the gravitational influence of the known planets \citep{Brown2004, Gomes2006}. Nevertheless, the existence of Planet Nine would be problematic if the detachment of the perihelia were the only evidence for it, since the proposed orbit of Planet Nine is itself detached from the giant planets \citep{Batygin2016a} and could not be explained by planet--planet scattering alone. In order to solve this problem, three mechanisms are proposed: i) \textit{in~situ} formation in an extended protoplanetary disc \citep{Kenyon2016}; ii) scattering from the region of giant planets followed by orbital circularisation \citep{Eriksson2018,Bromley2016}; iii) capture of a free-floating planet or a planet from another system during close stellar encounters \citep{Li2016}.

Each of these scenarios has its difficulties \citep[see][for a brief overview on this topic]{Batygin2019}. In the second scenario, for example, it was proposed that the planet was first scattered and then its orbit was circularised by planetesimals forming an extended disc up to $\sim 1000\,\au$ \citep{Eriksson2018}. The chain of proposed solutions to the ETNOs detachment problem is then as follows: To explain detached ETNOs -- a distant detached Planet Nine is proposed; to explain detached Planet Nine -- an extended detached disc of planetesimals is proposed. The addition of more and more objects only seems to shift the problem, and without more evidence, the Planet Nine hypothesis would be in trouble.

\begin{figure}
\centerline{
\hbox{\includegraphics[width=0.4\textwidth]{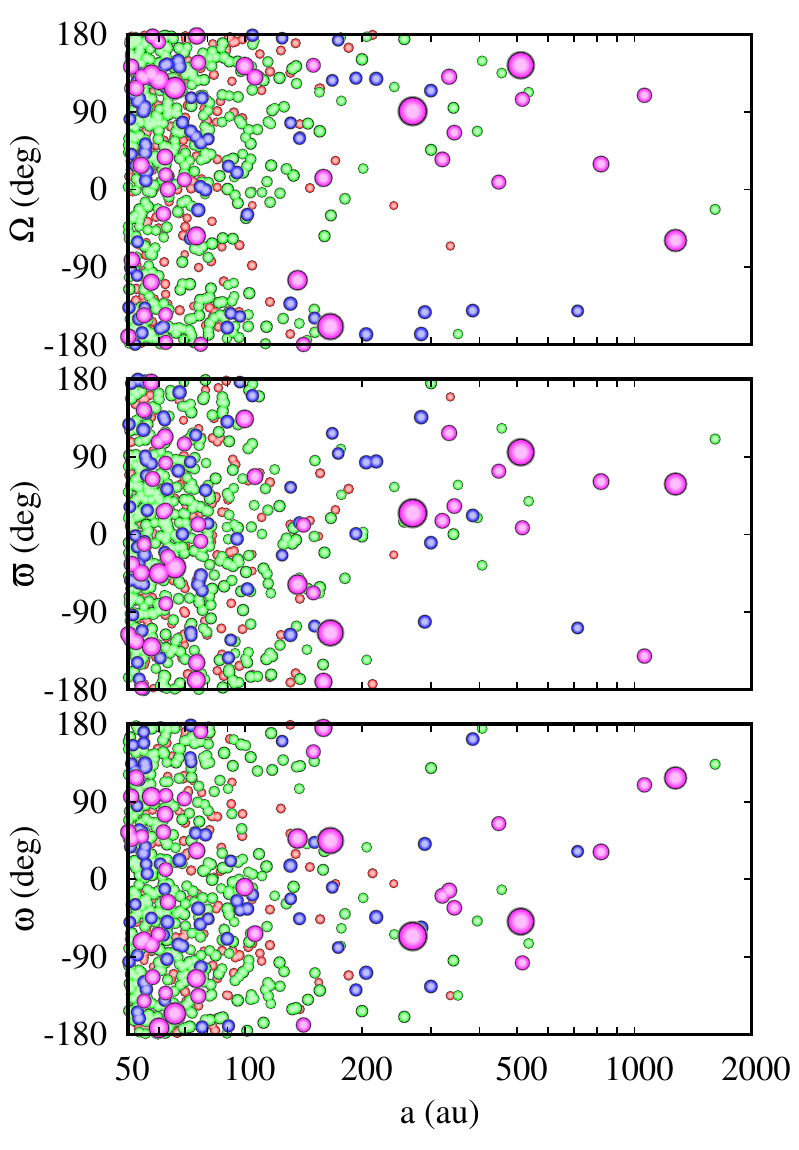}}
}
\caption{The statistics of the known TNOs with $a > 50\,\au$ and $q > 30\,\au$ are shown in the diagrams $(a, \Omega)$, $(a, \varpi)$ and $(a, \omega)$ respectively from top to bottom. Different colours denote different ranges of $q$, i.e. $q \in (30, 35]\,\au$ -- red, $q \in (35, 40]\,\au$ -- green, $q \in (40, 45]\,\au$ -- blue, $q > 45\,\au$ -- magenta. Furthermore, the sizes of the symbols increase with $q$.}
\label{fig:Fig8}
\end{figure}

However, there are other features of ETNOs that support this hypothesis. Planet Nine has been shown to explain the high inclinations of Centaurs \citep{Batygin2016b}, which cannot otherwise be reproduced due to scattering by known planets. However, the most telling feature of ETNO orbits is the clustering of orbital planes and the apsidal confinement reported for $a \gtrsim 250\,\au$ \citep{Batygin2016a}. As mentioned in the introduction, the robustness of these features has been questioned by several authors \citep{Bernardinelli2020,Napier2021,Clement2020}. In order to verify the features, we have presented the statistics of the TNO orbital elements in Fig.~\ref{fig:Fig8}. In the upper panel, the longitudes of the ascending node are plotted against the semi-major axes. The colours of the symbols as well as their size encode the perihelion distance (see caption). The grouping of $\Omega$ is not clear, although for $a \gtrsim 200\,\au$ it can be noted that the range $(0, 180)\,$deg is preferred over the range $(-180, 0)\,$deg, especially for the most detached orbits (magenta symbols with $q > 45\,\au$). The distribution of $\varpi$ (middle panel) shows a similar feature. The grouping of $\omega$, on the other hand, is more likely to be found in the range $(-90, 90)\,$deg.

\begin{figure*}
\centerline{
\includegraphics[width=0.95\textwidth]{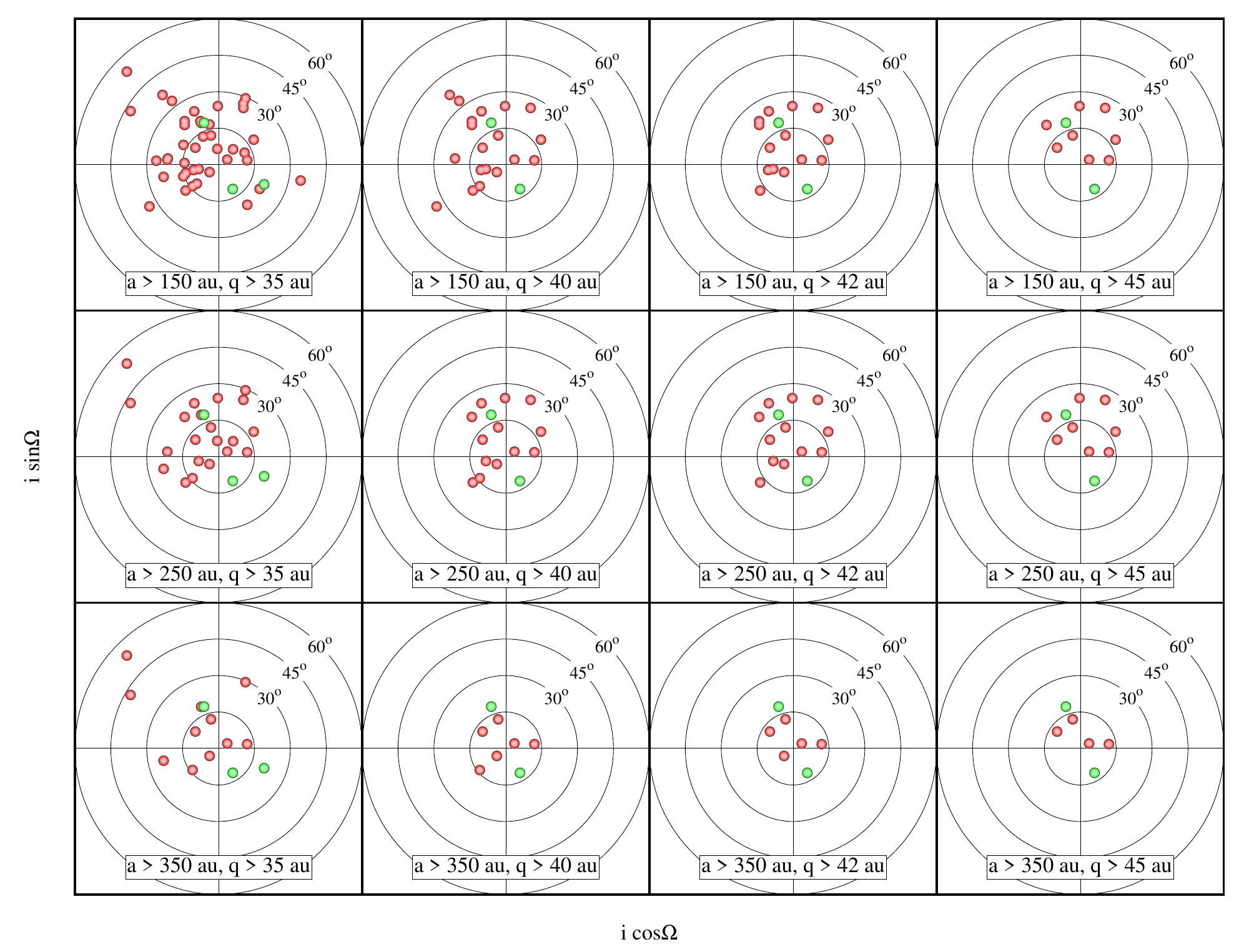}
}
\caption{The statistics of $(i \cos\Omega, i \sin\Omega)$ of the prograde ETNOs with different selection criteria. The red symbols represent orbits with $a < 1000\,\au$, while the green colour denotes $a \geq 1000\,\au$.}
\label{fig:Fig9}
\end{figure*}

By choosing certain criteria for $a$ and $q$, one can achieve a more or less clear grouping of the elements. \cite{Brown2021} suggested to consider only objects with $q > 42\,\au$ and $a \in (150, 1000)\,\au$ which are on the one hand sufficiently detached from Neptune and on the other hand not too far away to be influenced by Planet Nine. Figure~\ref{fig:Fig9} illustrates the clustering of orbital planes in the $(i \cos\Omega, i \sin\Omega)$--diagram for different selection criteria. Each row represents a different minimum $a$, while each column represents a different minimum $q$. The upper half-plane of each panel is favoured over the lower half-plane for all criteria, which means that the clustering of orbital planes is independent of a particular choice of $a$ and $q$ limits.

\begin{figure*}
\centerline{
\includegraphics[width=0.95\textwidth]{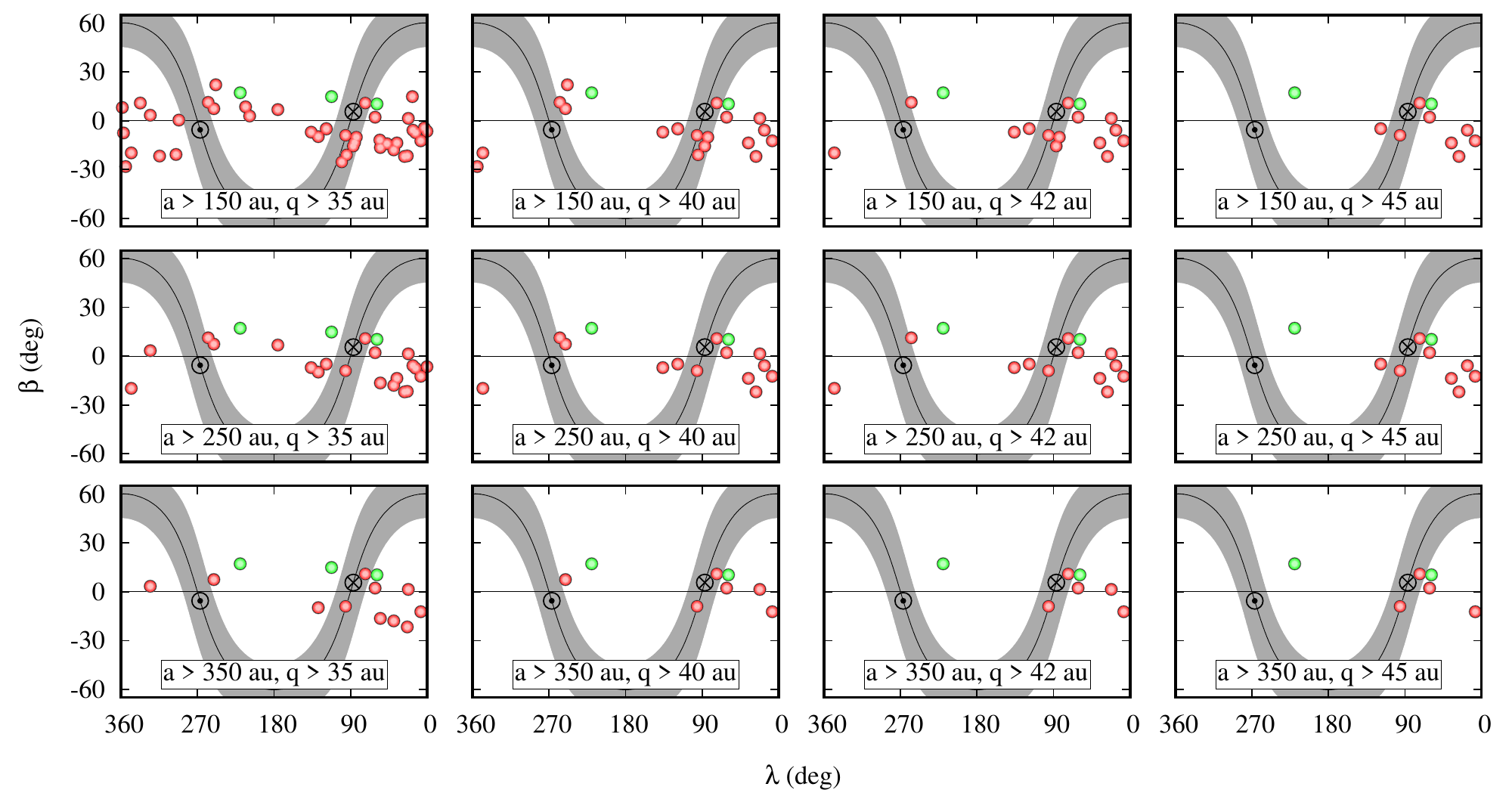}
}
\caption{The distribution of the ecliptic coordinates $(\lambda, \beta)$ of the perihelia of prograde ETNOs with different selection criteria. The red symbols represent orbits with $a < 1000\,\au$, while the green colour denotes $a \geq 1000\,\au$. The grey area denotes the galactic disc, while the black circle with a dot indicates the direction to the centre of the Galaxy and the circle with a cross marks the position of the anticentre.}
\label{fig:Fig10}
\end{figure*}

The apsidal confinement for prograde ETNOs is shown in Fig.~\ref{fig:Fig10} for different selection criteria. The perihelia group around the ecliptic, which can be attributed to observational bias. For $q > 40\,\au$, the ecliptic longitude in the range $(0, 180)\,$deg is preferred over the range $(180, 360)\,$deg. This corresponds to a grouping around the direction of the anticentre of the Galaxy, although there are also perihelia closer to the centre of the Galaxy.

\begin{figure}
\centerline{
\includegraphics[width=0.38\textwidth]{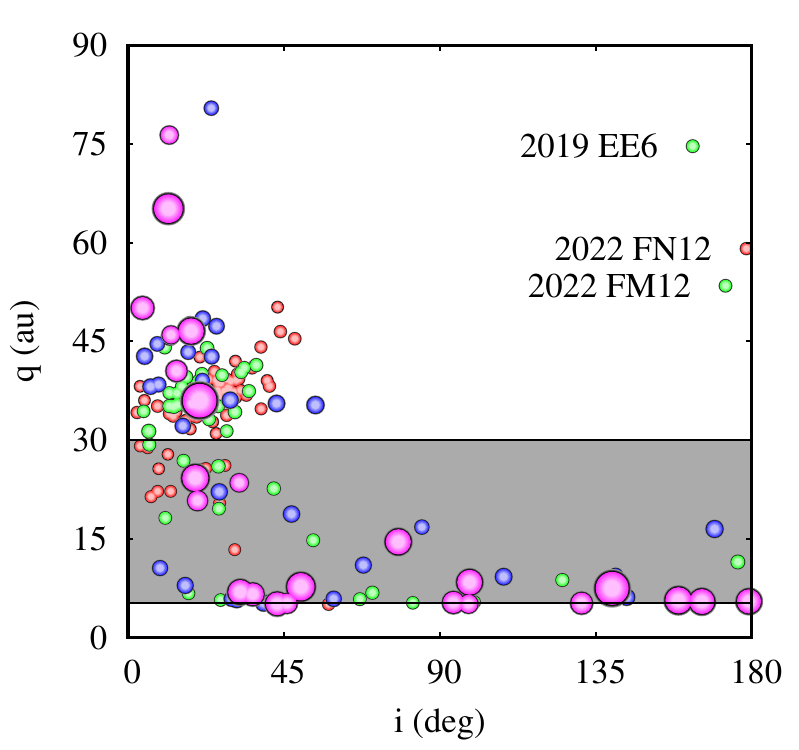}
}
\caption{The statistics of TNOs with $a \in (100, 2000)\,\au$, $q > 5.2\,\au$ shown in the $(i, q)$--diagram. Different colours indicate different ranges of the semi-major axes: $a \in [100, 150)\,\au$ -- red, $a \in [150, 250)\,\au$ -- green, $a \in [250, 500)\,\au$ -- blue, $a \in [500, 2000)\,\au$ -- magenta. Moreover, the size of the symbols increases with $a$. Three retrograde ETNOs are labeled.}
\label{fig:Fig11}
\end{figure}

The statistics of the perihelia and inclinations are shown in Fig.~\ref{fig:Fig11}. Apart from the previously selected objects, this set is extended by Centaurs, $q \in (5.2, 30)\,\au$ and objects with smaller $a$ down to $100\,\au$. Centaur statistics are limited to $a < 2000\,\au$. The inclinations of the Centaurs are distributed over the whole range, but for $i \gtrsim 45\,$deg only objects with $q \lesssim 10\,\au$ are observed. Moreover, for the objects with $a > 500\,\au$, the perihelion distances are close to $5.2\,\au$, indicating that these Centaurs were scattered by Jupiter. The objects with $q > 30\,\au$ have inclinations of $\lesssim 60\,$deg, but there are three recently discovered objects with $q > 50\,\au$ and retrograde orbits (they are labelled in the graph). Also, we can see that the objects with $q > 30\,\au$ and $a > 500\,\au$ (magenta dots) generally have lower inclinations than the objects with $a < 500\,\au$.

As shown in the previous section and already demonstrated by \cite{Pauco2016}, the detachment of the perihelia as well as highly inclined Centaurs can be naturally explained within EFE. The connection between Centaurs and ETNOs is also natural in EFE and was also shown by \cite{Pauco2016}. The lower left panel in Fig.~\ref{fig:Fig5} shows that they belong to the same population of objects, but at two different stages of the evolution. We can also see that the retrograde ETNOs can be explained within EFE as well.

The apsidal confinement as well as the clustering of the orbital planes are more difficult to explain. \cite{Pauco2017} concluded that the EFE quadrupole strength parameter required to reconstruct the observed properties of the orbits is not consistent with observational constraints\footnote{\corr{Recently, after submitting the original manuscript of this work, a new paper considering EFE as an alternative explanation to the Planet Nine hypothesis appeared \citep{Jones-Smith2023}. The authors notice that $6$ ETNOs belonging to the Sedna family have all $\omega_G \sim -\pi/2$ and connect this with one of the stable equilibria described in Section~\ref{sec:simple_model}.}}. As we mentioned in the introduction, a comparison between the synthetic and the observed orbits is problematic. The difficulty lies in the choice of the initial set of orbits and in the way the final set is constructed for comparison with the observations. To understand the first problem, in the next section we trace the evolution of the observed ETNOs backwards in time and show that they originate from the region of the giant planets of the solar system.

\section{Evolution of known ETNOs and Centaurs}
\label{sec:known_objects}

The restricted $N$-body model with MOND perturbation is used to study the evolution of the known ETNOs and Centaurs. The model of motion is the following. The system consists of the Sun, the four giant planets and a given number of massless asteroids. The model is thus restricted in the sense that the asteroids do not affect the motions of the planets. Apart from the Newtonian gravitational interaction between the bodies, each of them is subject to the perturbation $\apert$, which is determined for each position in space using bicubic interpolation, as described earlier in this paper.

The perturbation is computed on the assumption that there are no massive bodies other than the Sun and the point-like Galaxy, which is not necessarily fulfilled due to the existence of massive planets. However, a more general Milgromian model of motion is much more complex and would be impractical for the problem studied here. The simplified model can be validated in the following way. In the regime of giant planets, the planetary contribution to $\phi_N$ is not negligible, but the MOND perturbation, $\apert$, is very weak. On the other hand, MOND effects are important for large distances from the Sun, where $\phi_N$ used to compute $\apert$ can be well approximated by the Sun alone, the planetary contribution to $\phi_N$ is of the order of $10^{-3}$. In both ranges, an additional contribution to $\apert$ resulting from the presence of giant planets is a higher order effect and is omitted in this analysis.

\subsection{Prograde ETNOs}

Following the original selection criteria in \citep{Batygin2016a}, we selected $25$ ETNOs with $a > 250\,\au$ and $q > 30\,\au$ and randomly chose $100$ clones for each of them according to their orbital uncertainties. Such a set of $2525$ massless particles were numerically integrated backwards in time. All the asteroids evolve into orbits with perihelions well inside the orbit of Neptune. For ETNOs with initial $a \gtrsim 1000\,\au$ a few Myr were enough to reach the low $q$ range, while for the ones with $a \sim 250\,\au$ a few hundreds Myr was necessary.

Once a given object reaches the giant planets region, its evolution becomes strongly chaotic due to close encounters with the planets. Close encounters result in the energy and angular momentum gain or loss of the object, thus not only $e$ but also $a$ is modified. If the energy is lost, the asteroid orbit becomes smaller and may reach the region $a < 30\,\au$. If also the eccentricity decreases, the entire orbit of the object can reside in the giant planets region.

\begin{figure*}
\centerline{
\hbox{
\includegraphics[width=0.4\textwidth]{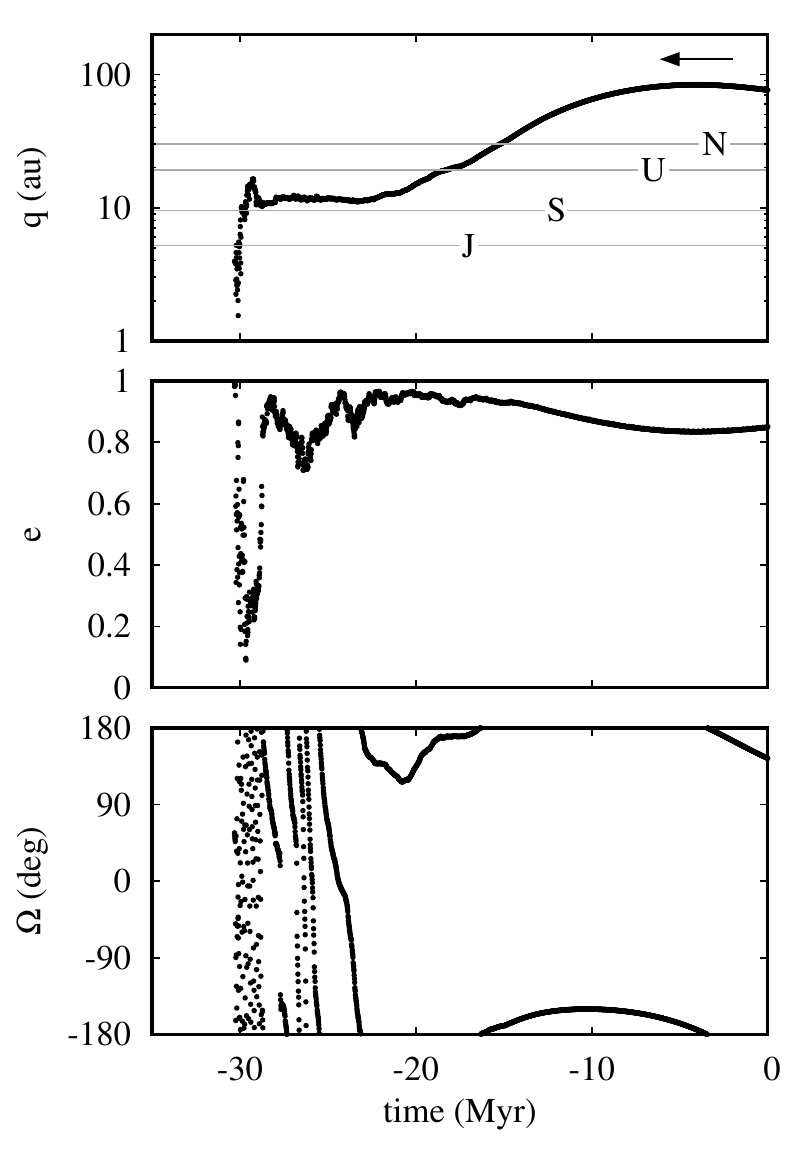}
\includegraphics[width=0.4\textwidth]{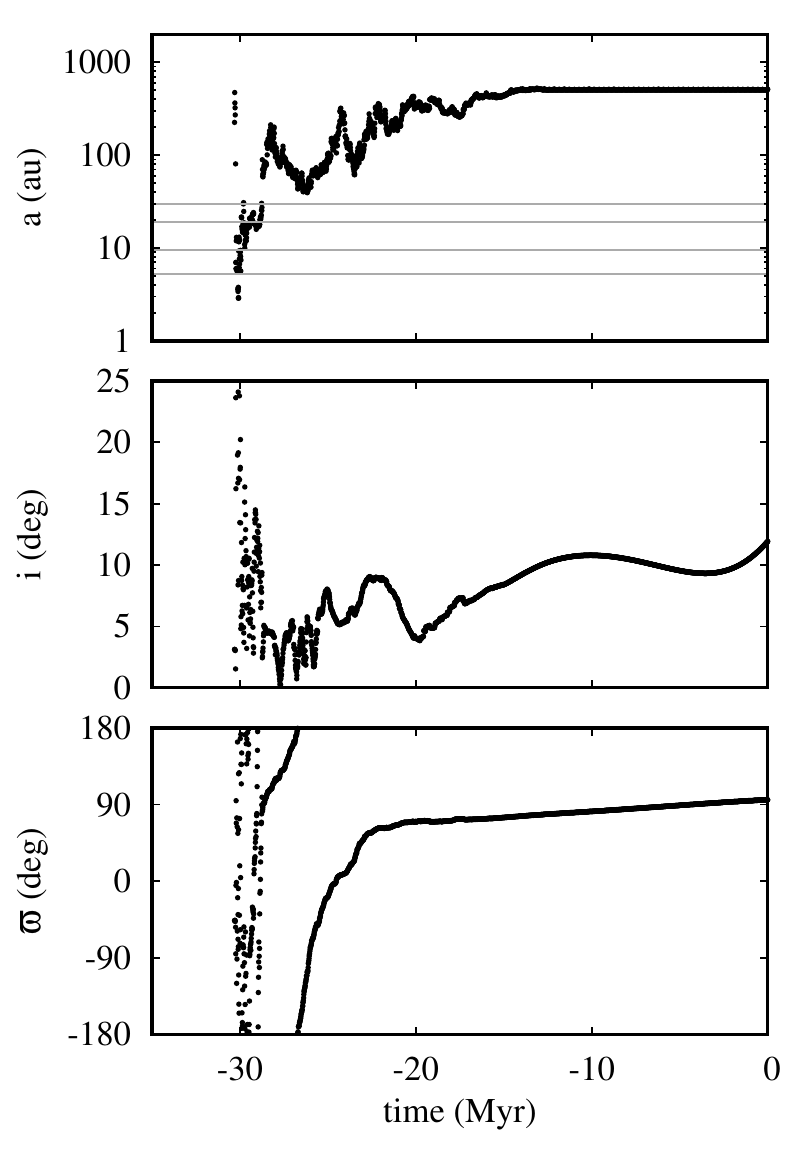}
}
}
\caption{Backward evolution of an example clone of Sedna. The arrow at the top-left panel shows the direction of evolution.}
\label{fig:Fig12}
\end{figure*}

The evolution of a particular object cannot be traced because of the strong chaos and one should only investigate the past orbits of ETNOs statistically. Nevertheless, to illustrate the possible origin of a particular ETNO, we present the evolution of Sedna (one of its clones) in Fig.~\ref{fig:Fig12}. Once its perihelion reaches the orbits of Neptune, Uranus, Saturn and Jupiter (upper left panel), the semi-major axis changes chaotically towards smaller values down to the sizes of the orbits of Saturn and Jupiter (upper right panel). This occurs in the epoch around $-30\,$Myr and is accompanied by a decrease in eccentricity (the middle left panel), while the inclination remains moderate (the middle right panel) and the $\Omega$ and $\omega$ angles vary throughout (the lower panels).

Further past evolution results in the ejection of the object. However, this does not mean that this object entered the solar system from outside, if the evolution of the clone was the real evolution of Sedna. It is the result of a strong chaos and sooner or later every object is ejected from the solar system, regardless of the direction of evolution.

\begin{figure*}
\centerline{
\vbox{
\hbox{
\includegraphics[width=0.3\textwidth]{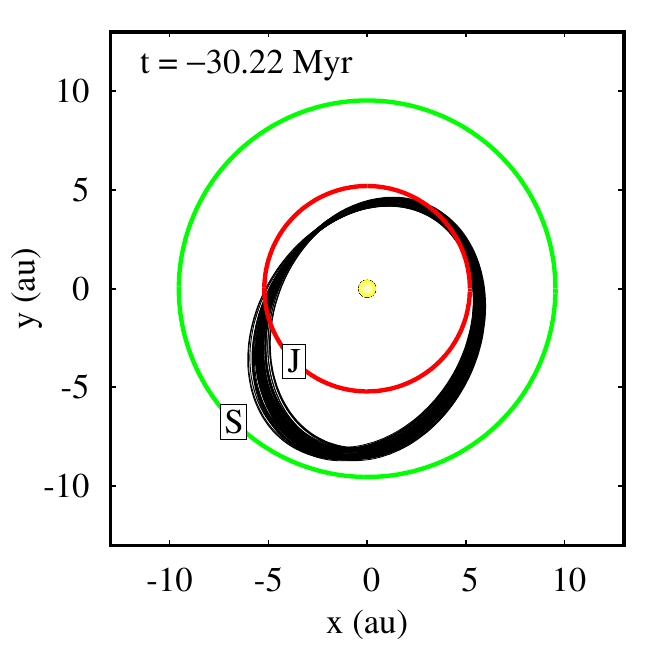}
\includegraphics[width=0.3\textwidth]{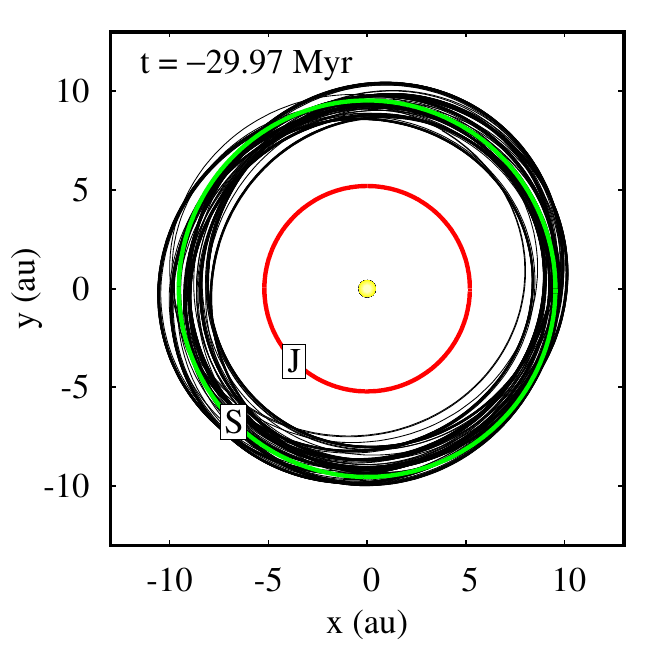}
\includegraphics[width=0.3\textwidth]{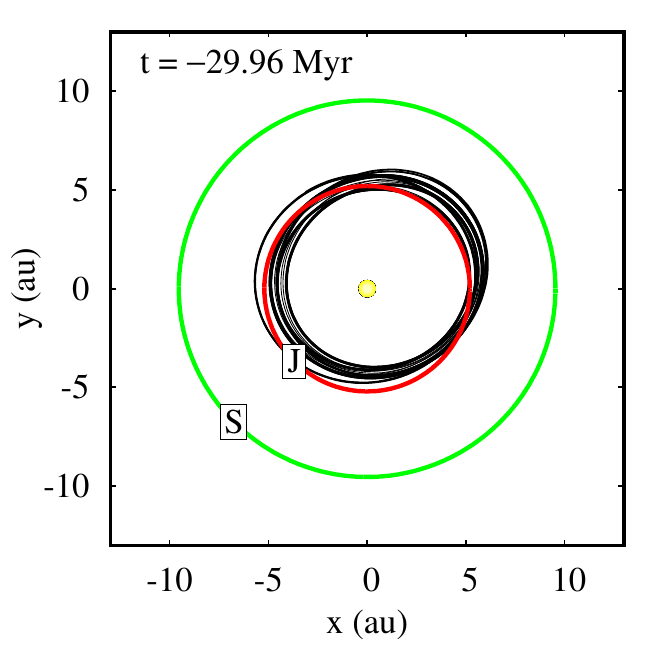}
}
\hbox{
\includegraphics[width=0.3\textwidth]{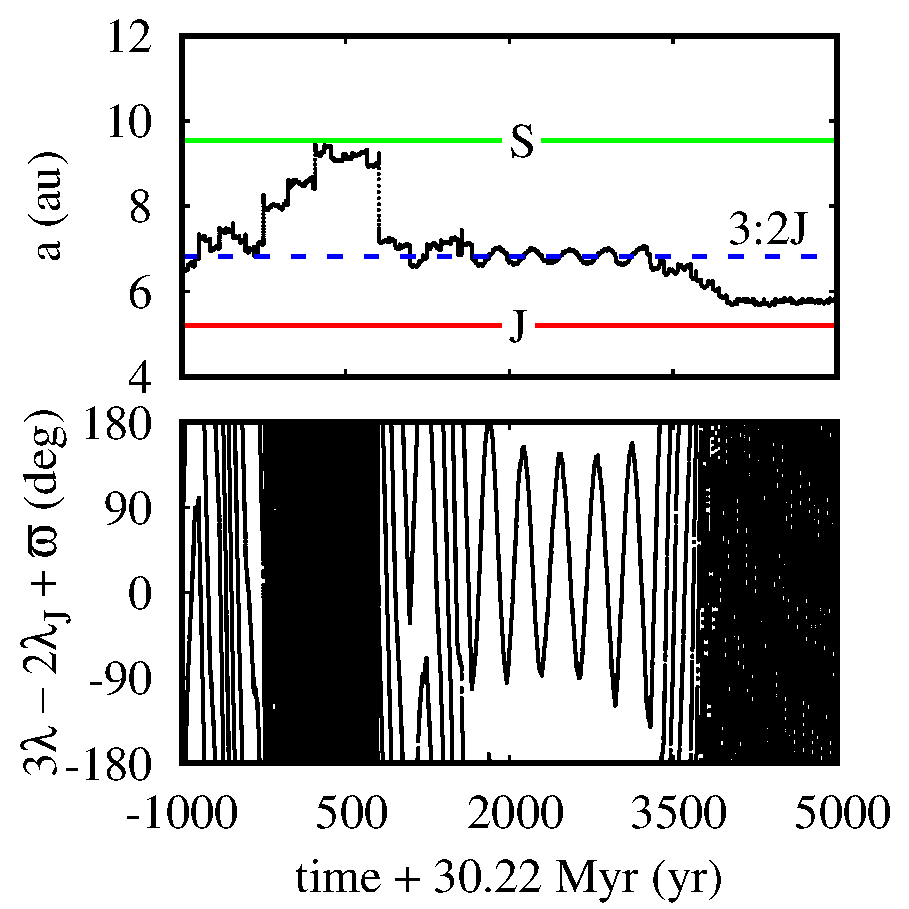}
\includegraphics[width=0.3\textwidth]{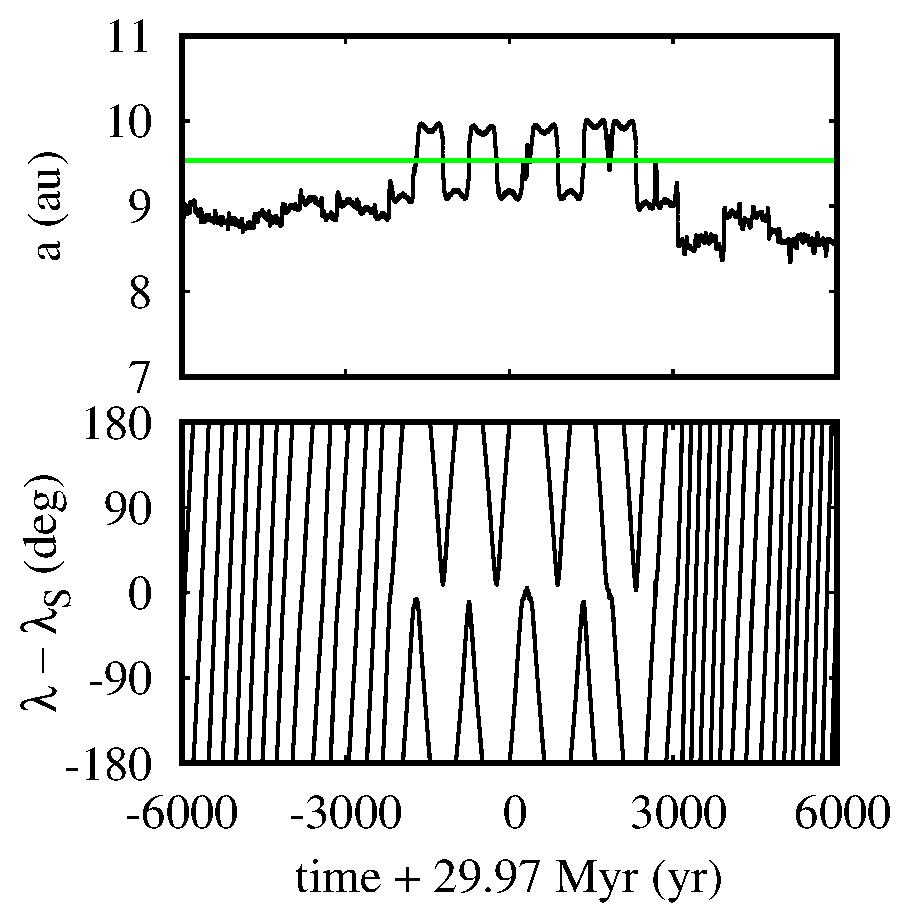}
\includegraphics[width=0.3\textwidth]{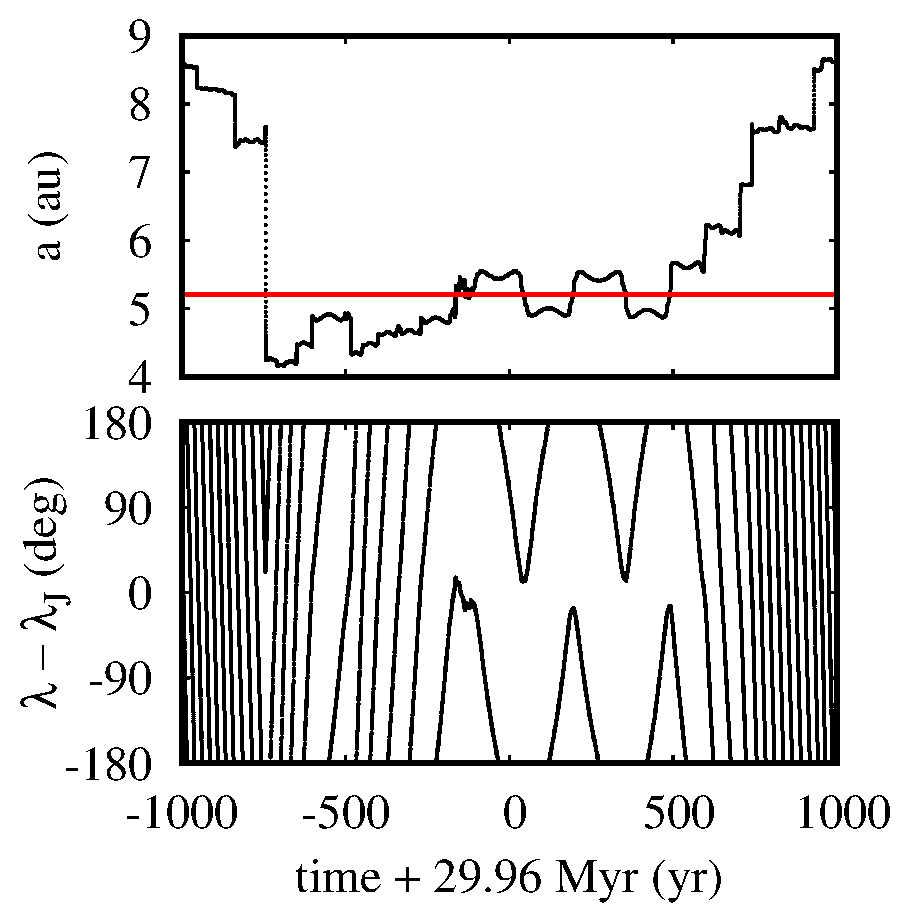}
}
}
}
\caption{Temporal evolution of Sedna (one of the clones) for selected past epochs around $-30\,$Myr, shown in the $(x,y)$-diagram in the upper panels and as temporal variation of resonant angles in the lower panels, from left to right: $\phi_{\idm{res}} = 3 \lambda - 2 \lambda_{\idm{J}} + \varpi$, $\phi_{\idm{res}} = \lambda - \lambda_{\idm{S}}$ and $\phi_{\idm{res}} = \lambda - \lambda_{\idm{J}}$, where the indices $S$ and $J$ refer to Saturn and Jupiter, while symbols without an index refer to the asteroid. The dashed blue line in the lower left panel denotes the 3:2~MMR with Jupiter.}
\label{fig:Fig13}
\end{figure*}

Figure~\ref{fig:Fig13} shows the temporary evolution of the same clone of Sedna about the epoch $-30\,$Myr. The clone remains close to the orbit of Jupiter or Saturn and is temporarily in a 1:1 mean motion resonance (MMR) with one or other of the giant planets. It may also be in a transient 3:2~MMR with Jupiter. Since Jupiter and Saturn are themselves near 5:2~MMR, the clone is in the region of overlapping resonances with the two giant planets. This is illustrated in the lower panels where the evolution of the resonant angles can be seen. The proximity to the orbits of Jupiter or Saturn means that the respective resonant angles oscillate around $180\,$deg, but with a half amplitude of $\sim 180\,$deg. The resonant angles for 1:1~MMR are simply differences between the mean longitudes of the asteroid and one of the planets. Such oscillations of the angles are characteristic of horseshoe orbits in the restricted three-body problem. For the proximity to 3:2~MMR with Jupiter, the resonant angle oscillates around $\sim 30\,$deg with a moderate amplitude of $\sim 120\,$deg. In all cases, the oscillations occur for only a few cycles.

If this particular clone of Sedna was the actual object, Sedna may even have come from the group of stable Jupiter Trojans after being perturbed by a collision with another member of the group. Naturally, due to the strong chaos, such predictions cannot be made and one can only get the statistical information from the backward integration experiment. Not all of the $25$ ETNOs reach orbits close to Jupiter or Saturn. Some of them reach the range of $a \sim 20-40\,\au$ and $e \gtrsim 0.3$ with moderate-to-high inclinations.

\begin{figure*}
\centerline{
\vbox{
\hbox{
\includegraphics[width=0.42\textwidth]{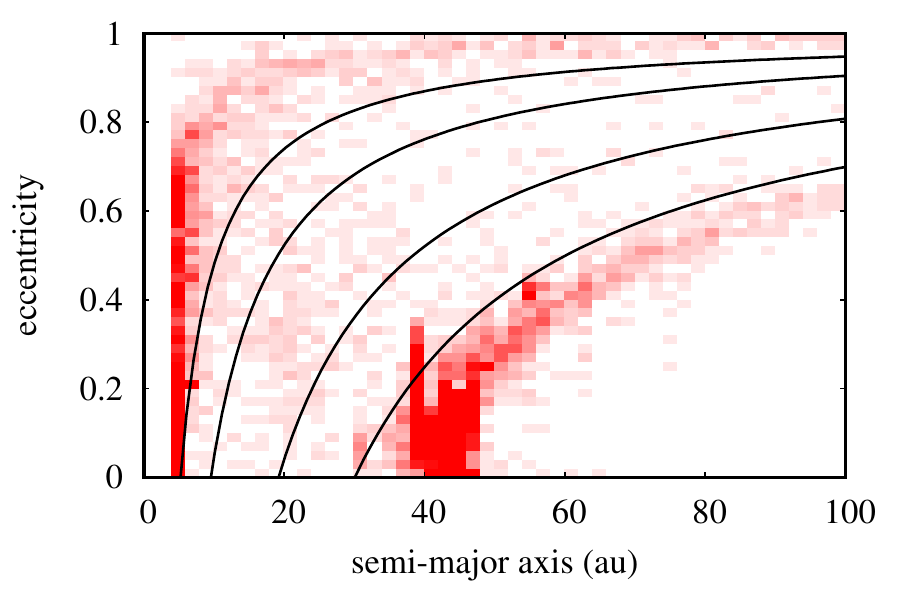}
\includegraphics[width=0.42\textwidth]{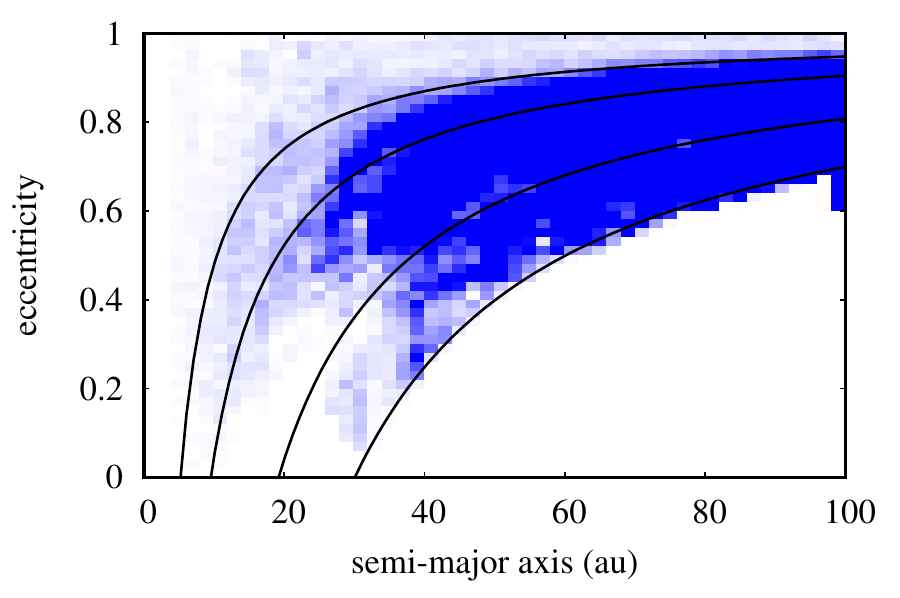}
}
\hbox{
\includegraphics[width=0.42\textwidth]{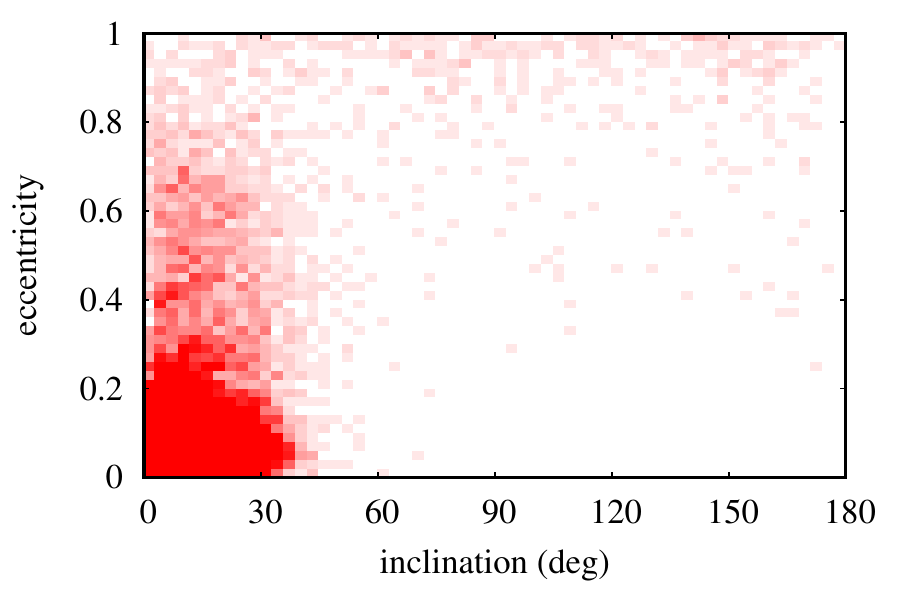}
\includegraphics[width=0.42\textwidth]{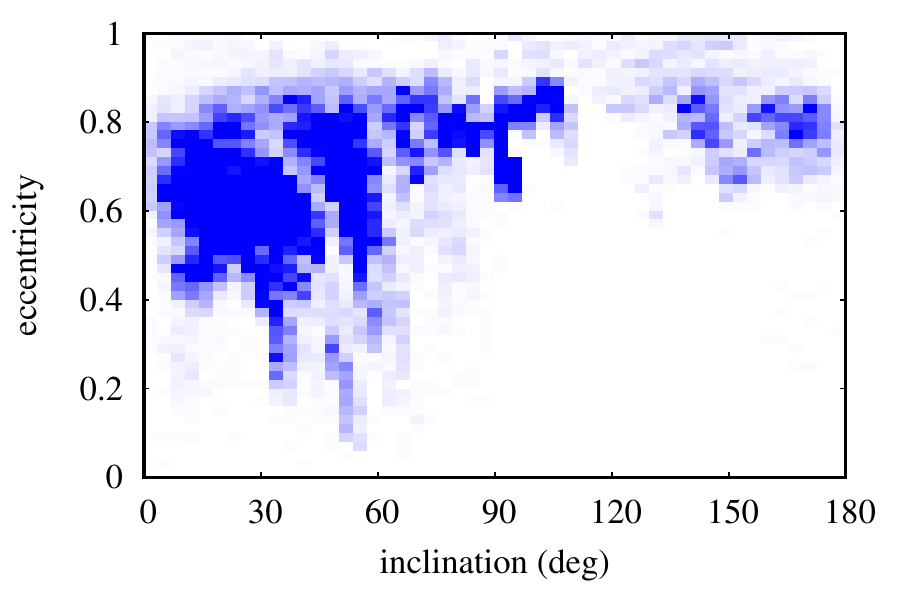}
}
}
}
\caption{The statistics of the orbital elements of the observed asteroids (left) and the backward integrated ETNOs with initial $a > 250\,\au$ and $q > 30\,\au$ (right), shown in the $(a,e)$--diagram (top) and $(i,e)$--diagram (bottom). In the bottom right plot, only objects with $a < 60\,\au$ were included in the statistics. For each ETNO, $100$ clones were selected according to the orbital uncertainties and backward integrated for $0.1-0.5\,$Gyr. See the text for details.}
\label{fig:Fig14}
\end{figure*}

Figure~\ref{fig:Fig14} shows the statistical results of the experiment and the comparison with the observed distribution of asteroids. On the left side, the known solar system objects are shown as two-dimensional scans mapping the number density of the objects. They are shown in two diagrams $(a,e)$ in the top row and $(i,e)$ in the bottom row. A darker shade of red indicates a higher number density of objects. Most asteroids are located in two regions: Trojans and Plutinos. There are also traces of scattered objects from both groups. Asteroids with perihelia between $5.2\,\au$ and $30\,\au$ (Centaurs) are much less numerous. There is an almost empty gap for $a > 40\,\au$ and $q \in (5.2, 30)\,\au$. The inclinations are mostly moderate $\lesssim 40\,$deg with less numerous objects with higher values of $i$ accompanied by higher values of $e$.

The right panel of Fig.~\ref{fig:Fig14} shows the results of backward integrations of $25$ ETNOs with $100$ clones for each. Clearly, most of the objects in the $(a,e)$--diagram are in the gap mentioned above. Therefore, ETNOs could have come from the disc between Jupiter and Neptune after being scattered. Depending on the semi-major axis reached after the scattering, the detachment from Neptune was slower or faster. If $a$ is moderate ($\sim 200-300\,\au$), the next visit to the region of giant planets occurs after one or more revolutions of the Sun around the centre of the Galaxy. If $a$ is large ($\sim 1000\,\au$), the next visit takes place even within a few Myr.

The $(i,e)$--diagram shows that there are very few objects with simultaneously low $i$ and $e$. This area is mainly occupied by Plutinos and Trojans (bottom left), which are less likely to get scattered and join the population of trans-Neptunians. Most of the past ETNOs have $i$ and $e$ at the boundary of the grouping of Plutinos and Trojans shown in the lower left panel of Fig.~\ref{fig:Fig14}. The majority of objects are in the $i \in (0, 60)\,$degree and $e \in (0.4, 0.9)$ region, therefore ETNOs more likely had prograde orbits in the past.

\begin{figure}
\centerline{
\includegraphics[width=0.4\textwidth]{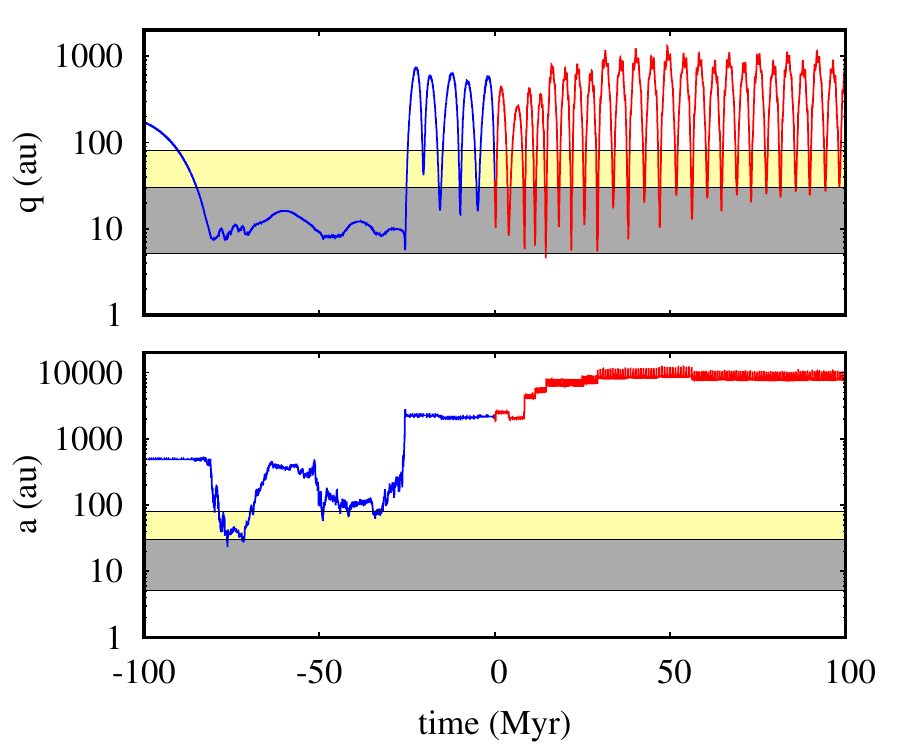}
}
\caption{Forward (red) and backward (blue) evolution of 2014~FE72 (one of the clones) for $\alpha = 2$. The grey area in the upper panel represents the distances between the orbits of Jupiter and Neptune, while the yellow colour means $q \in (30, 80)\,\au$.}
\label{fig:Fig15}
\end{figure}

Every time perihelion is in the planetary region, i.e. an ETNO becomes a Centaur, the evolution is chaotic, but the object is not necessarily ejected from the solar system. However, its orbit is significantly perturbed. An example of such an evolution is shown in Fig.~\ref{fig:Fig15} for 2014~FE72 (one of the clones). For $t = -100\,$Myr $a \sim 500\,\au$, while $q \sim 200\,\au$ and decreases. After reaching $q \sim 80\,\au$, it could be detected (according to current detection capabilities) and classified as ETNO. For $t \sim -75\,$Myr, the object has $a \sim 30-50\,\au$ and $q \sim 10\,\au$. After a series of close planetary encounters, the semi-major axis for $t \sim -25\,$Myr reached $1000-2000\,\au$, as observed today. After several more EFE cycles and close encounters with the planets, the semi-major axis grows to $\sim 10000\,\au$.

\begin{table*}
\caption{Orbital elements of retrograde ETNOs (JPL Small-Body Database, accessed 13 October 2022).}
\label{tab:tab1}
\begin{tabular}{l c c c c c c c}
\hline
Name & $a\,(\au)$ & $e$ & $i\,$(deg) & $\Omega\,$(deg) & $\omega\,$(deg) & $\mathcal{M}\,$(deg) & $q\,(\au)$\\
\hline
2019 EE6 & $165.5$ & $0.5488$ & $162.95$ & $201.04$ & $44.76$ & $355.77$ & $74.668$\\
2022 FM12 & $158.8$ & $0.6633$ & $172.45$ & $13.42$ & $175.64$ & $359.28$ & $53.465$\\
2022 FN12 & $135.9$ & $0.5651$ & $178.46$ & $254.65$ & $47.32$ & $1.15$ & $59.102$\\
\hline
\end{tabular}
\end{table*}

According to MOND, ETNOs are ephemeral objects rather than stable asteroids held together by a distant planet. Their semi-major axes, perihelion distances as well as inclinations vary significantly over time. A given object can evolve between a low-$a$ Centaur, a high-$a$ Centaur, an ETNO, in a prograde or retrograde orbit. Recently, the first three retrograde ETNOs were discovered and their orbits are explained in the next subsection within MOND.

\subsection{Retrograde ETNOs}

The orbital parameters of the three retrograde ETNOs are not well determined, since for two of them (2022~FN12 and 2022~FM12) there are only $6$ observations covering $2$ days, while for the third (2019~EE6) there are $7$ observations covering $63$ days. We use the orbital parameters listed in the JPL Small-Body Database (accessed 13 October 2022,  Tab.~\ref{tab:tab1}). All objects are detached from Neptune, $q > 50\,\au$, and retrograde, $i > 160\,$deg. The formal uncertainties of the orbital elements are not determined.

The question naturally arises whether or not these objects are dynamically connected with prograde ETNOs and Centaurs. To be related to the former, the inclinations should decrease, while to be related to the latter means that the perihelion distances should decrease. The lower left panel of Fig.~\ref{fig:Fig5} suggests that this is possible. The only potential problem lies in the time required to evolve into the low-$q$ or low-$i$ region of parameter space. The semi-major axes of all three asteroids are relatively low, i.e. $136.3\,\au$ (2022~FN12), $158.8\,\au$ (2022~FM12) and $165.5\,\au$ (2019~EE6). Their EFE-induced evolution thus takes place on timescales longer than the Sun's orbital period around the centre of the Galaxy (see Fig.~\ref{fig:Fig7}). For sufficiently low $a$, it could be that planetary perturbations strongly dominate the evolution and significant $q$ and $i$ variations are not possible.

\begin{figure*}
\centerline{
\hbox{
\includegraphics[width=0.32\textwidth]{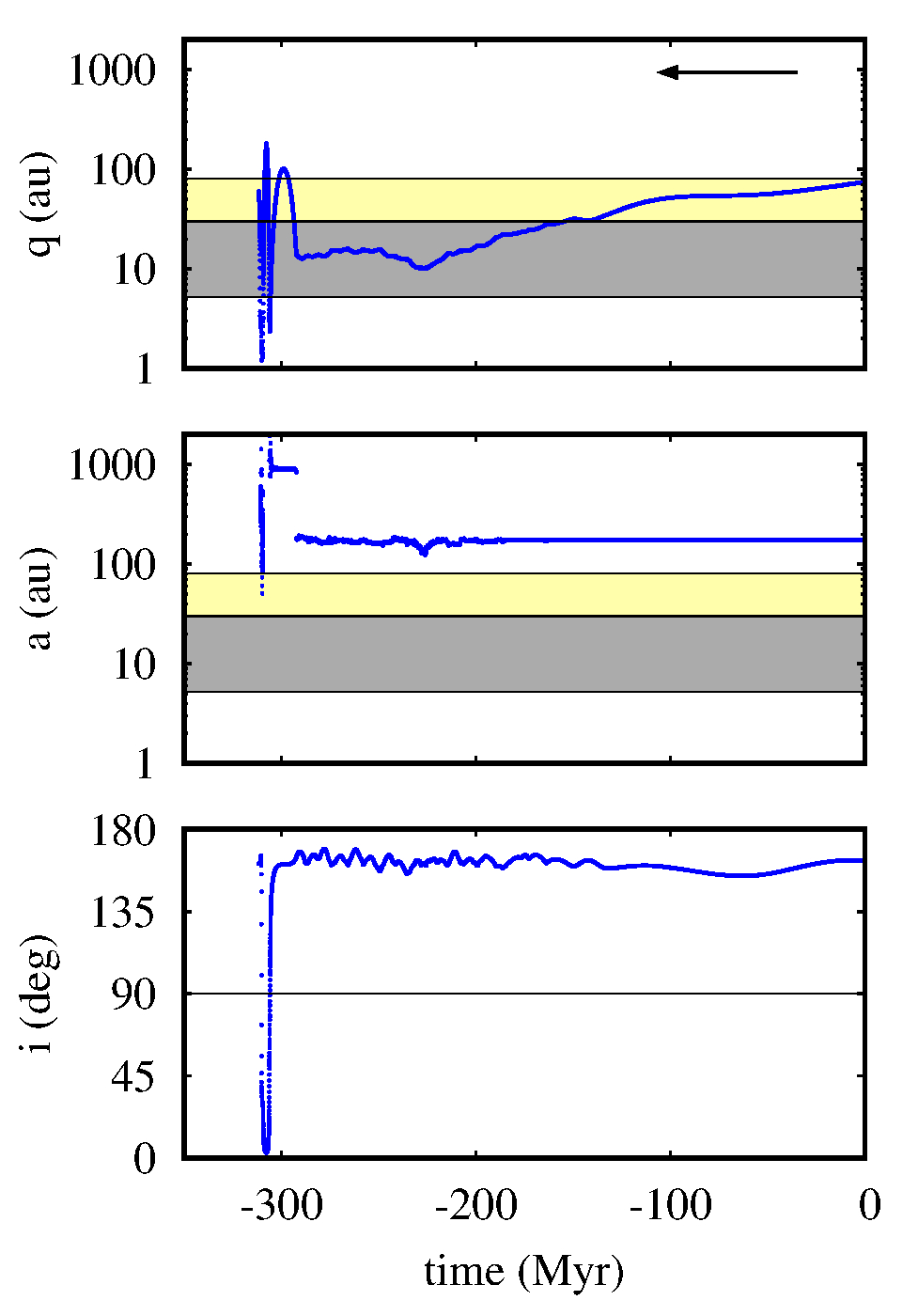}
\includegraphics[width=0.32\textwidth]{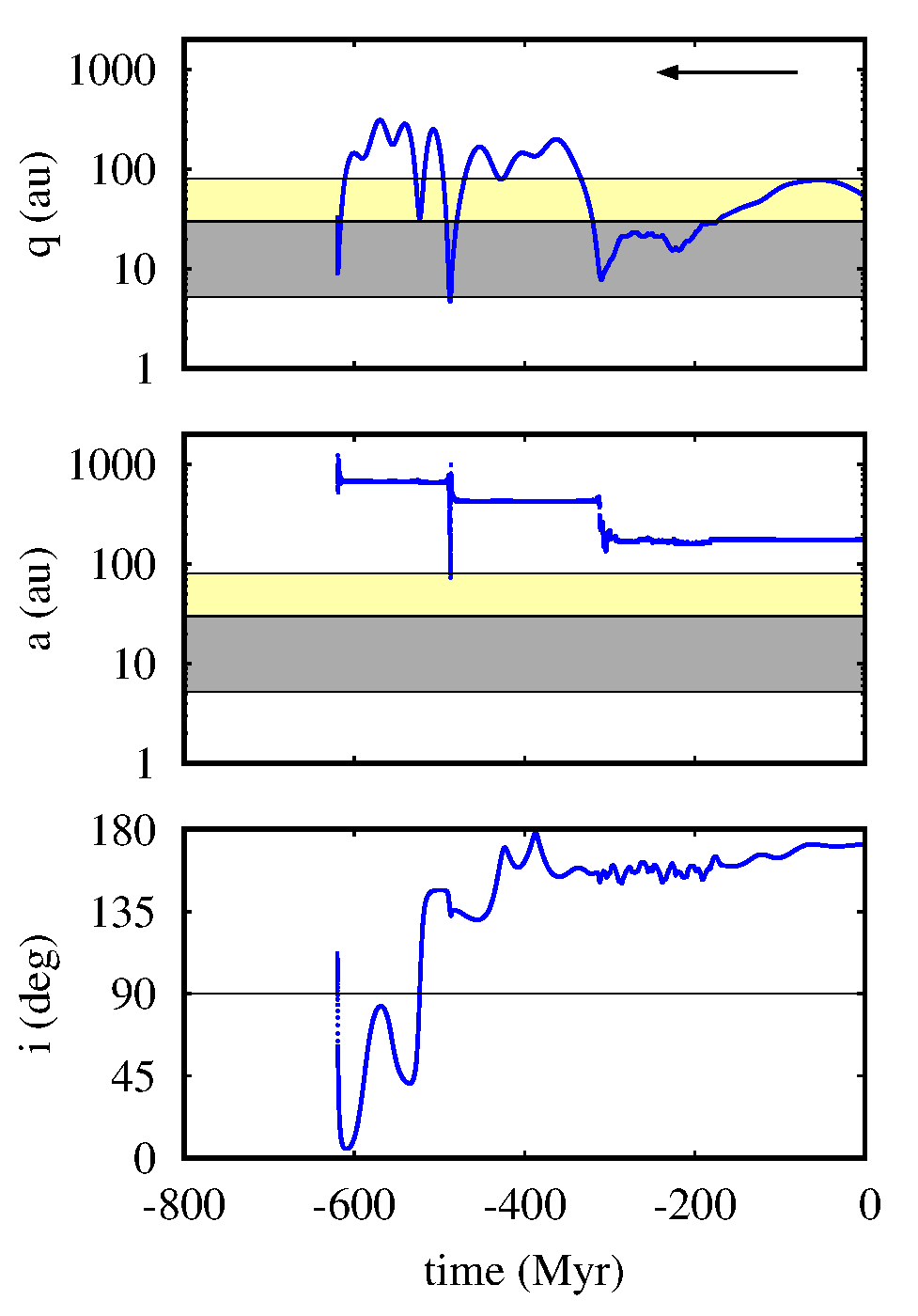}
\includegraphics[width=0.32\textwidth]{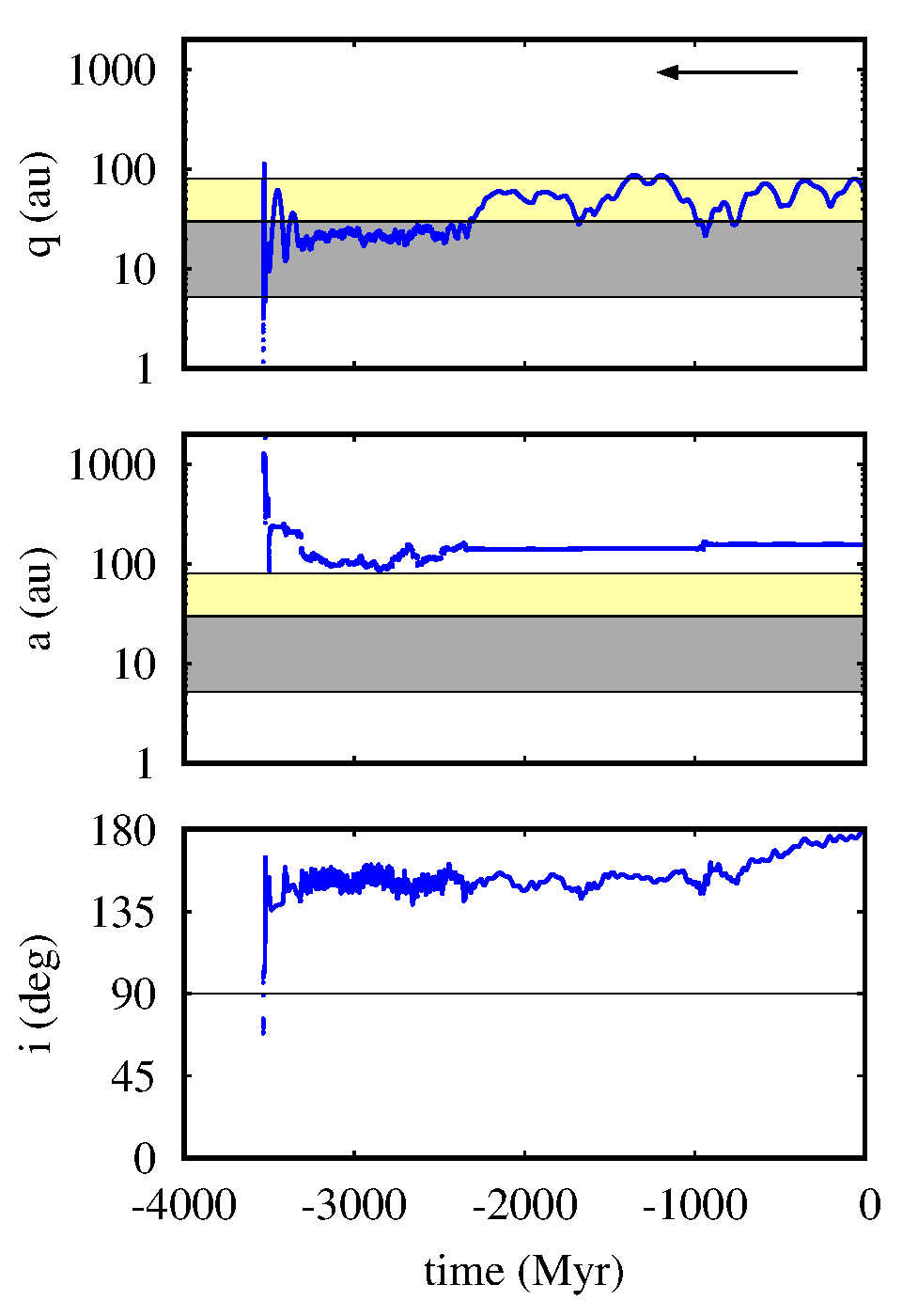}
}
}
\caption{Backward evolution of 2019~EE6 (left), 2022~FM12 (middle), 2022~FN12 (right). For each object, the evolution of one of its clones is shown. The results were obtained for $\alpha = 2$.}
\label{fig:Fig16}
\end{figure*}

\begin{figure*}
\centerline{
\hbox{
\includegraphics[width=0.32\textwidth]{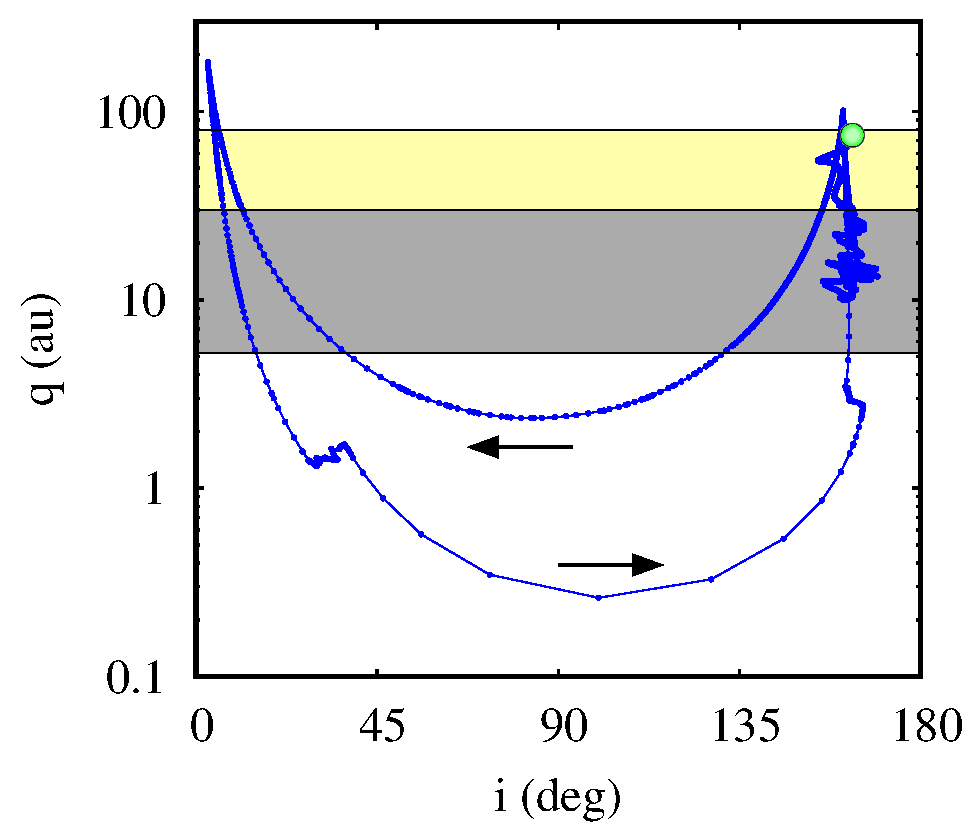}
\includegraphics[width=0.32\textwidth]{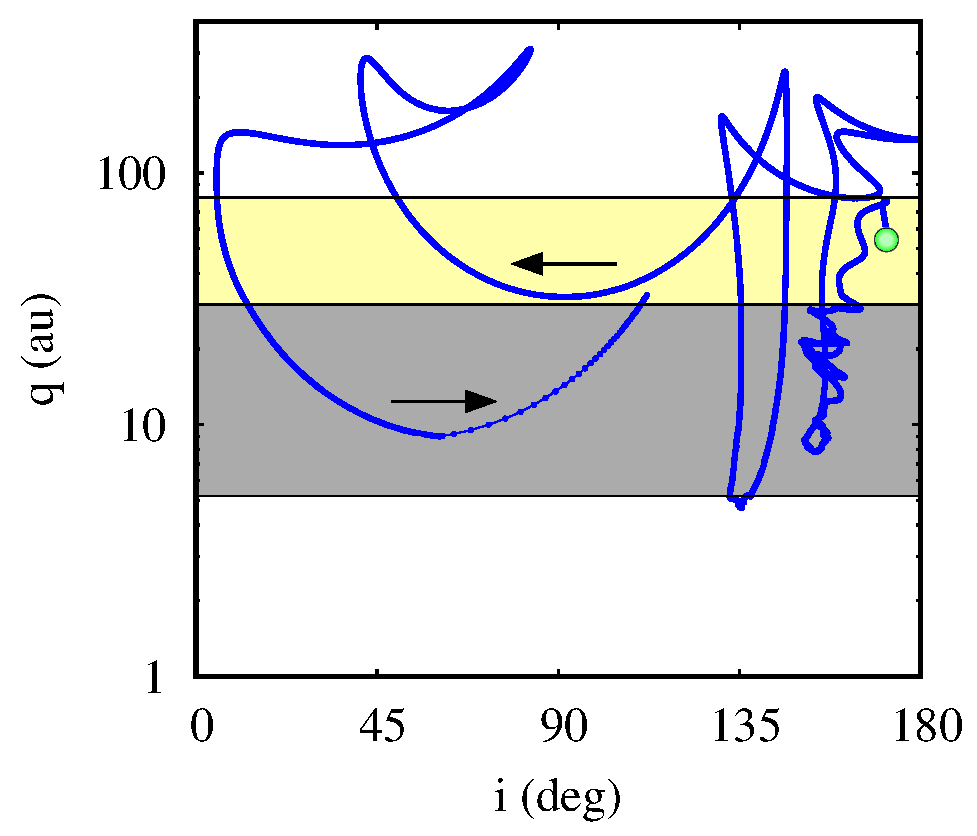}
\includegraphics[width=0.32\textwidth]{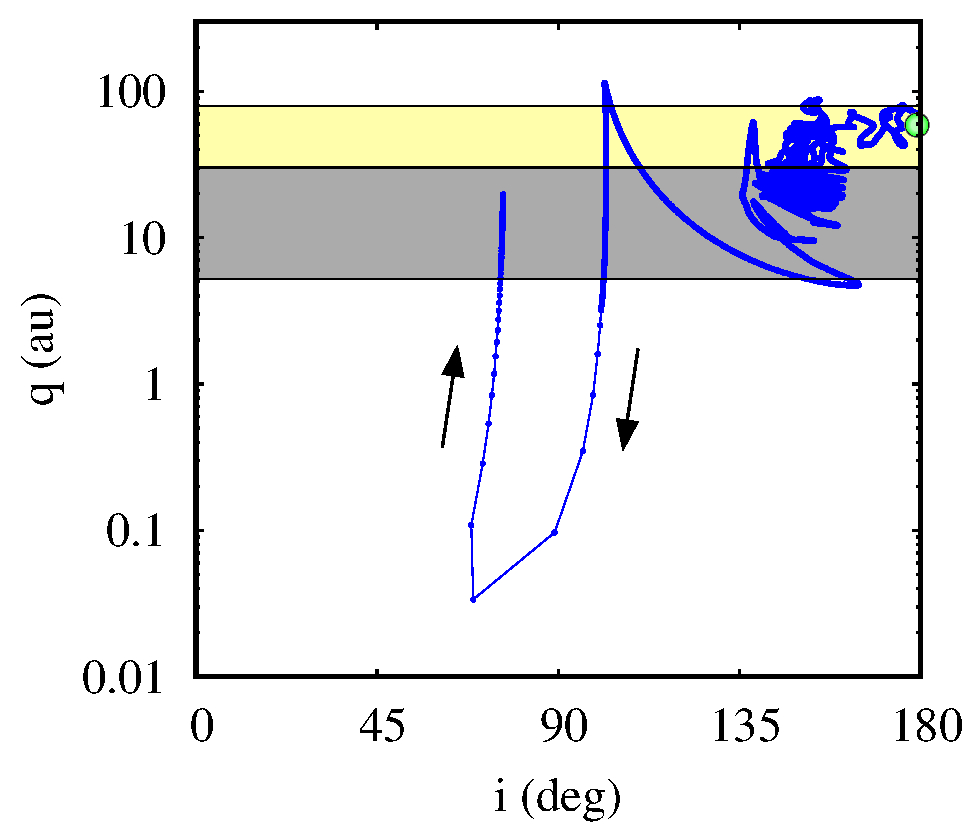}
}
}
\caption{Backward evolution of the same objects as in Fig.~\ref{fig:Fig16}, but shown on the $(i,q)$--diagrams. The arrows indicate the direction of the backward evolution.}
\label{fig:Fig17}
\end{figure*}

In order to verify that, we performed backward integrations with $10$ clones for each object chosen according to the Gaussian distribution with assumed uncertainties of the orbital elements of $\sigma_a = 20\,\au$, $\sigma_q = 1\,\au$ and $\sigma_{\idm{angle}} = 1\,$deg for the angles. Figure~\ref{fig:Fig16} illustrates the results of integrating selected clones of each retrograde ETNO.

The left column shows the results for 2019~EE6. The perihelion distance decreases and for $t \sim -150\,$Myr the asteroid becomes a Centaur. A further decrease of $q$ down to $10\,\au$ results in strong perturbations by Saturn. At time $t \sim -300\,$My the inclination decreases to $\sim 0$ and the orbit becomes prograde. Shortly thereafter, the orbit becomes retrograde again. The change in $i$ is accompanied by a significant change in $a$ and $q$. The left panel of Fig.~\ref{fig:Fig17} shows the evolution of the same object in the $(q, i)$--diagram, where the U-shaped trajectory can be seen, which is similar to that shown in the lower left panel of Fig.~\ref{fig:Fig5}. Clearly, 2019~EE6 could have been a prograde ETNO or a Centaur in the past.

In the middle column of Fig.~\ref{fig:Fig16} the results for 2022~FM12 are shown. At epoch $\sim -200\,$Myr the object becomes a Centaur. The orbit is perturbed by the planets. In particular, the perturbation by Saturn leads to a significant increase in $a$, which in turn leads to a faster evolution of $q$. The perihelion distance first increases up to $\sim 200\,\au$ and then decreases again down to the planetary region. At the same time, the inclination decreases to moderate values. The perihelion leaves the planetary region and returns, with the inclination decreasing further. In the epoch $\sim -600\,$Myr the inclination is as low as a few degrees. The evolution of this object can be followed in the middle panel of Fig.~\ref{fig:Fig17}. The trajectory is more complicated than that of the previous object, but a characteristic U-shape can still be seen. Similar to 2019~EE6, 2022~FM12 may also have originated in the Centaurs region and passed through the prograde ETNO phase, with the semi-major axis of a few hundred astronomical units.

The third object 2022~FN12 has the smallest semi-major axis among the retrograde ETNOs, i.e. $a = 135.9\,\au$, and therefore it is more difficult to experience a significant variation of $q$. More clones (we tested $50$ clones) were needed to find a configuration that evolves to a Centaur (the right column of Figs.~\ref{fig:Fig16} and \ref{fig:Fig17}). The initial semi-major axis of this clone is $162\,\au$, which is significantly larger than the nominal value. However, the uncertainties in the orbital parameters for the retrograde ETNOs are not known and it is possible that the orbit of 2022~FN12 is wider than currently reported in the database. About $800$~Myr are needed to reach $q < 30\,\au$ and about $3.5\,$Gyr to reach $i < 90~$deg. The clone reaches the region inside Earth's orbit and the perihelion distance increases again. The object is eventually scattered by Uranus.

Despite the difficulties of reaching a low-$q$/low-$i$ region by 2022~FN12, the tests described above suggest that EFE could in principle explain the origin of retrograde ETNOs. Future improvements of their orbits and possibly new objects of this type would help to test the MOND hypothesis even better.

\begin{figure*}
\centerline{
\hbox{
\includegraphics[width=0.32\textwidth]{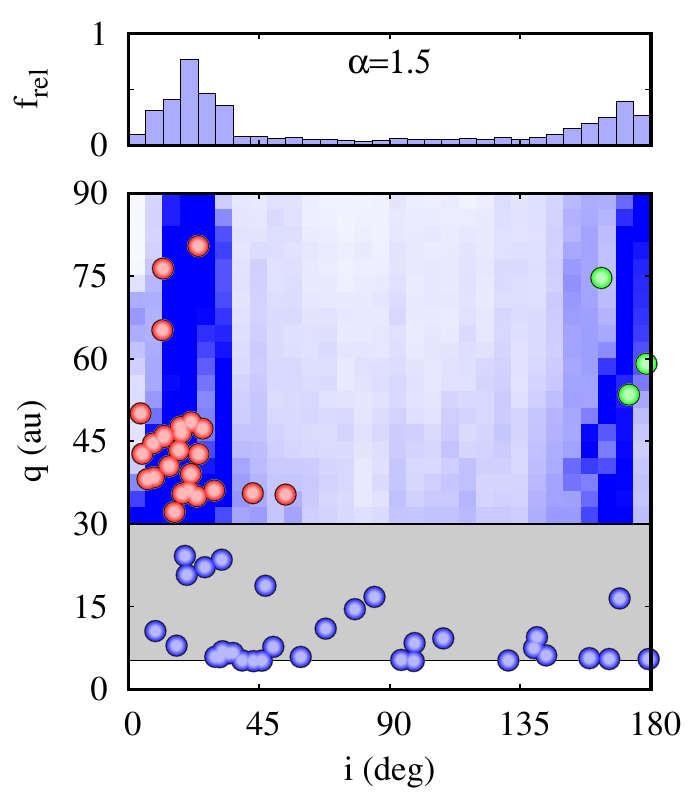}
\includegraphics[width=0.32\textwidth]{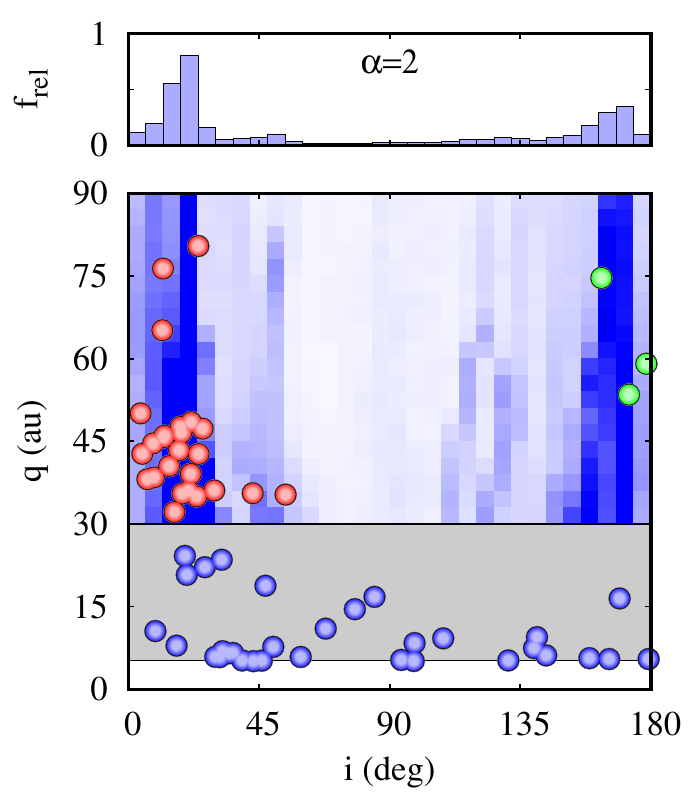}
\includegraphics[width=0.32\textwidth]{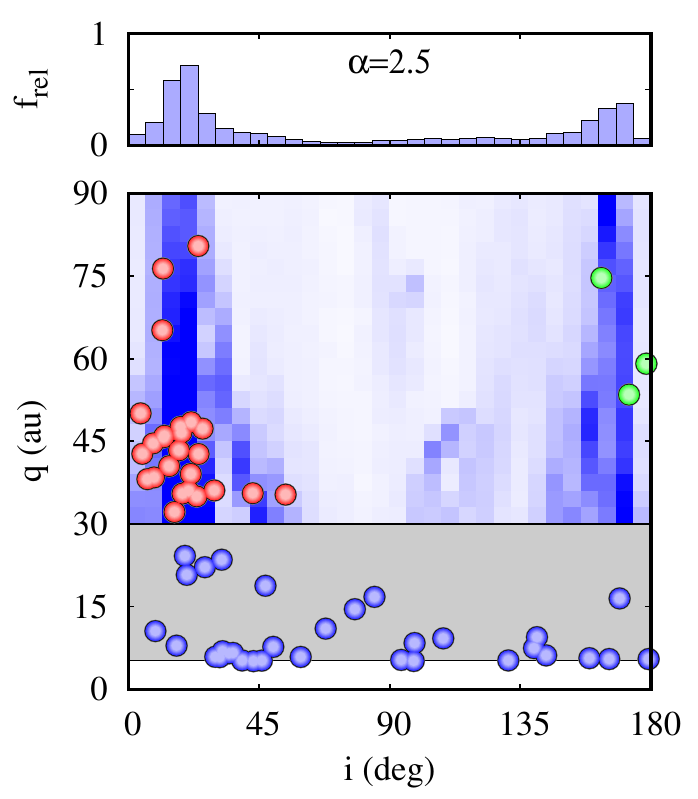}
}
}
\caption{Probability of visiting certain parts of the $(i, q)$--diagram by evolving known Centaurs with initial $a \in (250, 2000)\,\au$ for $\pm 20\,$Myr. Each column shows the results for different $\alpha$ ($1.5$, $2$ and $2.5$, from left to right).}
\label{fig:Fig18}
\end{figure*}

\subsection{Centaurs}

In the previous section it was shown that retrograde ETNOs belong to the same group of trans-Neptunians as prograde ETNOs and Centaurs, only at a different stage of evolution. In this section we show the dynamic relationship between the three subgroups through statistical analysis rather than by looking at the evolution of individual objects. We start with the orbits of $32$ Centaurs with $a \in (250, 2000)\,\au$. For each object, apart from the nominal configuration, we have chosen $10$ clones according to the uncertainties of the orbital elements. Such a set of $352$ objects is integrated forward and backward for $20\,$Myr. The choice of this relatively short period was dictated by the comparison between the synthetic and the observed ETNOs, which must be performed for a given epoch of the Sun's motion in the Galaxy. We performed the simulations for three different values of $\alpha = 1.5, 2, 2.5$.

The left panel of Fig.~\ref{fig:Fig18} shows the results for $\alpha = 1.5$. The shades of blue in the $q > 30\,\au$ part of the diagram depicts the density of the trajectories in phase space, where darker means higher density. We take into account the observational bias for the ecliptic latitudes $|b| < 30\,$deg and only orbits that meet this criterion are used to calculate the density. The densest region is clearly for $i < 40\,$deg, with a maximum around $20\,$deg, while the second peak is for $i > 140\,$deg. The smaller histogram above the $2$-dimensional scan shows density as a function of $i$, assuming $q \in (30, 90)\,\au$.

\begin{figure*}
\centerline{
\vbox{
\hbox{
\includegraphics[width=0.32\textwidth]{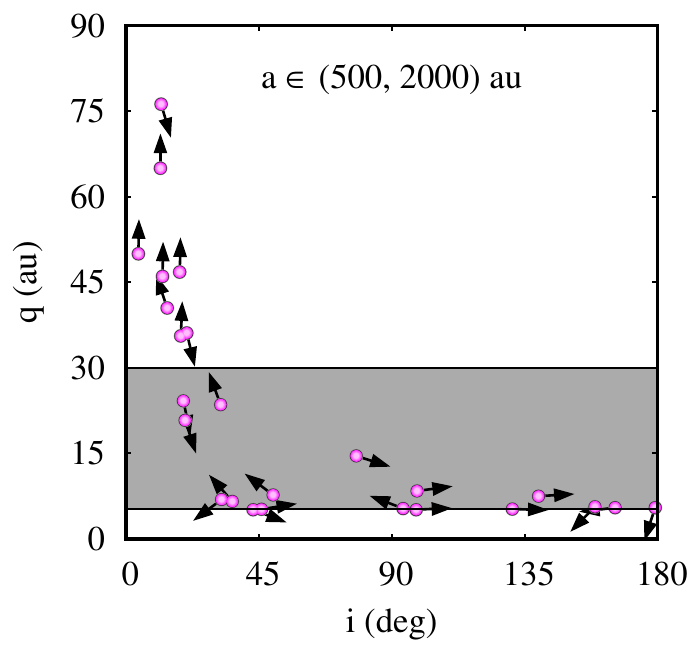}
\includegraphics[width=0.32\textwidth]{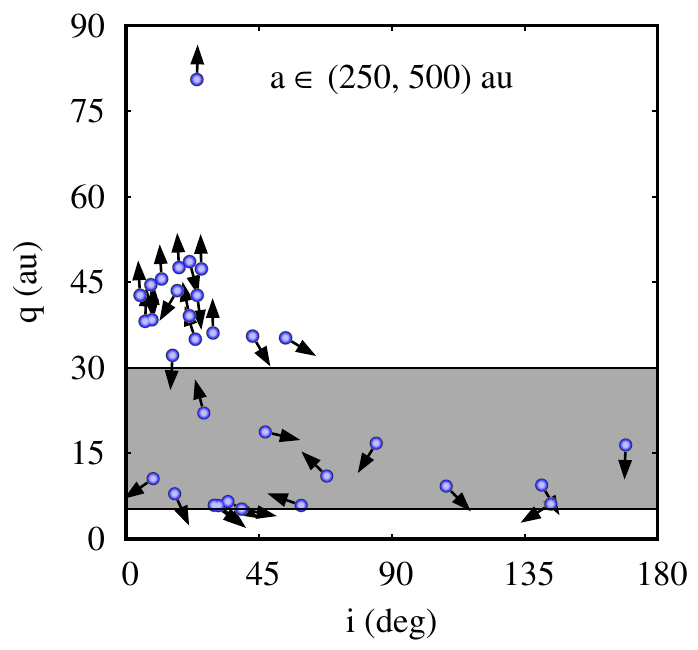}
\includegraphics[width=0.32\textwidth]{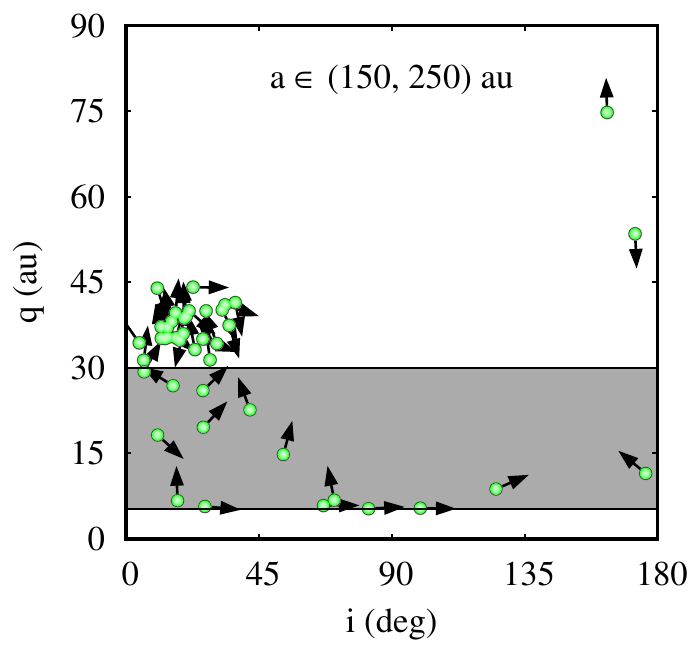}
}
\hbox{
\includegraphics[width=0.32\textwidth]{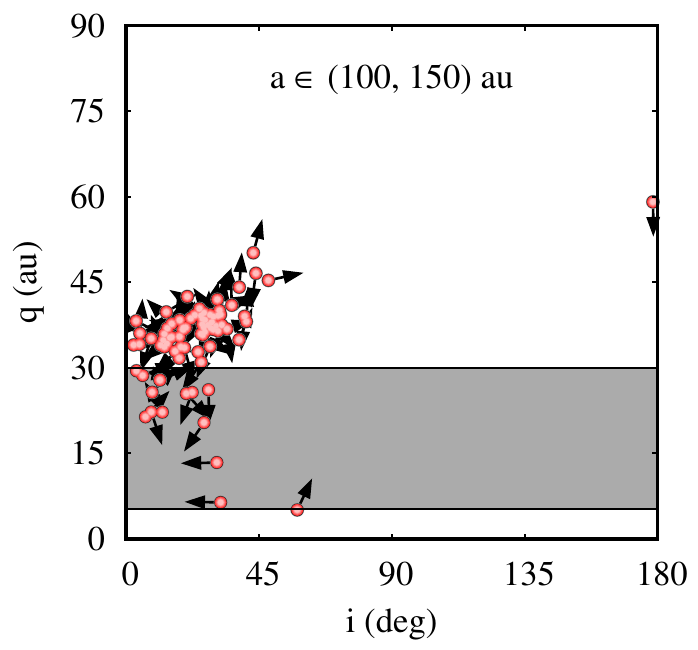}
\includegraphics[width=0.32\textwidth]{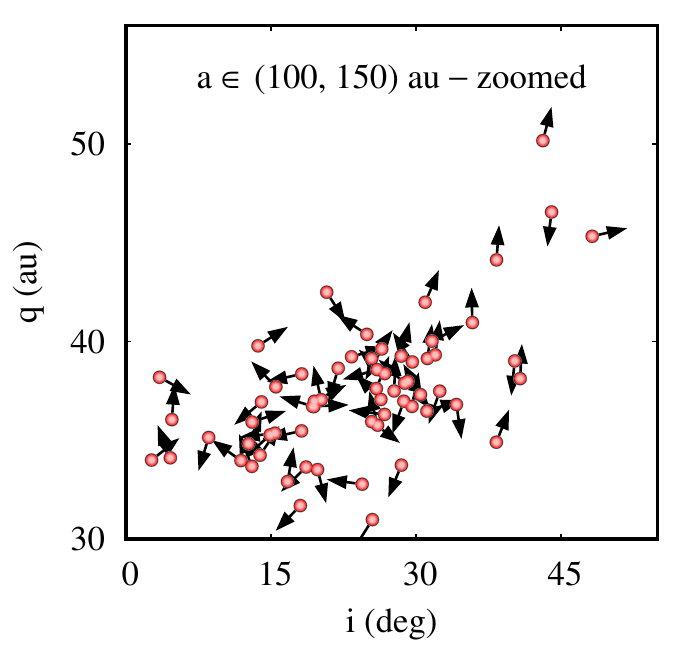}
\includegraphics[width=0.32\textwidth]{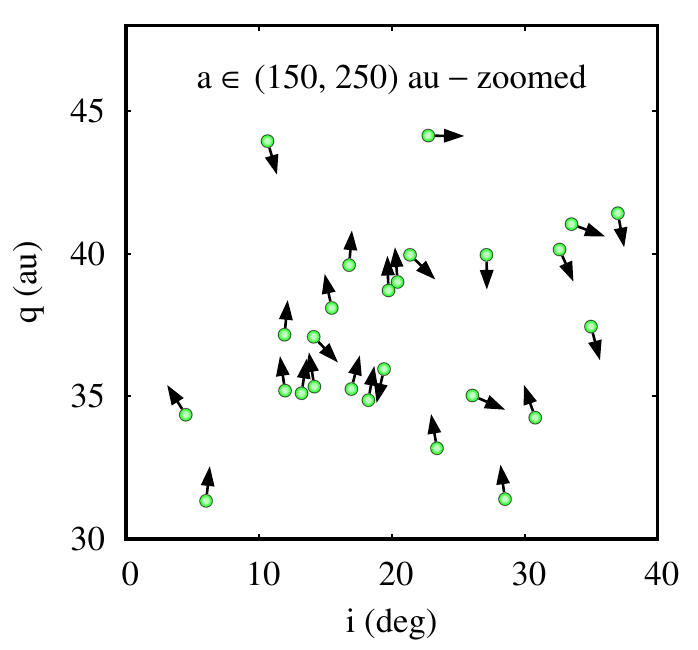}
}
}
}
\caption{The distribution of orbital elements in the $(i, q)$--diagram for different ranges of $a$ (given at each panel). The arrows show the direction of evolution under the planetary perturbations and the EFE with $\alpha = 2$.}
\label{fig:Fig19}
\end{figure*}

This is consistent with the observations of both the prograde and retrograde ETNOs. However, it is noticeable that the two ETNOs with $i \sim 45\,$deg are in the relatively low density region. The density is slightly higher in this region for $\alpha = 2$ and $2.5$ (see middle and right panels). However, the overall results are qualitatively the same for the three values of $\alpha$. This experiment shows that EFE can explain the transition between different subgroups of trans-Neptunians, but the statistics of the orbital elements cannot be used to distinguish between different values of the interpolation parameter $\alpha$.

\subsection{EFE--induced flows of trans-Neptunians}

The uneven distribution of the orbital elements $\Omega, \omega, \varpi$, which seem to cluster around certain values, have been interpreted as indicating the existence of an additional planet in the solar system \citep{Batygin2016a}. Before checking whether the Milgromian dynamics can explain these facts without Planet Nine, we show that there are additional features in the distribution of orbital elements that point to EFE as the main factor for the statistics of ETNOs.

Figure~\ref{fig:Fig11} shows the distribution of $i$ and $q$ for TNOs, while the colours and sizes of the symbols encode the values of $a$. After dividing the objects into four groups according to $a$, i.e. $a \in (500, 2000)\,\au$, $a \in (250, 500)\,\au$, $a \in (150, 250)\,\au$ and $a \in (100, 150)\,\au$, we find that the distributions for each group differ from each other (see Fig.~\ref{fig:Fig19}). In addition, we can add arrows indicating the direction of evolution of each object when both the EFE and planetary perturbations are considered.

The upper left panel of Fig.~\ref{fig:Fig19} shows the $(i, q)$-- diagram of the asteroids with the widest orbits, i.e. $a \in (500, 2000)\,\au$. We use the upper limit of $a$ to exclude extremely wide orbits of Centaurs and keep the maximum $a$ for them in the same order of magnitude as the maximum $a$ for ETNOs. The points form a narrow strip that is nearly vertical for $q > 30\,\au$ and nearly horizontal for $q < 10\,\au$, with a transition between these limiting cases. The arrows show that the flow of objects is consistent with the observed distribution. In the vertical part, the arrows point mainly in the up/down direction, while in the horizontal part, the arrows point mainly in the left/right direction. For wide orbits, the perturbations from the giant planets are much weaker than the EFE perturbations (see Fig.~\ref{fig:Fig7}) and the EFE should dominate. Such a correspondence between flow and distribution clearly indicates that EFE could be responsible for the distribution.

In the upper middle panel of Fig.~\ref{fig:Fig19} the $(i, q)$--diagram for $a \in (250, 500)\,\au$ is shown. The velocity distribution is less ordered, especially in the Centaurs region. In addition to the main flow, which is similar to the previous one, a considerable amount of random motion can be seen. The distribution of orbital elements is also less tight. In the ETNOs region ($q > 30\,\au$), both the distribution of orbital elements and the velocities are relatively well constrained, especially for $i < 30~$degrees. The picture shown in this panel is to be expected for the intermediate range of $a$. The EFE still dominates over the planetary perturbations for $a \in (250, 500)\,\au$, but for $a \sim 250\,\au$ the two perturbations are already of equal magnitude (see Fig.~\ref{fig:Fig7}).

For even smaller $a$, both the distribution and the flow become less confined. The top-right and bottom-left panels of Fig.~\ref{fig:Fig19} show the diagrams for $a \in (150, 250)\,\au$ and $(100, 150)\,\au$ respectively. The bottom-middle and bottom-right panels show the close-ups of the ETNOs region. For $a \in (150, 250)\,\au$ the points form a clump with random motions. This is to be expected in the planet--dominated range of $a$. For $a \in (100, 150)\,\au$ a positive correlation between $i$ and $q$ appears, consistent with the flow. In this range of $a$, the planetary perturbations strongly dominate over EFE and the correlation and the flow can be attributed to the conservation of the projection of angular momentum onto the $z$ axis. This is expected for the perturbation by the giant planets, which is almost axisymmetric. The absence of low $i$ and high-$q$ objects among TNOs with $a < 150\,\au$ is consistent with the formation scenario described in \citep{Nesvorny2016,Kaib2016,Anderson2021}, combining the crossing of mean motion resonances with Neptune during its early migration and the Lidov--Kozai mechanism \citep{Lidov1962,Kozai1962}.

The experiments described above show that the observed distributions of the orbits agree with the Milgromian model of gravity. In the next section we try to verify whether MOND can explain the non-uniform distribution of the longitudes of the ascending nodes as well as the longitudes of the perihelia of ETNOs.

\section{Orbital plane clustering and apsidal confinement}
\label{sec:clustering}

In the previous section we showed that the evolution of trans-Neptunians is strongly chaotic. Moreover, depending on its evolutionary phase, a given object can be classified as a Centaur or as a prograde/retrograde ETNO with smaller or greater semi-major axis. Some of the known ETNOs may have visited the giant planet region only a few Myr ago, resulting in a significant and unpredictable change in their orbits due to the chaos and uncertainties of the orbits. Other objects were probably scattered from the giant planet region at least a few hundred Myr ago.

This means that it is very difficult, if not impossible, to choose a specific initial epoch and a synthetic initial set of orbits for the numerical simulations, which are then compared with the observational data. Instead, we chose the currently observed objects as initial orbits. The orbits were then integrated forward and backward for $\pm 400\,$Myr. The difference from the previous experiment is that we selected all objects with $a > 50\,\au$ and $q > 5\,\au$. There were $978$ objects in our sample. The reason for this selection is that while we use the criteria $q \in [42, 85]\,\au$ and $a \in [150, 2000]\,\au$ for comparison with observations, the initial sample must include all objects whose orbits could possibly evolve into these ranges (at least for a limited time). Furthermore, we are not looking for a particular fixed cluster in parameter space, since the position of the cluster in $(\Omega, \varpi)$--space may vary over time and ETNO orbits at a particular epoch may nevertheless be clustered.

Another feature of the distribution of TNOs studied in this experiment was the clustering of orbital planes of objects with $q>30\,\au$ and relatively small semi-major axes, $a \in [50, 100]\,\au$ \citep{Volk2017}. Unfortunately, our simulations show that MOND does not prevent the randomisation of orbits with $a \in [50, 100]\,\au$ after a few Myr. Interestingly, \citep{Brown2018} show that this is also true for Planet Nine. We discuss a possible solution to this problem in the next section. Here we focus on the clustering of orbits with $a > 150\,\au$.

\begin{figure*}
\centerline{
\hbox{
\includegraphics[height=0.9\textwidth]{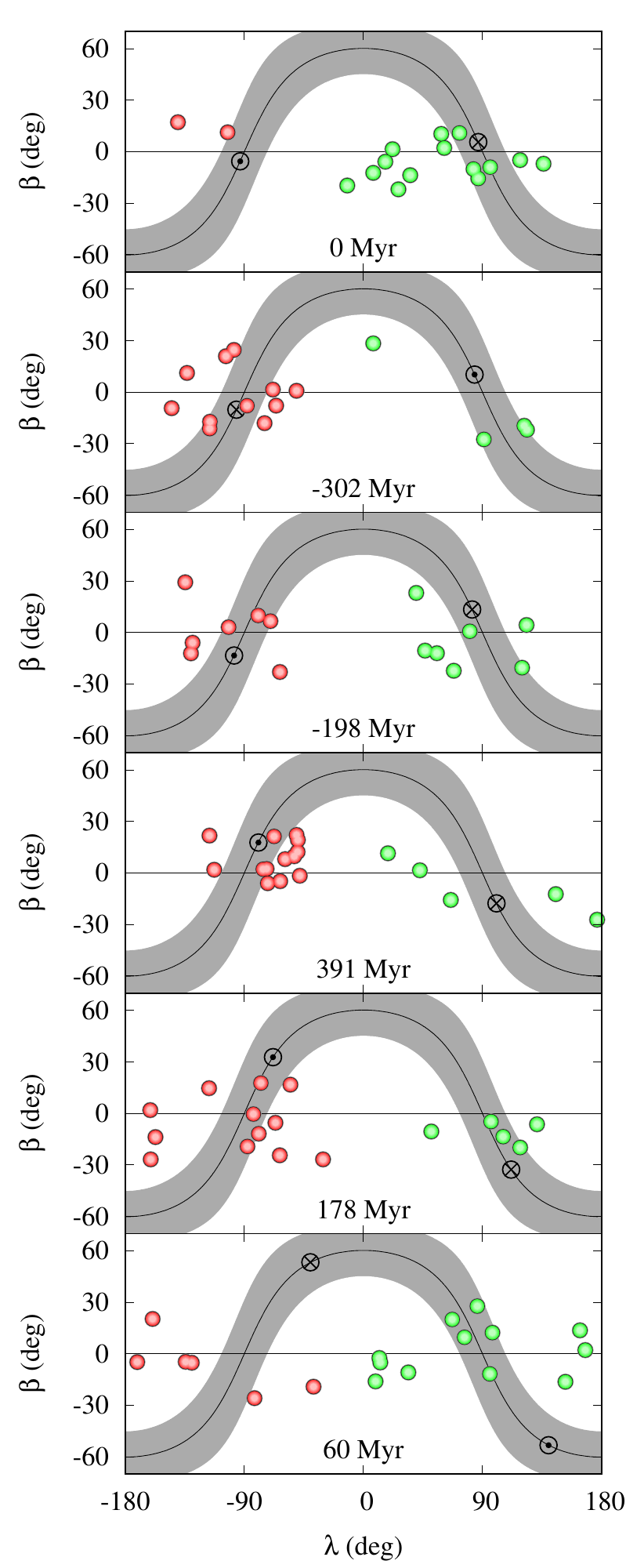}
\includegraphics[height=0.9\textwidth]{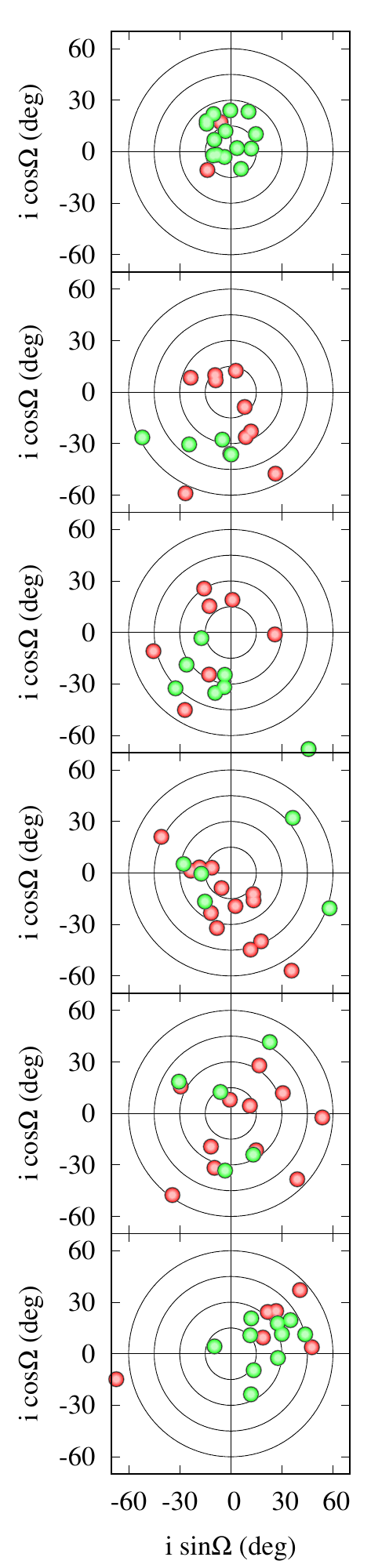}
}
}
\caption{Selected past and future epochs (labelled in the left column) of integrated objects shown in the $(\lambda, \beta)$-- and $(i \cos\Omega, i \sin\Omega)$--diagrams (left and right respectively). The selection criteria are $a \in (150, 2000)\,\au$, $q \in (42, 85)\,\au$, $|b| < 30\,$deg. The grey area in the left column denote the Galaxy disc, while the positions of the Galaxy centre and anticentre at the respective epochs are marked with a circle--dot and a circle--cross symbol, respectively. The division into two groups of objects (red and green) corresponds to the two currently observed modes of distribution, near the centre and the anticentre of the Galaxy.}
\label{fig:Fig20}
\end{figure*}

Figure~\ref{fig:Fig20} shows the distribution of orbital elements for a few selected epochs in the evolution of the trans-Neptunians. The top row shows the current epoch. Most of the objects have perihelia near the anticentre of the Galaxy (the cluster is shifted by $\sim 45\,$deg from this direction), only two of them have perihelia near the Galaxy centre. There are gaps between the two groups. A similar picture is repeated in different past and future epochs. The three selected epochs $-302, -198$ and $391\,$Myr correspond to the situation where the centre/anticentre of the Galaxy is close to the ecliptic. In each of these cases, the perihelia confinement is clearly visible. For higher ecliptic latitudes of the Galaxy centre/anticentre (the bottom two rows), the confinement may be less clear (especially for $t = 60\,$Myr).

The right column of Fig.~\ref{fig:Fig20} shows the distributions in the respective epochs in the $(i \cos\Omega, i \sin\Omega)$--diagram. Clearly, the inclinations in past/future epochs are generally larger than today. Nevertheless, the majority of orbits have $i \lesssim 60\,$deg. The experiments with Planet Nine show a similar excitation of inclinations \citep{Shankman2017,Batygin2021}. One can also observe that $\Omega$ is unevenly distributed, which is consistent with the observations.

\begin{figure}
\centerline{
\includegraphics[width=0.4\textwidth]{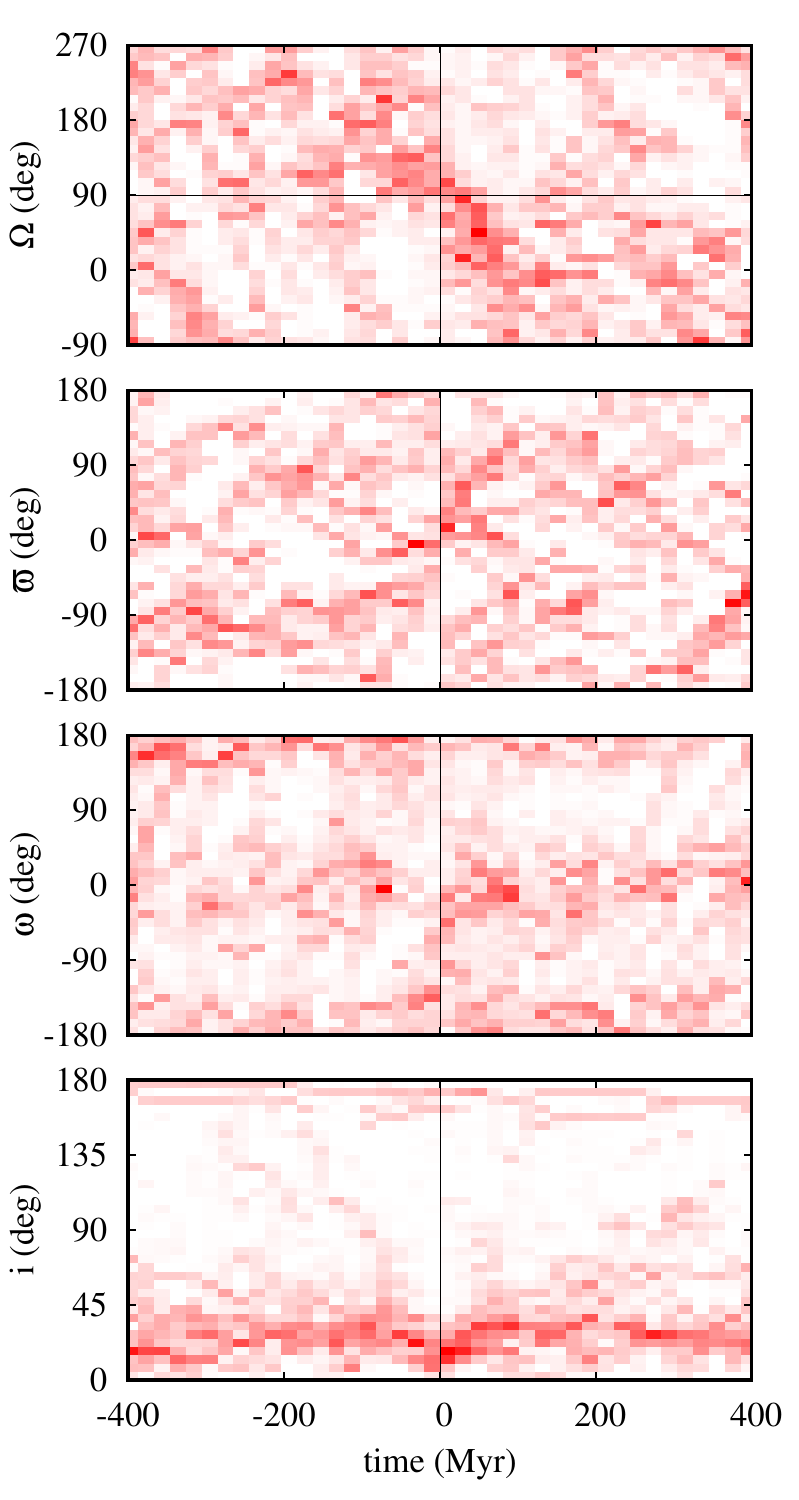}
}
\caption{Density maps for the Euler angles as a function of epoch resulting from the $N$-body integrations of all objects with current $a > 50\,\au$ and $q > 5\,\au$. The selection criteria for the plot are $a \in (150, 2000)\,\au$, $q \in (42, 85)\,\au$ and $|\beta| < 30\,$deg. The MOND parameter $\alpha = 2$.}
\label{fig:Fig21}
\end{figure}

The results presented in Fig.~\ref{fig:Fig20} show only selected epochs where both the clustering of orbits and the confinement of perihelia are relatively clear. Figure~\ref{fig:Fig21} shows the full results in the form of density maps for the angles $\Omega, \varpi, \omega, i$ as a function of epoch. The upper panel shows the map for $\Omega$. It is centred at $90\,$degrees, which is close to the current cluster. Within $\pm 100\,$Myr most ETNOs are clustered, but the position of the cluster centre varies with time. At more distant epochs, most objects have $\Omega \in (-\pi/2, +\pi/2)$, while at past epochs $\Omega \in (\pi/2, 3 \pi/2)$. For more distant epochs, the clustering is less clear. However, the sample of objects considered in this study cannot be complete, as some of the objects that are not observed today (e.g. due to $q \gtrsim 85\,\au$) may become detectable in a distant epoch and objects observed today are only temporarily detectable.

\begin{figure}
\centerline{
\includegraphics[width=0.44\textwidth]{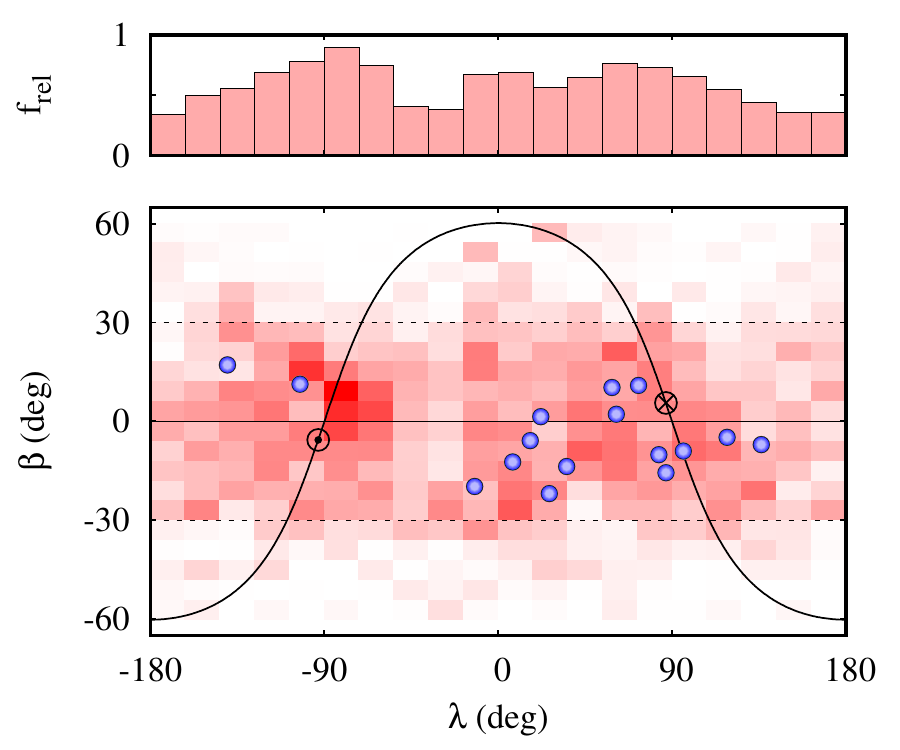}
}
\caption{A density map for the experiment presented in Fig.~\ref{fig:Fig21}, shown here as ecliptic coordinates of the perihelia. The top panel shows the one-dimensional histogram for $|\beta| < 30\,$deg.}
\label{fig:Fig22}
\end{figure}

The second panel from the top shows the density map for $\varpi$. It is less clear than that for $\Omega$, although configurations with $\varpi \sim \pm \pi/2$ are preferred. The confinement in $\varpi$ corresponds to the confinement in the ecliptic longitude of perihelia. Figure~\ref{fig:Fig22} shows a density map in the ecliptic coordinates. The confinement can also be analysed in a one-dimensional representation (upper panel). The configurations with $\lambda$ near the centre/anti-centre of the Galaxy are favoured, but there are also a significant number of objects with $\lambda \sim 0$, which is consistent with the observations (the blue symbols). The maximum near the Galaxy centre is at least as strong as that near the anticentre. The observed asymmetry is probably due to the observational bias.

The density map for the argument of perihelion is shown in the third panel of Fig.~\ref{fig:Fig21}. Configurations with $\omega$ near $0$ or $180\,$deg are preferred, due to the observational limitation $|\beta| < 30\,\deg$. In the lower panel of Fig.~\ref{fig:Fig21} the map for $i$ is shown. Most of the objects are in orbits of low-to-moderate inclination. However, there are some ETNOs whose inclination exceeds $90\,$degrees and whose orbits become retrograde. This is consistent with the observations and analysis presented in this paper.

\section{Discussion and future work}
\label{sec:discussion}

The problem of modelling the clustering of orbital planes as well as the apsidal confinement remains not fully solved. On the one hand, the observations are not yet completely conclusive. The features may still be artefacts or are weak. On the other hand, the simulations do not show beyond doubt that the model produces strong non-uniformities in the distribution of the orbital elements, although the results presented above show moderate clustering and confinement. More data are definitely needed to verify the properties of the ETNOs. The model should also be improved.

Such an improvement could be based on the consideration of non-zero asteroid masses. \cite{Madigan2016} have shown that the inclusion of interactions between the small masses significantly affects the dynamics. \corr{The significance of self-gravitation between the trans-Neptunians has been also demonstrated by \cite{Sefilian2019}, who showed that a massive and moderately eccentric disc of TNOs can balance the giant planets--induced apsidal precession of highly eccentric ETNOs, resulting in their apsidal confinement.} Such an extended model \corr{of massive TNOs} will be investigated within MOND in a future work.

The clustering of orbits is not the only feature that could possibly be explained by the self-gravitating TNOs. The obliquity of the Sun with respect to the invariant plane of the solar system is another puzzle to be solved. The solar spin is tilted by $I_{\odot} = 7.155 \pm 0.002\,$deg and the orientation of the solar equator is given by $\Omega_{\odot} = 73.5 \pm 0.1\,$deg \citep{Beck2005} with respect to the ecliptic-equinox J2000.0, while the invariant plane has the parameters $I_{\idm{inv}} = 1.58\,$deg, $\Omega_{\idm{inv}} = 107.58\,$deg \citep{Souami2012}.

There have been several attempts in the literature \citep{Bailey2016,Gomes2017} in which authors tried to reconstruct the obliquity as a result of the gravitational interaction of the giant planets with Planet Nine. If the giants are initially in the same plane as the solar equator, the obliquity can be achieved if Planet Nine is in an inclined orbit. The inclinations of the giant planets' orbits increase due to the exchange of angular momentum with Planet Nine. However, there is a problem with such a scenario. The initial invariant plane, defined as the total angular momentum of five planets, i.e. the known giants and Planet Nine, must initially be inclined with respect to the solar equator. However, it could work if Planet Nine was a free-floating planet captured by the Sun. If Planet Nine was formed together with the known planets in a common disc, the initial inclination between P9 and the four giants would be small and could not increase.

Another way of solving this puzzle is to notice that the planets alter the solar spin \citep{Lai2016}, which precesses around the vector of the total angular momentum of the planets. However, even in this case, the initial obliquity is required, which cannot be generated from initial values $\sim 0$. In contrast, Milgrom's gravity can lead to inclination excitation even if the inclinations are initially zero.

\begin{figure}
\centerline{
\hbox{
\includegraphics[height=0.21\textwidth]{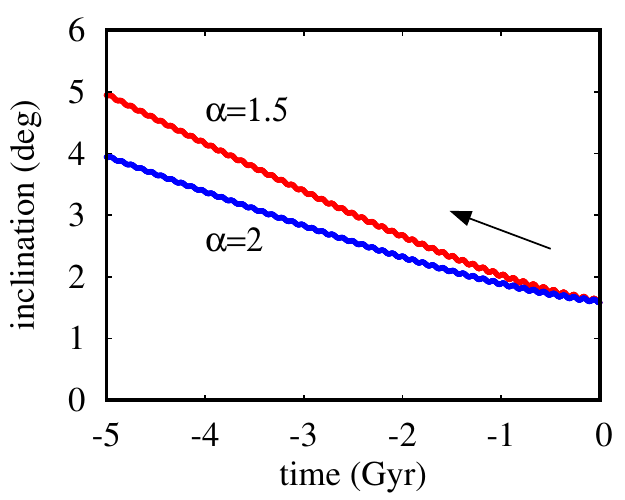}
\includegraphics[height=0.21\textwidth]{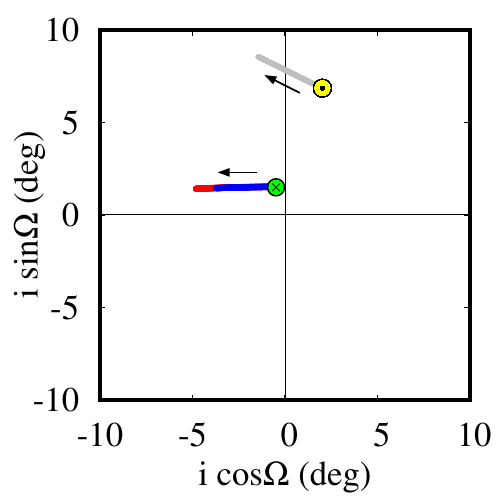}
}
}
\caption{The evolution of the invariant plane of the solar system (approximated by the total angular momentum of the four giant planets) resulting from the MOND perturbation for $\alpha = 1.5$ (red) and $\alpha = 2$ (blue). The left panel: the temporal evolution of the inclination; the right panel: the evolution shown in the $(i \cos\Omega, i \sin\Omega)$--diagram and compared with the evolution of the solar spin (yellow symbol and grey evolution trace).}
\label{fig:Fig23}
\end{figure}

As explained earlier in this paper, the change in inclination results from the precession of the orbit with respect to the axis of symmetry of EFE (the Galaxy centre-anticentre axis). The mechanism is very efficient for trans-Neptunians, but also works for giant planets, just much less efficiently. Figure~\ref{fig:Fig23} shows the backward evolution of $i$ and $\Omega$ of the Laplace plane of Jupiter, Saturn, Uranus and Neptune perturbed by EFE for $\alpha = 1.5$ (red) and $\alpha = 2$ (blue). Clearly, the inclination increases by $\sim 3\,$deg over the lifetime of the solar system. This is still not enough to reconstruct the observed solar obliquity, but more importantly, the evolution of $\Omega$ is not consistent with the observed solar system parameters. In the right panel of Fig.~\ref{fig:Fig23}, the evolution is shown in the $(i \cos\Omega, i \sin\Omega)$--diagram. The spin of the Sun also evolves due to the interaction with the planets \citep[we used a simple model in][]{Lai2016}. The obliquity does not decrease for past epochs, but even increases slightly.

Trans-Neptunian objects with non-zero mass could hopefully solve this problem, but non-restricted large-$N$-body simulations are beyond the scope of this article and we defer them to future studies. The scenario has a few elements: i) EFE effectively increases the inclinations of TNOs and much weaker the inclinations of giant planets; ii) TNOs interact with giant planets, especially strongly in epochs of low $q$; iii) giant planets alter the spin of the Sun. All elements are important and should be modelled in a self-consistent way.

Since MOND is presented here as an alternative to Planet Nine, one might ask whether both hypotheses can be correct. If gravity is Milgromian, then Planet Nine would obey chaotic evolution similar to other trans-Neptunian objects. Its perihelion would repeatedly return to the planetary region every few tens or hundreds of Myrs, depending on the semi-major axis, showing behaviour similar to that in Fig.~\ref{fig:Fig15}. The semi-major axis of Planet Nine would change each time perihelion $q \lesssim 30\,\au$, but the giant planets would also be significantly perturbed. If both hypotheses are correct, the stability of the solar system would be in question.

\begin{figure}
\centerline{
\includegraphics[width=0.4\textwidth]{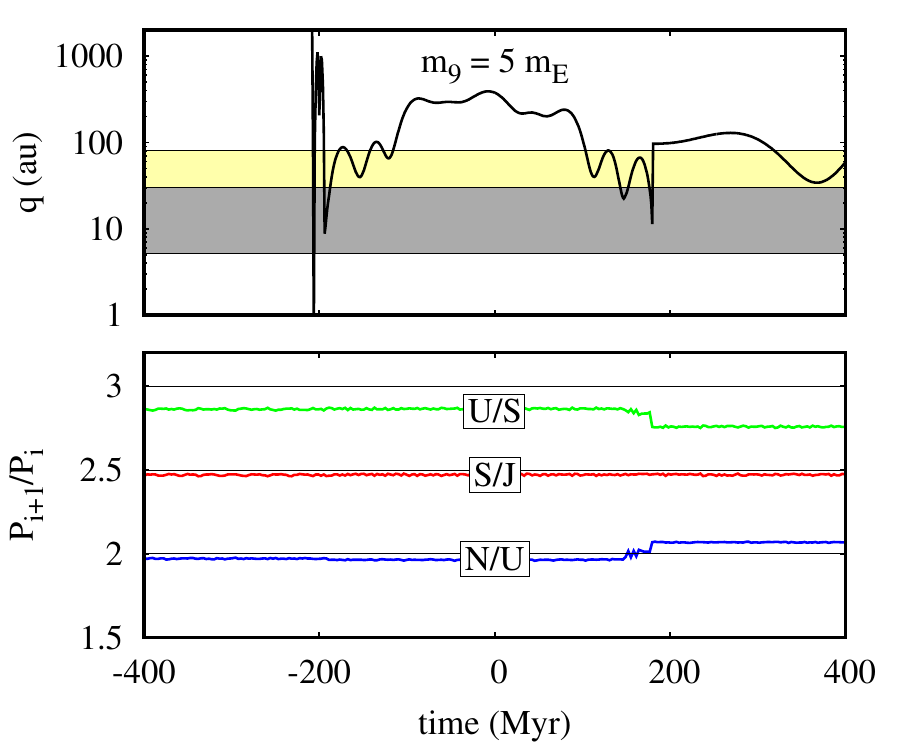}
}
\caption{Forward (red) and backward (blue) evolution of the solar system with Planet Nine with the following parameters $m=5\,\mE$, $a = 500\,\au$, $e = 0.25$, $i = 20\,$deg, $\Omega = 82\,$deg, $\omega = 158\,$deg and $\mathcal{M} = 180\,$deg. The upper panel shows the evolution of the perihelion distance of Planet Nine, while the lower panel illustrates the evolution of the period ratios of the giant planets.}
\label{fig:Fig24}
\end{figure}

An example of the backward and forward evolution of the five-planet solar system within Milgromian dynamics is shown in Fig.~\ref{fig:Fig24}. Planet Nine visits the inner solar system in epochs $\pm 200\,$Myr, which leads to significant changes in its orbit (destabilisation in epoch $-200\,$Myr) and/or the orbits of the giant planets, i.e. in epoch $+200\,$Myr the orbits of Uranus and Neptune are strongly perturbed.

\corr{However, the stability problem in the past could be overcome if Planet Nine was a free-floating planet \citep{Li2016} or a primordial black hole \citep{Scholtz2020} captured by the Sun relatively recently, i.e., less than $\sim 200\,$Myr ago. On the other hand, the future fate of the solar system could still be potentially catastrophic in this scenario.}

\section{Conclusions}
\label{sec:conclusions}

The hunt for Planet Nine continues by selecting the most likely positions in the sky \citep{Brown2021,Socas-Navarro2023}, calculating theoretical fluxes in various bands \citep{Ginzburg2016,Wright2010,PlanckColl2014,PlanckColl2016,Cowan2016,Linder2016} and conducting/planning surveys \citep{Trilling2018,Naess2021,RowanRobinson2022}. So far, Planet Nine has not been discovered. According to the flux models, the planet is detectable, but a large area of the sky must be searched with large telescopes to detect or rule out the putative planet of our solar system.

In this paper we have investigated the effects of Modified Newtonian Dynamics on the evolution and dynamical origin of extreme trans-Neptunian objects. We found that the observed objects could originate from the planetary region of the solar system, with semi-major axes on the order of a few to a few tens of astronomical units. A striking example is Sedna, which could have been in a horseshoe orbit with Jupiter and Saturn about $30\,$Myr ago. According to this scenario, ETNOs were scattered by the giant planets into orbits with high-$a$ and $q \in (5, 30)\,\au$, which were then detached from the influence of the planets by EFE. According to MOND, both ETNOs and Centaurs are ephemeral objects whose evolution is strongly chaotic.

Milgromian dynamics explains the formation of ETNOs in a natural way. Particularly interesting objects are the newly discovered retrograde ETNOs with relatively small $a$ and large $q$, which have been shown to belong to the same dynamical class as the already known prograde ETNOs and Centaurs. These objects are dynamically unconnected when Newtonian dynamics is used. MOND produces the clustering of orbital planes as well as the apsidal confinement, but these features are not particularly strong in this model. On the other hand, the observations are still inconclusive. More new objects are clearly needed to enrich the sample and confirm or rule out the anomalous features of ETNOs.

We have shown that MOND can produce solar obliquity, but the observed parameters cannot be fully reconstructed if the giant planets are only modified by EFE. We suggest that accounting for the masses of the TNOs could hopefully resolve this discrepancy.

We have considered whether the two alternative explanations for the distribution of ETNOs, namely Planet Nine and MOND, can be correct at the same time. We have shown that the dynamical stability of the solar system with Planet Nine can be problematic if the gravity is Milgromian. On the other hand, Planet Nine could have disrupted the orbits of the known giant planets and broken the chain of mean-motion resonances that are thought to have formed early in the history of the solar system, but not necessarily destabilise the entire system. A scenario in which the initially resonant and more compact solar system has been significantly disrupted by planetesimals has been proposed in the {\em Nice} model \citep{Tsiganis2005,Morbidelli2005,Gomes2005}. Naturally, if gravity is Milgromian and Planet Nine exists, the history of the solar system would have to be reconsidered. In any case, the Milgromian dynamics of the solar system with Planet Nine is much more complicated than without it and requires further investigation.

Modified Newtonian Dynamics has been considered here as an alternative for the Planet Nine hypothesis. However, MOND is also a well-known alternative to the dark matter hypothesis. Since falsification of dark matter is problematic when considering galactic and extra-galactic systems, the dynamics of the solar system can serve as a testing ground for the dark matter hypothesis. In other words: If Planet Nine is not found and the anomalous features of ETNOs are confirmed by future observations and fully explained by MOND, then the dark matter hypothesis could be in trouble.

\section*{Acknowledgements}

I would like to thank Dan Fabrycky for bringing the Planet Nine hypothesis to my attention \corr{and Dr Man Ho Chan for helpful comments that improved the paper}.

\section*{Data availability}

The orbital parameters of TNOs used in this work were taken from the JPL Small-Body Database \verb+https://ssd.jpl.nasa.gov/tools/sbdb_query.html+ (accessed 13 October 2022). The actual values used in the simulations as well as other data underlying this article will be shared on reasonable request to the corresponding author. \corr{The uncertainties of the solar system planets masses were taken from the JPL database, }\verb+https://ssd.jpl.nasa.gov/planets/phys_par.html+ \corr{ (accessed 6 July 2023).}

\bibliographystyle{mnras}
\bibliography{ms}

\begin{thebibliography}{}
\makeatletter
\relax
\def\mn@urlcharsother{\let\do\@makeother \do\$\do\&\do\#\do\^\do\_\do\%\do\~}
\def\mn@doi{\begingroup\mn@urlcharsother \@ifnextchar [ {\mn@doi@}
  {\mn@doi@[]}}
\def\mn@doi@[#1]#2{\def\@tempa{#1}\ifx\@tempa\@empty \href
  {http://dx.doi.org/#2} {doi:#2}\else \href {http://dx.doi.org/#2} {#1}\fi
  \endgroup}
\def\mn@eprint#1#2{\mn@eprint@#1:#2::\@nil}
\def\mn@eprint@arXiv#1{\href {http://arxiv.org/abs/#1} {{\tt arXiv:#1}}}
\def\mn@eprint@dblp#1{\href {http://dblp.uni-trier.de/rec/bibtex/#1.xml}
  {dblp:#1}}
\def\mn@eprint@#1:#2:#3:#4\@nil{\def\@tempa {#1}\def\@tempb {#2}\def\@tempc
  {#3}\ifx \@tempc \@empty \let \@tempc \@tempb \let \@tempb \@tempa \fi \ifx
  \@tempb \@empty \def\@tempb {arXiv}\fi \@ifundefined
  {mn@eprint@\@tempb}{\@tempb:\@tempc}{\expandafter \expandafter \csname
  mn@eprint@\@tempb\endcsname \expandafter{\@tempc}}}

\bibitem[\protect\citeauthoryear{{Anderson} \& {Kaib}}{{Anderson} \&
  {Kaib}}{2021}]{Anderson2021}
{Anderson} K.~E.,  {Kaib} N.~A.,  2021, \mn@doi [\apjl]
  {10.3847/2041-8213/ac26ca}, \href
  {https://ui.adsabs.harvard.edu/abs/2021ApJ...920L...9A} {920, L9}

\bibitem[\protect\citeauthoryear{{Arnold}, {Kozlov}  \& {Neishtadt}}{{Arnold}
  et~al.}{2006}]{Arnold2006}
{Arnold} V.~I.,  {Kozlov} V.~V.,   {Neishtadt} A.~I.,  2006, {Dynamical systems
  III. Mathematical aspects of classical and celestial mechanics}.
{Springer-Verlag, Berlin and Heidelberg}

\bibitem[\protect\citeauthoryear{{Bailey}, {Batygin}  \& {Brown}}{{Bailey}
  et~al.}{2016}]{Bailey2016}
{Bailey} E.,  {Batygin} K.,   {Brown} M.~E.,  2016, \mn@doi [\aj]
  {10.3847/0004-6256/152/5/126}, \href
  {https://ui.adsabs.harvard.edu/abs/2016AJ....152..126B} {152, 126}

\bibitem[\protect\citeauthoryear{{Batygin} \& {Brown}}{{Batygin} \&
  {Brown}}{2016a}]{Batygin2016a}
{Batygin} K.,  {Brown} M.~E.,  2016a, \mn@doi [\aj]
  {10.3847/0004-6256/151/2/22}, \href
  {https://ui.adsabs.harvard.edu/abs/2016AJ....151...22B} {151, 22}

\bibitem[\protect\citeauthoryear{{Batygin} \& {Brown}}{{Batygin} \&
  {Brown}}{2016b}]{Batygin2016b}
{Batygin} K.,  {Brown} M.~E.,  2016b, \mn@doi [\apjl]
  {10.3847/2041-8205/833/1/L3}, \href
  {https://ui.adsabs.harvard.edu/abs/2016ApJ...833L...3B} {833, L3}

\bibitem[\protect\citeauthoryear{{Batygin} \& {Brown}}{{Batygin} \&
  {Brown}}{2021}]{Batygin2021}
{Batygin} K.,  {Brown} M.~E.,  2021, \mn@doi [\apjl]
  {10.3847/2041-8213/abee1f}, \href
  {https://ui.adsabs.harvard.edu/abs/2021ApJ...910L..20B} {910, L20}

\bibitem[\protect\citeauthoryear{{Batygin}, {Adams}, {Brown}  \&
  {Becker}}{{Batygin} et~al.}{2019}]{Batygin2019}
{Batygin} K.,  {Adams} F.~C.,  {Brown} M.~E.,   {Becker} J.~C.,  2019, \mn@doi
  [\physrep] {10.1016/j.physrep.2019.01.009}, \href
  {https://ui.adsabs.harvard.edu/abs/2019PhR...805....1B} {805, 1}

\bibitem[\protect\citeauthoryear{{Beck} \& {Giles}}{{Beck} \&
  {Giles}}{2005}]{Beck2005}
{Beck} J.~G.,  {Giles} P.,  2005, \mn@doi [\apjl] {10.1086/429224}, \href
  {https://ui.adsabs.harvard.edu/abs/2005ApJ...621L.153B} {621, L153}

\bibitem[\protect\citeauthoryear{{Bekenstein} \& {Milgrom}}{{Bekenstein} \&
  {Milgrom}}{1984}]{Bekenstein1984}
{Bekenstein} J.,  {Milgrom} M.,  1984, \mn@doi [\apj] {10.1086/162570}, \href
  {https://ui.adsabs.harvard.edu/abs/1984ApJ...286....7B} {286, 7}

\bibitem[\protect\citeauthoryear{{Bernardinelli} et~al.,}{{Bernardinelli}
  et~al.}{2020}]{Bernardinelli2020}
{Bernardinelli} P.~H.,  et~al., 2020, \mn@doi [\psj] {10.3847/PSJ/ab9d80},
  \href {https://ui.adsabs.harvard.edu/abs/2020PSJ.....1...28B} {1, 28}

\bibitem[\protect\citeauthoryear{{Blanchet} \& {Novak}}{{Blanchet} \&
  {Novak}}{2011}]{Blanchet2011}
{Blanchet} L.,  {Novak} J.,  2011, \mn@doi [\mnras]
  {10.1111/j.1365-2966.2010.18076.x}, \href
  {https://ui.adsabs.harvard.edu/abs/2011MNRAS.412.2530B} {412, 2530}

\bibitem[\protect\citeauthoryear{{Brada} \& {Milgrom}}{{Brada} \&
  {Milgrom}}{1999}]{Brada1999}
{Brada} R.,  {Milgrom} M.,  1999, \mn@doi [\apj] {10.1086/307402}, \href
  {https://ui.adsabs.harvard.edu/abs/1999ApJ...519..590B} {519, 590}

\bibitem[\protect\citeauthoryear{{Bromley} \& {Kenyon}}{{Bromley} \&
  {Kenyon}}{2016}]{Bromley2016}
{Bromley} B.~C.,  {Kenyon} S.~J.,  2016, \mn@doi [\apj]
  {10.3847/0004-637X/826/1/64}, \href
  {https://ui.adsabs.harvard.edu/abs/2016ApJ...826...64B} {826, 64}

\bibitem[\protect\citeauthoryear{{Brouwer} \& {Clemence}}{{Brouwer} \&
  {Clemence}}{1961}]{Brouwer1961}
{Brouwer} D.,  {Clemence} G.~M.,  1961, {Methods of Celestial Mechanics}.
{Academic Press, New York and London}

\bibitem[\protect\citeauthoryear{{Brown}}{{Brown}}{2017}]{Brown2017}
{Brown} M.~E.,  2017, \mn@doi [\aj] {10.3847/1538-3881/aa79f4}, \href
  {https://ui.adsabs.harvard.edu/abs/2017AJ....154...65B} {154, 65}

\bibitem[\protect\citeauthoryear{{Brown} \& {Batygin}}{{Brown} \&
  {Batygin}}{2019}]{Brown2019}
{Brown} M.~E.,  {Batygin} K.,  2019, \mn@doi [\aj] {10.3847/1538-3881/aaf051},
  \href {https://ui.adsabs.harvard.edu/abs/2019AJ....157...62B} {157, 62}

\bibitem[\protect\citeauthoryear{{Brown} \& {Batygin}}{{Brown} \&
  {Batygin}}{2021}]{Brown2021}
{Brown} M.~E.,  {Batygin} K.,  2021, \mn@doi [\aj] {10.3847/1538-3881/ac2056},
  \href {https://ui.adsabs.harvard.edu/abs/2021AJ....162..219B} {162, 219}

\bibitem[\protect\citeauthoryear{{Brown} \& {Dahlke}}{{Brown} \&
  {Dahlke}}{2018}]{Brown2018}
{Brown} R.~B.,  {Dahlke} S.~R.,  2018, \mn@doi [International Journal of
  Aeronautical and Space Sciences] {10.11648/j.ijass.20180601.14}, \href
  {https://ui.adsabs.harvard.edu/abs/2018IJASS...6...38B} {6, 38}

\bibitem[\protect\citeauthoryear{{Brown}, {Trujillo}  \& {Rabinowitz}}{{Brown}
  et~al.}{2004}]{Brown2004}
{Brown} M.~E.,  {Trujillo} C.,   {Rabinowitz} D.,  2004, \mn@doi [\apj]
  {10.1086/422095}, \href
  {https://ui.adsabs.harvard.edu/abs/2004ApJ...617..645B} {617, 645}

\bibitem[\protect\citeauthoryear{{Clement} \& {Kaib}}{{Clement} \&
  {Kaib}}{2020}]{Clement2020}
{Clement} M.~S.,  {Kaib} N.~A.,  2020, \mn@doi [\aj]
  {10.3847/1538-3881/ab9227}, \href
  {https://ui.adsabs.harvard.edu/abs/2020AJ....159..285C} {159, 285}

\bibitem[\protect\citeauthoryear{{Cowan}, {Holder}  \& {Kaib}}{{Cowan}
  et~al.}{2016}]{Cowan2016}
{Cowan} N.~B.,  {Holder} G.,   {Kaib} N.~A.,  2016, \mn@doi [\apjl]
  {10.3847/2041-8205/822/1/L2}, \href
  {https://ui.adsabs.harvard.edu/abs/2016ApJ...822L...2C} {822, L2}

\bibitem[\protect\citeauthoryear{{Eriksson}, {Mustill}  \&
  {Johansen}}{{Eriksson} et~al.}{2018}]{Eriksson2018}
{Eriksson} L. E.~J.,  {Mustill} A.~J.,   {Johansen} A.,  2018, \mn@doi [\mnras]
  {10.1093/mnras/sty111}, \href
  {https://ui.adsabs.harvard.edu/abs/2018MNRAS.475.4609E} {475, 4609}

\bibitem[\protect\citeauthoryear{{Famaey} \& {McGaugh}}{{Famaey} \&
  {McGaugh}}{2012}]{Famaey2012}
{Famaey} B.,  {McGaugh} S.~S.,  2012, \mn@doi [Living Reviews in Relativity]
  {10.12942/lrr-2012-10}, \href
  {https://ui.adsabs.harvard.edu/abs/2012LRR....15...10F} {15, 10}

\bibitem[\protect\citeauthoryear{{Fienga}, {Laskar}, {Manche}  \&
  {Gastineau}}{{Fienga} et~al.}{2016}]{Fienga2016}
{Fienga} A.,  {Laskar} J.,  {Manche} H.,   {Gastineau} M.,  2016, \mn@doi
  [\aap] {10.1051/0004-6361/201628227}, \href
  {https://ui.adsabs.harvard.edu/abs/2016A&A...587L...8F} {587, L8}

\bibitem[\protect\citeauthoryear{{Ginzburg}, {Sari}  \& {Loeb}}{{Ginzburg}
  et~al.}{2016}]{Ginzburg2016}
{Ginzburg} S.,  {Sari} R.,   {Loeb} A.,  2016, \mn@doi [\apjl]
  {10.3847/2041-8205/822/1/L11}, \href
  {https://ui.adsabs.harvard.edu/abs/2016ApJ...822L..11G} {822, L11}

\bibitem[\protect\citeauthoryear{{Gomes}, {Levison}, {Tsiganis}  \&
  {Morbidelli}}{{Gomes} et~al.}{2005}]{Gomes2005}
{Gomes} R.,  {Levison} H.~F.,  {Tsiganis} K.,   {Morbidelli} A.,  2005, \mn@doi
  [\nat] {10.1038/nature03676}, \href
  {https://ui.adsabs.harvard.edu/abs/2005Natur.435..466G} {435, 466}

\bibitem[\protect\citeauthoryear{{Gomes}, {Matese}  \& {Lissauer}}{{Gomes}
  et~al.}{2006}]{Gomes2006}
{Gomes} R.~S.,  {Matese} J.~J.,   {Lissauer} J.~J.,  2006, \mn@doi [\icarus]
  {10.1016/j.icarus.2006.05.026}, \href
  {https://ui.adsabs.harvard.edu/abs/2006Icar..184..589G} {184, 589}

\bibitem[\protect\citeauthoryear{{Gomes}, {Deienno}  \& {Morbidelli}}{{Gomes}
  et~al.}{2017}]{Gomes2017}
{Gomes} R.,  {Deienno} R.,   {Morbidelli} A.,  2017, \mn@doi [\aj]
  {10.3847/1538-3881/153/1/27}, \href
  {https://ui.adsabs.harvard.edu/abs/2017AJ....153...27G} {153, 27}

\bibitem[\protect\citeauthoryear{{Hees}, {Folkner}, {Jacobson}  \&
  {Park}}{{Hees} et~al.}{2014}]{Hees2014}
{Hees} A.,  {Folkner} W.~M.,  {Jacobson} R.~A.,   {Park} R.~S.,  2014, \mn@doi
  [\prd] {10.1103/PhysRevD.89.102002}, \href
  {https://ui.adsabs.harvard.edu/abs/2014PhRvD..89j2002H} {89, 102002}

\bibitem[\protect\citeauthoryear{{Hees}, {Famaey}, {Angus}  \&
  {Gentile}}{{Hees} et~al.}{2016}]{Hees2016}
{Hees} A.,  {Famaey} B.,  {Angus} G.~W.,   {Gentile} G.,  2016, \mn@doi
  [\mnras] {10.1093/mnras/stv2330}, \href
  {https://ui.adsabs.harvard.edu/abs/2016MNRAS.455..449H} {455, 449}

\bibitem[\protect\citeauthoryear{{Hoffman}}{{Hoffman}}{2001}]{Hoffman2001}
{Hoffman} J.~D.,  2001, {Numerical Methods for Engineers and Scientists}.
{CRC Press}

\bibitem[\protect\citeauthoryear{{Holman} \& {Payne}}{{Holman} \&
  {Payne}}{2016a}]{Holman2016}
{Holman} M.~J.,  {Payne} M.~J.,  2016a, \mn@doi [\aj]
  {10.3847/0004-6256/152/4/80}, \href
  {https://ui.adsabs.harvard.edu/abs/2016AJ....152...80H} {152, 80}

\bibitem[\protect\citeauthoryear{{Holman} \& {Payne}}{{Holman} \&
  {Payne}}{2016b}]{Holman2016b}
{Holman} M.~J.,  {Payne} M.~J.,  2016b, \mn@doi [\aj]
  {10.3847/0004-6256/152/4/94}, \href
  {https://ui.adsabs.harvard.edu/abs/2016AJ....152...94H} {152, 94}

\bibitem[\protect\citeauthoryear{{Hunt}, {Bovy}  \& {Carlberg}}{{Hunt}
  et~al.}{2016}]{Hunt2016}
{Hunt} J. A.~S.,  {Bovy} J.,   {Carlberg} R.~G.,  2016, \mn@doi [\apjl]
  {10.3847/2041-8205/832/2/L25}, \href
  {https://ui.adsabs.harvard.edu/abs/2016ApJ...832L..25H} {832, L25}

\bibitem[\protect\citeauthoryear{{Iorio}}{{Iorio}}{2010}]{Iorio2010}
{Iorio} L.,  2010, \mn@doi [The Open Astronomy Journal]
  {10.2174/1874381101003010156}, \href
  {https://ui.adsabs.harvard.edu/abs/2010OAJ.....3..156I} {3, 156}

\bibitem[\protect\citeauthoryear{{Iorio}}{{Iorio}}{2017}]{Iorio2017}
{Iorio} L.,  2017, \mn@doi [\apss] {10.1007/s10509-016-2993-8}, \href
  {https://ui.adsabs.harvard.edu/abs/2017Ap&SS.362...11I} {362, 11}

\bibitem[\protect\citeauthoryear{{Jones-Smith} \& {Mathur}}{{Jones-Smith} \&
  {Mathur}}{2023}]{Jones-Smith2023}
{Jones-Smith} K.,  {Mathur} H.,  2023, \mn@doi [arXiv e-prints]
  {10.48550/arXiv.2304.00576}, \href
  {https://ui.adsabs.harvard.edu/abs/2023arXiv230400576J} {p. arXiv:2304.00576}

\bibitem[\protect\citeauthoryear{{Kaib} \& {Sheppard}}{{Kaib} \&
  {Sheppard}}{2016}]{Kaib2016}
{Kaib} N.~A.,  {Sheppard} S.~S.,  2016, \mn@doi [\aj]
  {10.3847/0004-6256/152/5/133}, \href
  {https://ui.adsabs.harvard.edu/abs/2016AJ....152..133K} {152, 133}

\bibitem[\protect\citeauthoryear{{Karim} \& {Mamajek}}{{Karim} \&
  {Mamajek}}{2017}]{Karim2017}
{Karim} T.,  {Mamajek} E.~E.,  2017, \mn@doi [\mnras] {10.1093/mnras/stw2772},
  \href {https://ui.adsabs.harvard.edu/abs/2017MNRAS.465..472K} {465, 472}

\bibitem[\protect\citeauthoryear{{Kenyon} \& {Bromley}}{{Kenyon} \&
  {Bromley}}{2016}]{Kenyon2016}
{Kenyon} S.~J.,  {Bromley} B.~C.,  2016, \mn@doi [\apj]
  {10.3847/0004-637X/825/1/33}, \href
  {https://ui.adsabs.harvard.edu/abs/2016ApJ...825...33K} {825, 33}

\bibitem[\protect\citeauthoryear{{Kozai}}{{Kozai}}{1962}]{Kozai1962}
{Kozai} Y.,  1962, \mn@doi [\aj] {10.1086/108790}, \href
  {https://ui.adsabs.harvard.edu/abs/1962AJ.....67..591K} {67, 591}

\bibitem[\protect\citeauthoryear{{Lai}}{{Lai}}{2016}]{Lai2016}
{Lai} D.,  2016, \mn@doi [\aj] {10.3847/0004-6256/152/6/215}, \href
  {https://ui.adsabs.harvard.edu/abs/2016AJ....152..215L} {152, 215}

\bibitem[\protect\citeauthoryear{{Lelli}, {McGaugh}, {Schombert}  \&
  {Pawlowski}}{{Lelli} et~al.}{2017}]{Lelli2017}
{Lelli} F.,  {McGaugh} S.~S.,  {Schombert} J.~M.,   {Pawlowski} M.~S.,  2017,
  \mn@doi [\apj] {10.3847/1538-4357/836/2/152}, \href
  {https://ui.adsabs.harvard.edu/abs/2017ApJ...836..152L} {836, 152}

\bibitem[\protect\citeauthoryear{{Li} \& {Adams}}{{Li} \&
  {Adams}}{2016}]{Li2016}
{Li} G.,  {Adams} F.~C.,  2016, \mn@doi [\apjl] {10.3847/2041-8205/823/1/L3},
  \href {https://ui.adsabs.harvard.edu/abs/2016ApJ...823L...3L} {823, L3}

\bibitem[\protect\citeauthoryear{{Lidov}}{{Lidov}}{1962}]{Lidov1962}
{Lidov} M.~L.,  1962, \mn@doi [\planss] {10.1016/0032-0633(62)90129-0}, \href
  {https://ui.adsabs.harvard.edu/abs/1962P&SS....9..719L} {9, 719}

\bibitem[\protect\citeauthoryear{{Linder} \& {Mordasini}}{{Linder} \&
  {Mordasini}}{2016}]{Linder2016}
{Linder} E.~F.,  {Mordasini} C.,  2016, \mn@doi [\aap]
  {10.1051/0004-6361/201628350}, \href
  {https://ui.adsabs.harvard.edu/abs/2016A&A...589A.134L} {589, A134}

\bibitem[\protect\citeauthoryear{{L{\'o}pez-Corredoira}, {Betancort-Rijo},
  {Scarpa}  \& {Chrob{\'a}kov{\'a}}}{{L{\'o}pez-Corredoira}
  et~al.}{2022}]{LopezCorredoira2022}
{L{\'o}pez-Corredoira} M.,  {Betancort-Rijo} J.~E.,  {Scarpa} R.,
  {Chrob{\'a}kov{\'a}} {\v{Z}}.,  2022, \mn@doi [\mnras]
  {10.1093/mnras/stac3117}, \href
  {https://ui.adsabs.harvard.edu/abs/2022MNRAS.517.5734L} {517, 5734}

\bibitem[\protect\citeauthoryear{{Madigan} \& {McCourt}}{{Madigan} \&
  {McCourt}}{2016}]{Madigan2016}
{Madigan} A.-M.,  {McCourt} M.,  2016, \mn@doi [\mnras]
  {10.1093/mnrasl/slv203}, \href
  {https://ui.adsabs.harvard.edu/abs/2016MNRAS.457L..89M} {457, L89}

\bibitem[\protect\citeauthoryear{{Matese}, {Whitman}, {Innanen}  \&
  {Valtonen}}{{Matese} et~al.}{1995}]{Matese1995}
{Matese} J.~J.,  {Whitman} P.~G.,  {Innanen} K.~A.,   {Valtonen} M.~J.,  1995,
  \mn@doi [\icarus] {10.1006/icar.1995.1124}, \href
  {https://ui.adsabs.harvard.edu/abs/1995Icar..116..255M} {116, 255}

\bibitem[\protect\citeauthoryear{{McGaugh}, {Lelli}  \& {Schombert}}{{McGaugh}
  et~al.}{2016}]{McGaugh2016}
{McGaugh} S.~S.,  {Lelli} F.,   {Schombert} J.~M.,  2016, \mn@doi [\prl]
  {10.1103/PhysRevLett.117.201101}, \href
  {https://ui.adsabs.harvard.edu/abs/2016PhRvL.117t1101M} {117, 201101}

\bibitem[\protect\citeauthoryear{{Milgrom}}{{Milgrom}}{1983a}]{Milgrom1983}
{Milgrom} M.,  1983a, \mn@doi [\apj] {10.1086/161130}, \href
  {https://ui.adsabs.harvard.edu/abs/1983ApJ...270..365M} {270, 365}

\bibitem[\protect\citeauthoryear{{Milgrom}}{{Milgrom}}{1983b}]{Milgrom1983b}
{Milgrom} M.,  1983b, \mn@doi [\apj] {10.1086/161131}, \href
  {https://ui.adsabs.harvard.edu/abs/1983ApJ...270..371M} {270, 371}

\bibitem[\protect\citeauthoryear{{Milgrom}}{{Milgrom}}{1989}]{Milgrom1989}
{Milgrom} M.,  1989, \mn@doi [\apj] {10.1086/167184}, \href
  {https://ui.adsabs.harvard.edu/abs/1989ApJ...338..121M} {338, 121}

\bibitem[\protect\citeauthoryear{{Milgrom}}{{Milgrom}}{2009}]{Milgrom2009}
{Milgrom} M.,  2009, \mn@doi [\mnras] {10.1111/j.1365-2966.2009.15302.x}, \href
  {https://ui.adsabs.harvard.edu/abs/2009MNRAS.399..474M} {399, 474}

\bibitem[\protect\citeauthoryear{{Milgrom}}{{Milgrom}}{2010}]{Milgrom2010}
{Milgrom} M.,  2010, \mn@doi [\mnras] {10.1111/j.1365-2966.2009.16184.x}, \href
  {https://ui.adsabs.harvard.edu/abs/2010MNRAS.403..886M} {403, 886}

\bibitem[\protect\citeauthoryear{{Morbidelli}, {Levison}, {Tsiganis}  \&
  {Gomes}}{{Morbidelli} et~al.}{2005}]{Morbidelli2005}
{Morbidelli} A.,  {Levison} H.~F.,  {Tsiganis} K.,   {Gomes} R.,  2005, \mn@doi
  [\nat] {10.1038/nature03540}, \href
  {https://ui.adsabs.harvard.edu/abs/2005Natur.435..462M} {435, 462}

\bibitem[\protect\citeauthoryear{{Murray} \& {Dermott}}{{Murray} \&
  {Dermott}}{1999}]{Murray1999}
{Murray} C.~D.,  {Dermott} S.~F.,  1999, {Solar System Dynamics}.
{Cambridge University Press, New York}

\bibitem[\protect\citeauthoryear{{Naess} et~al.,}{{Naess}
  et~al.}{2021}]{Naess2021}
{Naess} S.,  et~al., 2021, \mn@doi [\apj] {10.3847/1538-4357/ac2307}, \href
  {https://ui.adsabs.harvard.edu/abs/2021ApJ...923..224N} {923, 224}

\bibitem[\protect\citeauthoryear{{Napier} et~al.,}{{Napier}
  et~al.}{2021}]{Napier2021}
{Napier} K.~J.,  et~al., 2021, \mn@doi [\psj] {10.3847/PSJ/abe53e}, \href
  {https://ui.adsabs.harvard.edu/abs/2021PSJ.....2...59N} {2, 59}

\bibitem[\protect\citeauthoryear{{Nesvorn{\'y}}, {Vokrouhlick{\'y}}  \&
  {Roig}}{{Nesvorn{\'y}} et~al.}{2016}]{Nesvorny2016}
{Nesvorn{\'y}} D.,  {Vokrouhlick{\'y}} D.,   {Roig} F.,  2016, \mn@doi [\apjl]
  {10.3847/2041-8205/827/2/L35}, \href
  {https://ui.adsabs.harvard.edu/abs/2016ApJ...827L..35N} {827, L35}

\bibitem[\protect\citeauthoryear{{Pau{\v{c}}o}}{{Pau{\v{c}}o}}{2017}]{Pauco2017}
{Pau{\v{c}}o} R.,  2017, \mn@doi [\aap] {10.1051/0004-6361/201630335}, \href
  {https://ui.adsabs.harvard.edu/abs/2017A&A...603A..11P} {603, A11}

\bibitem[\protect\citeauthoryear{{Pau{\v{c}}o} \& {Kla{\v{c}}ka}}{{Pau{\v{c}}o}
  \& {Kla{\v{c}}ka}}{2016}]{Pauco2016}
{Pau{\v{c}}o} R.,  {Kla{\v{c}}ka} J.,  2016, \mn@doi [\aap]
  {10.1051/0004-6361/201527713}, \href
  {https://ui.adsabs.harvard.edu/abs/2016A&A...589A..63P} {589, A63}

\bibitem[\protect\citeauthoryear{{Pau{\v{c}}o} \& {Kla{\v{c}}ka}}{{Pau{\v{c}}o}
  \& {Kla{\v{c}}ka}}{2017}]{Pauco2017b}
{Pau{\v{c}}o} R.,  {Kla{\v{c}}ka} J.,  2017, \mn@doi [arXiv e-prints]
  {10.48550/arXiv.1705.09273}, \href
  {https://ui.adsabs.harvard.edu/abs/2017arXiv170509273P} {p. arXiv:1705.09273}

\bibitem[\protect\citeauthoryear{{Planck Collaboration} et~al.,}{{Planck
  Collaboration} et~al.}{2014}]{PlanckColl2014}
{Planck Collaboration} et~al., 2014, \mn@doi [\aap]
  {10.1051/0004-6361/201321524}, \href
  {https://ui.adsabs.harvard.edu/abs/2014A&A...571A..28P} {571, A28}

\bibitem[\protect\citeauthoryear{{Planck Collaboration} et~al.,}{{Planck
  Collaboration} et~al.}{2016}]{PlanckColl2016}
{Planck Collaboration} et~al., 2016, \mn@doi [\aap]
  {10.1051/0004-6361/201526914}, \href
  {https://ui.adsabs.harvard.edu/abs/2016A&A...594A..26P} {594, A26}

\bibitem[\protect\citeauthoryear{{Rowan-Robinson}}{{Rowan-Robinson}}{2022}]{RowanRobinson2022}
{Rowan-Robinson} M.,  2022, \mn@doi [\mnras] {10.1093/mnras/stab3212}, \href
  {https://ui.adsabs.harvard.edu/abs/2022MNRAS.510.3716R} {510, 3716}

\bibitem[\protect\citeauthoryear{{Sanders}}{{Sanders}}{1999}]{Sanders1999}
{Sanders} R.~H.,  1999, \mn@doi [\apjl] {10.1086/311865}, \href
  {https://ui.adsabs.harvard.edu/abs/1999ApJ...512L..23S} {512, L23}

\bibitem[\protect\citeauthoryear{{Sanders}}{{Sanders}}{2010}]{Sanders2010}
{Sanders} R.~H.,  2010, {The Dark Matter Problem: A Historical Perspective}.
{Cambridge University Press}

\bibitem[\protect\citeauthoryear{{Scholtz} \& {Unwin}}{{Scholtz} \&
  {Unwin}}{2020}]{Scholtz2020}
{Scholtz} J.,  {Unwin} J.,  2020, \mn@doi [\prl]
  {10.1103/PhysRevLett.125.051103}, \href
  {https://ui.adsabs.harvard.edu/abs/2020PhRvL.125e1103S} {125, 051103}

\bibitem[\protect\citeauthoryear{{Sefilian} \& {Touma}}{{Sefilian} \&
  {Touma}}{2019}]{Sefilian2019}
{Sefilian} A.~A.,  {Touma} J.~R.,  2019, \mn@doi [\aj]
  {10.3847/1538-3881/aaf0fc}, \href
  {https://ui.adsabs.harvard.edu/abs/2019AJ....157...59S} {157, 59}

\bibitem[\protect\citeauthoryear{{Shankman}, {Kavelaars}, {Lawler}, {Gladman}
  \& {Bannister}}{{Shankman} et~al.}{2017}]{Shankman2017}
{Shankman} C.,  {Kavelaars} J.~J.,  {Lawler} S.~M.,  {Gladman} B.~J.,
  {Bannister} M.~T.,  2017, \mn@doi [\aj] {10.3847/1538-3881/153/2/63}, \href
  {https://ui.adsabs.harvard.edu/abs/2017AJ....153...63S} {153, 63}

\bibitem[\protect\citeauthoryear{{Sheppard} \& {Trujillo}}{{Sheppard} \&
  {Trujillo}}{2016}]{Sheppard2016}
{Sheppard} S.~S.,  {Trujillo} C.,  2016, \mn@doi [\aj]
  {10.3847/1538-3881/152/6/221}, \href
  {https://ui.adsabs.harvard.edu/abs/2016AJ....152..221S} {152, 221}

\bibitem[\protect\citeauthoryear{{Socas-Navarro}}{{Socas-Navarro}}{2023}]{Socas-Navarro2023}
{Socas-Navarro} H.,  2023, \mn@doi [\apj] {10.3847/1538-4357/acb817}, \href
  {https://ui.adsabs.harvard.edu/abs/2023ApJ...945...22S} {945, 22}

\bibitem[\protect\citeauthoryear{{Souami} \& {Souchay}}{{Souami} \&
  {Souchay}}{2012}]{Souami2012}
{Souami} D.,  {Souchay} J.,  2012, \mn@doi [\aap]
  {10.1051/0004-6361/201219011}, \href
  {https://ui.adsabs.harvard.edu/abs/2012A&A...543A.133S} {543, A133}

\bibitem[\protect\citeauthoryear{{Torres-Flores}, {Epinat}, {Amram}, {Plana}
  \& {Mendes de Oliveira}}{{Torres-Flores} et~al.}{2011}]{TorresFlores2011}
{Torres-Flores} S.,  {Epinat} B.,  {Amram} P.,  {Plana} H.,   {Mendes de
  Oliveira} C.,  2011, \mn@doi [\mnras] {10.1111/j.1365-2966.2011.19169.x},
  \href {https://ui.adsabs.harvard.edu/abs/2011MNRAS.416.1936T} {416, 1936}

\bibitem[\protect\citeauthoryear{{Trilling}, {Bellm}  \& {Malhotra}}{{Trilling}
  et~al.}{2018}]{Trilling2018}
{Trilling} D.~E.,  {Bellm} E.~C.,   {Malhotra} R.,  2018, \mn@doi [\aj]
  {10.3847/1538-3881/aabfc0}, \href
  {https://ui.adsabs.harvard.edu/abs/2018AJ....155..243T} {155, 243}

\bibitem[\protect\citeauthoryear{{Trujillo} \& {Sheppard}}{{Trujillo} \&
  {Sheppard}}{2014}]{Trujillo2014}
{Trujillo} C.~A.,  {Sheppard} S.~S.,  2014, \mn@doi [\nat]
  {10.1038/nature13156}, \href
  {https://ui.adsabs.harvard.edu/abs/2014Natur.507..471T} {507, 471}

\bibitem[\protect\citeauthoryear{{Tsiganis}, {Gomes}, {Morbidelli}  \&
  {Levison}}{{Tsiganis} et~al.}{2005}]{Tsiganis2005}
{Tsiganis} K.,  {Gomes} R.,  {Morbidelli} A.,   {Levison} H.~F.,  2005, \mn@doi
  [\nat] {10.1038/nature03539}, \href
  {https://ui.adsabs.harvard.edu/abs/2005Natur.435..459T} {435, 459}

\bibitem[\protect\citeauthoryear{{Tully} \& {Fisher}}{{Tully} \&
  {Fisher}}{1977}]{Tully1977}
{Tully} R.~B.,  {Fisher} J.~R.,  1977, \aap, \href
  {https://ui.adsabs.harvard.edu/abs/1977A&A....54..661T} {54, 661}

\bibitem[\protect\citeauthoryear{{Volk} \& {Malhotra}}{{Volk} \&
  {Malhotra}}{2017}]{Volk2017}
{Volk} K.,  {Malhotra} R.,  2017, \mn@doi [\aj] {10.3847/1538-3881/aa79ff},
  \href {https://ui.adsabs.harvard.edu/abs/2017AJ....154...62V} {154, 62}

\bibitem[\protect\citeauthoryear{{Wright} et~al.,}{{Wright}
  et~al.}{2010}]{Wright2010}
{Wright} E.~L.,  et~al., 2010, \mn@doi [\aj] {10.1088/0004-6256/140/6/1868},
  \href {https://ui.adsabs.harvard.edu/abs/2010AJ....140.1868W} {140, 1868}

\bibitem[\protect\citeauthoryear{{de Blok} \& {McGaugh}}{{de Blok} \&
  {McGaugh}}{1997}]{deBlok1997}
{de Blok} W.~J.~G.,  {McGaugh} S.~S.,  1997, \mn@doi [\mnras]
  {10.1093/mnras/290.3.533}, \href
  {https://ui.adsabs.harvard.edu/abs/1997MNRAS.290..533D} {290, 533}

\makeatother
\end{thebibliography}

\bsp	
\label{lastpage}
\end{document}